\documentclass[a4paper,11pt]{article}

\usepackage[margin=1in]{geometry} 

\usepackage{amsmath}
\usepackage{amsthm}
\usepackage{amssymb}

\usepackage{graphicx}
\usepackage{bm}
\usepackage{underoverlap}
\usepackage{lineno}
\usepackage{caption}
\usepackage{subcaption}

\usepackage[utf8]{inputenc}
\usepackage{hyperref}
\hypersetup{
	unicode,
	colorlinks=true,
	linkcolor=blue,
	filecolor=magenta,      
	citecolor=blue,
	urlcolor=cyan,
	pdfauthor={M.A. Mendez, D. Hess, B.B. Watz, J.-M. Buchlin},
	pdftitle={Multiscale Proper Orthogonal Decomposition (mPOD) of TR-PIV data-- a Case Study on Stationary and Transient Cylinder Wake Flows},
	pdfsubject={Multiscale Proper Orthogonal Decomposition (mPOD) of TR-PIV data-- a Case Study on Stationary and Transient Cylinder Wake Flows},
	pdfkeywords={article, template, simple},
	pdfproducer={LaTeX},
	pdfcreator={pdflatex}
	colorlinks=true,
	citecolor=blue,
	linkcolor=blue,
	filecolor=magenta,      
	urlcolor=blue,
}

\usepackage[sort&compress,numbers,round]{natbib}
\setcitestyle{authoryear,open={(},close={)}}

\usepackage{algorithm, algpseudocode} 
\usepackage{mathrsfs} 

\title{Multiscale Proper Orthogonal Decomposition (mPOD) of TR-PIV data-- a Case Study on Stationary and Transient Cylinder Wake Flows}
\author{M A Mendez$^1$ \and D Hess$^2$ \and 
B B Watz$^2$ \and J-M Buchlin$^1$}

\date{
	$^1$von Karman Institute for Fluid Dynamics, Sint-Genesius-Rode, Belgium \\ \texttt{mendez@vki.ac.be}\\%
	$^2$Dantec Dynamics AS, Denmark%
}

\begin{document}
	\maketitle
	
	\begin{abstract}
	Data-driven decompositions of Particle Image Velocimetry (PIV) measurements are widely used for a variety of purposes, including the detection of coherent features (e.g., vortical structures), filtering operations (e.g., outlier removal or random noise mitigation), data reduction and compression. This work presents the application of a novel decomposition method, referred to as Multiscale Proper Orthogonal Decomposition (mPOD, Mendez \emph{et al} 2019) to Time-Resolved PIV (TR-PIV) measurement. This method combines Multiresolution Analysis (MRA) and standard Proper Orthogonal Decomposition (POD) to achieve a compromise between decomposition convergence and spectral purity of the resulting modes.
	The selected test case is the flow past a cylinder in both stationary and transient conditions, producing a frequency-varying Karman vortex street. The results of the mPOD are compared to the standard POD, the Discrete Fourier Transform (DFT) and the Dynamic Mode Decomposition (DMD). The mPOD is evaluated in terms of decomposition convergence and time-frequency localization of its modes. The multiscale modal analysis allows for revealing beat phenomena in the stationary cylinder wake, due to the three-dimensional nature of the flow, and to correctly identify the transition from various stationary regimes in the transient test case. 
		
		\vspace{7mm}
		
	\end{abstract}


\section{Introduction and Motivation}

Data-driven decompositions of PIV measurements are nowadays part of the standard toolbox for post-processing techniques used to extract knowledge from data. Initially developed for identifying coherent structures in turbulent flows \citep{Lumley1,Berkooz1993,Siro1} and for reduced-order modeling \citep{Holmes,Holmes1996}, these decompositions have found applications also as random noise removal \citep{Raiola2015}, image pre-processing tools \citep{Mendez2017} and for validation of numerical simulations \citep{Meyer}.

The scope of any decomposition is to describe the data as a linear combination of elementary portions referred to as \emph{modes}. Each mode has its own spatial structure and temporal evolution and is potentially capable of describing essential features of the data. The removal of irrelevant modes enables filtering and data compression, while the dominant modes offer a basis for constructing reduced models capable of significantly reducing the computational burdens in the simulations of large dynamical systems. Modal decompositions and reduced-order models are thus fundamentals in the development of control laws \citep{Berger_2014,Brunton2015}.

Data-driven decompositons and modal analysis are reviewed by \cite{Taira,Rowley,Amor2019,Benner2015}. The most common approaches can be broadly classified into two main classes: energy-based and frequency-based.

Energy-based decompositions arise from the Proper Orthogonal Decomposition (POD, \cite{Siro1,Siro2,Lumley1}) and its variants. The POD provides the most energetic modes with no constraints on their frequency content. Variants of the POD can be constructed from different choices of the inner product or in the use of different averaging procedures in the computation of the data correlations.
Examples of the first variants are proposed by \cite{Maurel12001,Rowley2004,Lumley1997}, where multiple quantities are involved in the inner product. Examples of the second variants are proposed by \cite{CITRINITI2000} and \cite{Towne2018}, where the correlation matrix is computed in the frequency domain using time averaging over short windows, following the popular Welch's periodogram method \cite{Welch1967}. IThe POD is known in other fields as Empirical Orthogonal Functions (EOF) decomposition or Principal Component Analysis (PCA). A review of these formulations in climatology is provided by  \cite{Hannachi2007} and \cite{Ghil2002}.

Frequency-based methods arise from the assumption that a linear dynamical system can represent the dataset. The most common formulation in the fluid mechanics community is the Dynamic Mode Decomposition (DMD) introduced by \cite{Schmid} and \cite{Rowley2}, although analogous formulations (see also \citealt{Kutz2014}) were introduced in the late '80s in climatology under the names of Principal Oscillation Patterns (POP, \citealt{Hasselmann1988,Storch1990}) or Linear Inverse Modeling (LIM, \citealt{Penland1996,LIM1}). POP also inspired Oscillating Pattern Decomposition (OPD, \citealt{Vlacav1, Vlacav2, Vlacav3}) implemented in the Dantec software DynamicStudio (\citealt{Bo1,Bo2}). 

Both constraints of energy optimality and spectral purity can become unnecessarily extreme, as discussed by \cite{Mendez2019}. Energy-based methods fail to separate phenomena that occur at widely different frequencies (scales) and yet have comparable energetic contributions. Frequency-based methods face problems for datasets that are far from stationary and feature nonlinear phenomena such as frequency modulation, jitter, or impulsive events. 

Significant effort has been placed in the development of hybrid methods, to complement the limitations of energy-based and frequency-based formulations, as recently discussed by \cite{Focus}. Examples of successful hybrid methods are the Spectral Proper Orthogonal Decomposition proposed by \cite{SPOD}, the multiresolution Dynamic Mode Decomposition by \cite{MultiDMD}, the Recursive Dynamic Mode Decomposition by \citep{Noack_RDMD} or Cronos-Koopman analysis by \citep{Camilleri}. These methods propose ingenious combinations of POD and DMD and are designed for data sets that are statistically stationary (hinging on the time-frequency duality of the POD) or have short-duration departures from fixed points (hinging on the linearization of the dynamics in the DMD). 

For datasets featuring transient evolution between different states and generally nonlinear dynamics, these methods lose their theoretical foundations and yield poor feature detection capabilities. The Multiscale Proper Orthogonal Decomposition (mPOD) proposed by \cite{Mendez2019,Mendez_Journal_2} is a hybrid decomposition that requires neither assumptions of stationary data nor linear dynamics. The mPOD combines Multi-resolution Analysis (MRA) via filter banks and standard POD to produce modes that are optimal within a certain range of frequencies (scales). A brief description of the decomposition is presented in Section \ref{mPOD}.

The aim of this work is to test the mPOD on two challenging experimental test cases and compare it to classical energy-based and frequency-based tools. These investigated cases consist of Time-Resolved Particle Image Velocimetry (TR-PIV) of the turbulent flow past a cylinder in stationary and transient conditions, with varying free stream velocity and hence varying frequency of the vortex shedding. While the stationary test case is one of the most classical paradigms for wake flows exhibiting large scale vortex shedding (see \cite{Williamson1996} for a review), the transient configuration in turbulent conditions has received considerably less attention. Previous investigations on data-driven decomposition of the transient cylinder focus on the onset of the vortex shedding, both for reduced-order modeling \citep{NOACK2003,SIEGEL2008,Noack_RDMD,Murata2019,Pawar2019} and flow control applications \citep{Bergmann2008,Gronskis2009,Rabault2019}. 

It is worth highlighting that the cylinder wake flow at the Reynolds number investigated ($Re=[2600-4000]$) is inherently three-dimensional, characterized by the interaction of quasi two-dimensional structures with predominantly span-wise vorticity and longitudinal structures in the transverse plane (see \cite{Hussain1987,Wu1996,Chen2017}). A detailed experimental characterization of this interaction is out of the scope of this work (interested readers are referred to \cite{Sung2001,Zhou2003,Huang2006,Zhang2000}), which focuses on the assessment of the enhanced feature detection capabilities of the proposed mPOD.

The rest of the article is structured as follows. Section \ref{mPOD} briefly reviews the general formulation of energy-based and frequency-based methods, and introduces the mPOD. Section \ref{EXP} presents the experimental setup and selected test cases, while Section \ref{RES} collects the results. Conclusions and perspectives are reported in Section \ref{CON}.

\section{Data-Driven Decompositions: POD, DMD/OPD and mPOD}\label{mPOD}

The investigated dataset is a set of 2D velocity fields ${\vec{\bm{u}}}(\bm{x}_i,t_k)=(\bm{u}(\bm{x}_i,t),\bm{v}(\bm{x}_i,t))$ over a uniform grid $\bm{x}_i\in\mathbb{R}^{n_x\times n_y}$, with $i$ a matrix linear index, and a temporal discretization $t_k=\{(k-1)\Delta t\}^{k=n_t}_{k=1}$ with $\Delta t=1/f_s$ the time step and $f_s$ the sampling frequency. Bold fonts are used to indicate matrices.

All the analyzed data-driven decompositions break a discrete dataset into a linear combination of modes. These have a spatial structure $\bm{\vec{\phi}}_{k} (\bm{x}_i)$, a temporal structure $\psi_{r}(t_k)$, and an amplitude $\sigma_{r}$. The sampled velocity field is thus expanded as 

\begin{equation}
\label{EQ1}
\vec{\bm{u}}(\bm{x}_i,t_k)=\sum^{R}_{r=1}\,\sigma_{r}\,\bm{\vec{\phi}}_{r} (\bm{x}_i)\,\psi_{r} (t_k)\,
\end{equation}

A truncation of the summation to $\tilde{R}<min(n_s,n_t)$ modes, assuming that these are sorted in descending order of energy contribution, defines the best $\tilde{R}$ approximation of the data for a specific decomposition. The convergence of the decomposition as a function of the number of modes included can be monitored in terms of relative $L_2$ error in both space and time:

\begin{equation}
\label{ERROR}
\mathcal{E}(\tilde{R})=\frac{||\vec{u}(\mathbf{x}_i,t_k)-\sum^{\tilde{R}}_{r=1}\,\sigma_{r}\,\bm{\vec{\phi}}_{r} (\bm{x}_i)\,\psi_{r} (t_k)||_2}{||\vec{u}(\mathbf{x}_i,t_k)||_2}    
\end{equation}

The decompositions used in this work, namely the POD, the DMD/OPD, and the mPOD are hereinafter distinguished using the subscripts $\mathcal{P}$, $\mathcal{D}$ and $\mathcal{M}$ respectively.

\subsection{Energy-Based Formalism: the POD}

The Proper Orthogonal Decomposition arises by minimizing the error in \eqref{ERROR} $\forall \tilde{R}\in[1,R]$ or, equivalently, by maximizing the energy contribution of every mode.  
Both optimization problems are solved (see \cite{Holmes} or \cite{Bishop2011} in the framework of Principal Component Analysis, PCA) by taking the temporal structures as eigenvectors of the temporal correlation matrix. This matrix is defined as $\bm{K}[k,n]=\langle \vec{\bm{u}}(\bm{x}_i,t_k),\vec{\bm{u}}(\bm{x}_i,t_n)\rangle_T$, with $\langle \bullet\rangle_T$ the inner product in the time domain. Hence, the temporal structures of the POD satisfy the eigenvalue problem

\begin{equation}
\label{EIG}
\bm{K}\,\psi_{\mathcal{P}r} (t_k)=\lambda_r\,\psi_{\mathcal{P}r} (t_k)\quad \forall r\,\in[1,R]\longrightarrow \bm{K}=\sum^{R}_{r=1} \lambda_r\,\psi_{\mathcal{P}r} (t_k)\psi^T_{\mathcal{P}r} (t_k)\,
\end{equation} having considered each temporal structure as a column vector, and having recalled, in the last summation, that the correlation matrix is symmetric and positive definite. The POD structures are, therefore, orthonormal.

Furthermore, one could show that ${\lambda}_r=\sigma_{\mathcal{P}r}^2$ and $\mathcal{E}(\tilde{R})=\sigma_{\mathcal{P}_{\tilde{R}+1}}$. The normalization step in the calculation of the spatial structures in \eqref{EQ1} is thus no longer necessary, and the corresponding spatial structures can be computed, hinging on the orthonormality of the temporal basis, using the inner product in the time domain, that is 

\begin{equation}
\label{Siro}
\bm{\vec{\phi}}_{k} (\bm{x}_i)=\frac{1}{\sigma_r}\biggl \langle \vec{\bm{u}}(\bm{x}_i,t_k), \psi_{\mathcal{P}r}(t_k) \biggr \rangle_T=\frac{1}{\sigma_r} \sum^{n_t}_{k=1} \vec{\bm{u}}(\bm{x}_i,t_k)\,\psi_{\mathcal{P}r}(t_k)\,.
\end{equation}

It is possible to show that these spatial structures are by construction also orthonormal and eigenvectors of the spatial correlation matrix $\mathbf{C}[{i},n]=\langle \vec{\bm{u}}(\bm{x}_i,t_k),\vec{\bm{u}}(\bm{x_n},t_k)\rangle_S$, with $\langle \bullet\rangle_S$ the inner product in the space domain. The computation of the POD modes from the eigenvalue decomposition of $\mathbf{C}$ was originally presented by \cite{Lumley1} and is often referred to as \emph{standard} POD. The computation of the spatial structures from the temporal ones via \eqref{EIG} and \eqref{Siro} was proposed by \cite{Siro1}  and is often referred to as \emph{snapshot} POD.
The equivalence of the two approaches was first shown by \cite{Aubry1991}.

If the Euclidian inner product is used in both space and time, the POD is equivalent to a Singular Value Decomposition (SVD) of the dataset matrix $\bm{D}=[d_1,d_2,\dots d_{n_t}]\in\mathbb{R}^{n_s\times n_t}$, obtained by flattening each of the $k$ snapshots into a column vector $d_k\in \mathbb{R}^{n_s\times 1}$. Here $n_s=2 n_x n_y$ denotes the dimension of the snapshot and the factor $2$ accounts for the vector nature of the 2D velocity field. The SVD is thus introduced as $\bm{D}=\bm{\Phi}_{\mathcal{P}}\,\bm{\Sigma}_{\mathcal{P}}\,\bm{\Psi}_{\mathcal{P}}^T$, where $\bm{\Phi}_{\mathcal{P}}=[\phi_{\mathcal{P}1}[\mathbf {i}],\dots, \phi_{\mathcal{P}R}[\mathbf{i}]]\in \mathbb{R}^{n_s\times R}$ is the matrix of spatial structures, $\bm{\Sigma}_\mathcal{P}=diag[\sigma_{\mathcal{P}1},\dots,\sigma_{\mathcal{P}R}]$ is a diagonal matrix containing the amplitudes of each mode, and $\bm{\Psi}_\mathcal{P}=[\psi_{\mathcal{P}}[k],\dots, \psi_{\mathcal{P}R}[k]]\in \mathbb{R}^{n_t\times R}$ is the matrix of temporal structures. In what follows, a tilde is used to denote basis matrices that are reduced, i.e. include only the columns corresponding to the first $\tilde{R}$ dominant modes.

\subsection{Frequency-Based Formalism: The DMD and the OPD}\label{DMD_OPD}

In the Dynamic Mode Decomposition (DMD) and the Oscillatory Pattern Decomposition (OPD), the temporal structures are complex exponentials with complex frequencies. The temporal structures are thus constructed as powers of a complex number $\lambda_r\in\mathbb{C}$. Starting the time discretization  such that $t_0=0$ for $k=1$, these structures can be written as follows:

\begin{equation}
\label{DMD_STRUCT}
\begin{split}
\psi_{\mathcal{D}r }(t_k)=\lambda_r^{k-1}=\bigl(|\lambda_r|e^{\mathrm{i}\theta_r}\bigr)^{(k-1)}=|\lambda_r|^{(k-1)}e^{\mathrm{i}\theta_r (k-1)}=\\=|\lambda_r|^{(k-1)}\,e^{\mathrm{i}2\pi\,f_r \Delta t\, (k-1)}=|\lambda_r|^{(k-1)}\,e^{\mathrm{i}2\pi\,f_r t_k}\,,
\end{split} 
\end{equation} where  $\theta_r=arg(\lambda_r)$ is the argument of the complex number $\lambda_r$ and the oscillation frequency associated to each mode is $f_r=\theta_r/(2\pi\,\Delta t)$. 

In the Discrete Fourier Transform (DFT), the temporal basis is obtained by choosing the temporal structures as complex numbers of unitary modulus and frequency taken as multiple of a fundamental tone, that is $\psi_r[k]=exp(\mathrm{i} \, 2\pi r \Delta f/ f_s \,k)$ with $f_s=1/\Delta t$ the sampling frequency. The classical DFT (see, for instance, \cite{DFT}), the fundamental tone is taken as $\Delta f=1/T$ with $T$ the time span of the dataset. This is equivalent to assume periodicity of the data, and the resulting matrix of the temporal structure is the well known Fourier Matrix. The temporal DFT is equivalent to the DMD based on the Companion matrix (see \cite{Rowley2}) from which the temporal mean has been subtracted (see \cite{Chen2012}).

In DMD/OPD, aiming at approximating the dataset as a linear dynamical system, the complex numbers $\lambda_r$ are eigenvalues of the propagator $\bm{P}$ that advances one snapshot $\vec{\bm{u}}(\bm{x}_i,t_k)$ to the following --that is $\vec{\bm{u}}(\bm{x}_i,t_{k+1})=\bm{P}\,\vec{\bm{u}}(\bm{x}_i,t_k)$. Flattening each snapshot into a column vector $d_k$, the linear dynamical system is defined as $d_{k+1}=\bm{P}\,d_k$, with the propagator being a square real matrix $\bm{P}\in \mathbb{R}^{n_s\times n_s}$.

Since the least square calculation of such propagator is computationally prohibitive in most real applications, the fitting of the linear dynamical system is carried out on a reduced space: the one spanned by the first $\tilde{R}$ POD modes of the dataset. Splitting the dataset matrix into two shifted portions, that is $\bm{D}_1=[d_1,\dots, d_{n_t-1}]$ and $\bm{D}_2=[d_2,\dots,d_{n_t}]$, the propagator $\bm{P}$ and its reduced counter part $\tilde{\bm{S}}$ are:

\begin{equation}
\label{PROP}
\bm{D}_2=\bm{P}\,\bm{D_1}\longrightarrow    \underbrace{\bm{\tilde{\Phi}}^{T}_{\mathcal{P}}\bm{D}_2}_{\tilde{\large{\bm{V}}_2}} \approx \underbrace{\bm{\tilde{\Phi}}^{T}_{\mathcal{P}} \bm{P} \bm{\tilde{\Phi}}_{\mathcal{P}}}_{\large{\tilde{\bm{S}}}} \,\underbrace{\bm{\tilde{\Phi}}_{\mathcal{P}}^{T}\,\bm{D}_1}_{\tilde{\large{\bm{V_1}}}} \longrightarrow \tilde{\bm{V_2}}\approx \tilde{\bm{S}} \,\tilde{\bm{V}_1}
\end{equation} where the $L_2$ approximation lies in the second step, i.e. in the assumption that $\bm{\tilde{\Phi}}_{\mathcal{P}}\,\bm{\tilde{\Phi}}^T_{\mathcal{P}}\approx \bm{I}$ with $\bm{I}$ the identity matrix of size $\tilde{R}\times{\tilde{R}}$.

The frequency-based formalism for data-driven decomposition is therefore based on the computation of the eigenvalues $\lambda_r$ of the reduced propagator $\tilde{\bm{S}}$. In the classical DMD \citep{Schmid,Kutz2014} this propagator is defined as follows:

\begin{equation}
\label{prop1}\begin{split}
\bm{D}_2=\bm{P} \bm{D}_1 \rightarrow \bm{P}\approx \bm{D}_2 \bm{D}^{+}_1= \bm{D}_2 \,\bm{\tilde{\Psi}}_{\mathcal{P}} \bm{\tilde{\Sigma}}^{-1}_{\mathcal{P}} \bm{\tilde{\Phi}}^T_{\mathcal{P}}\, \\\rightarrow \tilde{\bm{S}}=\bm{\tilde{\Phi}}^{T}_{\mathcal{P}} \bm{P} \bm{\tilde{\Phi}}_{\mathcal{P}}=\bm{\tilde{\Phi}}^{T}_{\mathcal{P}} \bm{D}_2 \bm{\tilde{\Psi}}_{\mathcal{P}} \bm{\tilde{\Sigma}}^{-1}_{\mathcal{P}}\,,
\end{split}
\end{equation} having introduced the Moore-Penrose inverse $\bm{D}_1^+=\bm{\tilde{\Psi}}_{\mathcal{P}} \bm{\tilde{\Sigma}}^{-1}_{\mathcal{P}} \bm{\tilde{\Phi}}^T_{\mathcal{P}}$.

In the literature of Principal Oscillation Patterns (POP, \cite{Hasselmann1988,Storch1990,Penland1996}), the propagator is defined in terms of spatial covariance matrix $\bm{D}_1\,\bm{D}^T_1$. Multiplying both sides by $\bm{D}^T_1$ and inverting the spatial covariance matrix gives:

\begin{equation}
\label{prop2}
\bm{D}_2=\bm{P} \bm{D}_1\rightarrow \bm{D}_2 \bm{D}^T_1=\bm{P} \bm{D}_1 \bm{D}^T_1 \Longrightarrow \bm{P}\approx \bm{D}_2 \bm{D}^T_1 \bigl( \bm{D}_1 \bm{D}^T_1)^{-1}\,\,.
\end{equation} 

\subsection{The Multiscale Proper Orthogonal Decomposition (mPOD)}

The Multiscale Proper Orthogonal Decomposition \citep{Mendez2019,Mendez_Journal_2} presents a compromise between the POD and the DMD/OPD by adding spectral constraints to the energy optimality of the POD.

These constraints consist of breaking the datasets into the contributions of various scales using Multi-Resolution Analysis (MRA), as commonly proposed in Wavelet theory \citep{Wavelet1}, and in performing the POD for each scale.

A scale $m$ corresponds to a range of frequencies, say $[f_{m1},f_{m2}]$, identified by a transfer function, here indicated as a row vector $H_m\in\mathbb{C}^{1\times n_t }$. This transfer function is ideally unitary (`all pass') within the range of interest and null (`all stop') otherwise. In the formulation presented in \cite{Mendez_Journal_2}, the transfer function of a given scale is applied in an equal manner to all the spatial realizations: to this end, the transfer function is copied row-wise in order to create a matrix of appropriate size $H'_m\in\mathbb{C}^{n_s\times n_t}$, later applied via direct matrix multiplication. Moreover, these transfer functions are taken as complementary, that is $\sum^{M}_m {H}'_m=\underline{1}$, and ${H}_i\odot{H}_j=0\, \forall i\neq j$, where $\underline{1}\in \mathbb{R}^{n_s\times n_t}$ is the unitary matrix (`all-pass') and $\odot$ is the Schur (entry by entry) product.

The dataset decomposition into the contributions $\bm{D}_m$ of $M$ scales is:  

\begin{equation}
\label{MRA_D}
\bm{D}=\sum^{M}_{m=1}\bm{D}_m=\sum^{M}_{m=1}\overbrace{{\Bigr[{\bigl(\underbrace{\bm{D}\, \overline{\Psi}_\mathcal{F}}_{\widehat{D}}}\bigr)  \odot{H}'_m\Bigr]}}^{\widehat{\bm{D}}_m}{\Psi}_\mathcal{F}
\end{equation} where $\Psi_\mathcal{F}\in\mathbb{C}^{n_t\times n_t}$ is the Fourier matrix $\Psi_{\mathcal{F}}[i,j]=exp(2\pi \mathrm{j}/n_t)^{(i-1)\times(j-1)}$, with $i,j=[1,n_t]$, the over bar denotes complex conjugation. Note that right multiplication by $\Psi_\mathcal{F}$ produce the row-wise Fourier transform of a matrix (i.e. in the time domain for $\bm{D}$) while multiplication by its conjugation (i.e. inverse. since $\Psi_{\mathcal{F}}\overline{\Psi}_\mathcal{F}=\bm{I}$) produce the row-wise inverse Fourier transform. 
The hat is used to denote the Fourier transform in time of the data ($\widehat{\bm{D}}=\bm{D}\, \overline{\Psi}_\mathcal{F}$) and that of each contribution ($\widehat{\bm{D}}_m=\bm{D}_m\, \overline{\Psi}_\mathcal{F}$).

Every contribution has its own POD, which could be used to construct suitable approximation of each scale:

\begin{equation}
\label{MRA_POD}
\bm{D}=\sum^{M}_{m=1}\bm{D}_m=\sum^{M}_{m=1} \bm{\Phi}^{(m)}_{\mathcal{P}}\bm{\Sigma}^{(m)}_{\mathcal{P}}\bm{\Psi}^{(m)\,T}_{\mathcal{P}}=\sum^{M}_{m=1}\bm{D}_m\,\bm{\Psi}^{(m)}_{\mathcal{P}}\,\bm{\Psi}^{(m)\,T}_{\mathcal{P}} \,.
\end{equation}

The Multiscale POD is constructed by assembling the final mPOD temporal basis from all bases of all the scales. The mPOD algorithm saves computational time by operating the MRA on the correlation matrix and not on the dataset matrix- the equivalence between the two is revealed by introducing \eqref{MRA_D} in the definition $\bm{K}=\bm{D}_m^{\dag} \bm{D}_m$. In particular, the correlation matrix is split into the contribution of all the scales using an appropriate filter bank, i.e.:

\begin{equation}
\label{mPOD_K}
\bm{K}=\sum^{M}_{m=1}\,\bm{K}^{(m)}= \sum^{M}_{m=1} \Psi_{\mathcal{F}}\,\Bigl [\widehat{K}\odot \mathcal{H}_m\Bigr] \Psi_{\mathcal{F}}=\sum^{M}_{m=1}\Biggl(\sum^{n_m}_{r=1}\,\lambda^{(m)}_{r}\,\psi^{(m)}_{\mathcal{P}r}\,{\psi^{(m)}_{\mathcal{P}r}}^T\Biggr)
\end{equation} where $\widehat{\bm{K}}=\overline{\Psi}_{\mathcal{F}}\,\bm{K} \,\overline{\Psi}_{\mathcal{F}}$ is the 2D Fourier transform of the correlation matrix, $n_m$ the number of non-zero eigenvalues at each scale, and $\mathcal{H}_m=H'^{T}_m\,H'_m$ an appropriate 2D transfer function.

The eigenvalue decomposition of each contribution is introduced on the right-hand side of \eqref{mPOD_K}. The filters preserve the symmetry of the correlation matrices at each scale and hence the orthogonality of their eigenvectors.  Moreover, avoiding frequency overlapping between different contributions to the correlation matrix, it is possible to keep the eigenvectors of different scales mutually orthogonal (see \cite{Mendez2019}) and hence orthogonal complements spanning the entire $\mathbb{R}^{n_t}$ space (that is $\sum^{M}_{m=1}{n_m}\approx n_t$). This is achieved by constructing the 2D transfer functions, including only `pure' terms, that is, the correlation of a transfer function with itself. In wavelet terminology, this corresponds to performing a Wavelet decomposition of the correlation matrix $\bm{K}$ and retaining only the `approximation' and `diagonal details'.

More information on the MRA formulation via dyadic wavelet decomposition is available in \cite{Mendez_ICNAM,Mendez_Journal_2}.
In this work, the MRA is performed using a generalized filter bank, as in \cite{Mendez2019}, constructed using standard FIR filters designed via windowing method \citep{Signal_1}. 

The resulting decomposition lets the mPOD recover the energy optimality of the POD at the limit of a single scale (spanning the entire frequency range), and the spectral purity of the DMD at the limit of $n_t$ scales (each taking one frequency bin). 

\section{Experimental Set Up and Selected Test Case}\label{EXP}

The selected test case is the flow past a cylinder of $d=5$ mm diameter and $L=20cm$ length in transient conditions, with varying free stream velocity. The experiments were carried out in the L10 low-speed wind tunnel of the von Karman Institute, instrumented with a TR-PIV system from Dantec Dynamics. The tunnel has a cross-section of $20$cm$\times$ $20$cm and is equipped with a piezoresistive pressure transducer AMS5812 to monitor the pressure in the honeycomb chamber. A Laskin Nozzle PIVTEC45-M, operating with mineral oil Ondina Shell 91, is used to produce seeding particles of about $1.5\mu m$ in diameter, injected in the intake manifold of the wind tunnel fan. 

\begin{figure}[h]
	\centering
	\includegraphics[width=14.1cm]{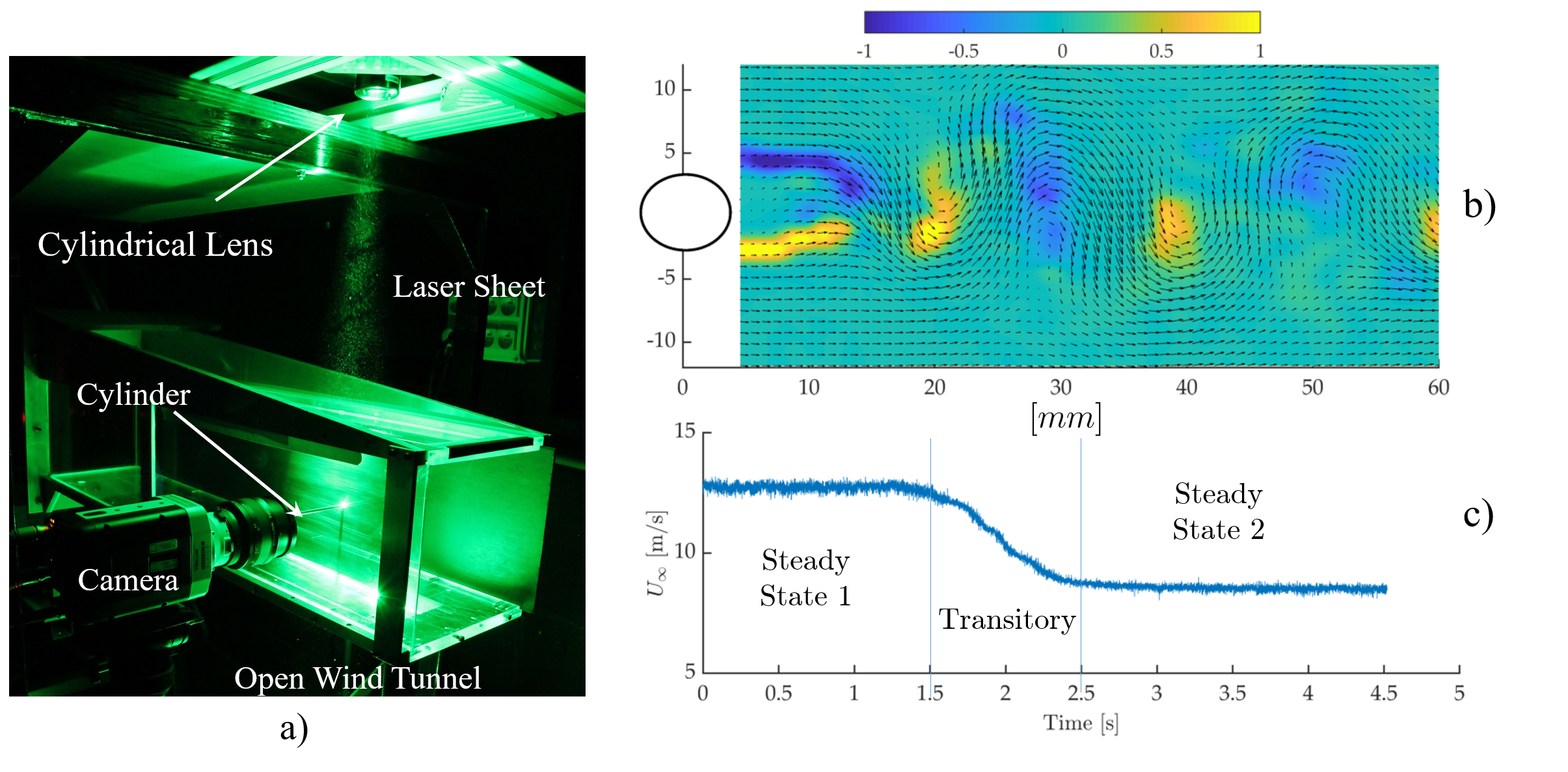} 
	\caption{Fig a) Picture of the experimental set up during a TR-PIV measurement. Fig b) Instantaneous velocity field and normalized vorticity (that is scaled between $[-1, 1]$). Fig c) time evolution of the free stream velocity $U_{\infty}$, taken from a point far from the cylinder surface.}
	\label{FIG1}
\end{figure}

The particles are illuminated with a Photonics Industries DM20-527DH Nd:YLF laser offering 20 mJ/pulse at 1 kHz. The exposed scene of about $70 \times 26$ mm is recorded by a SpeedSense 9090 camera offering 7500 fps at a resolution of $1280 \times 800$ px. To extend the measurement duration, the sensor is cropped to $1280 \times 488$ px resulting in a resolution of $18.3$ px/mm and about $n_t=13500$ double frames, sampled at a frequency of $f_s=3kHz$. At this frequency, the available light intensity is of the order of $6$ mJ/pulse. The full dataset covers about 4.5 seconds. A picture of the experimental set-up is shown in Figure \ref{FIG1}a).

The DynamicStudio software is used for acquisition and analysis. For the initial data processing, standard adaptive PIV (see \cite{RAF}) is used with an initial interrogation area of $96 \times 96$ px and final size of $24\times 24$ px, the vectors are calculated every 12 px resulting in an overlap of $50 \%$. Before the mPOD analysis, the only filter applied is the universal outlier detection with a filter kernel of $3 \times 3$ vectors, as suggested by \cite{Westerweel2005}.

An example of instantaneous velocity is shown in Figure \ref{FIG1}b, together with the vorticity field normalized by the maximum and minimum values so as to have $\omega\in[-1,1]$. In the investigated test case, the velocity of the free stream $U_{\infty}$ evolves through two steady state conditions, namely from $U_\infty=12.1 \pm 3\%$ to $U_\infty=7.9 \pm3\%$ m/s, as shown in Figure \ref{FIG1}c). The transition between these is a smooth step of approximately $1s$. The Reynolds number varies from $Re\approx4000$ to $Re=2600$ and the frequency of the vortex shedding varies from $450 Hz$ to $303 Hz$, corresponding to a Strouhal number of about $St=f\,d/U_{\infty}\approx 0.19$ in both stationary regimes. The entire range of Reynolds number experienced during the test falls in the three-dimensional vortex shedding regime (see \cite{Williamson1996}).

\section{Results and Discussion}\label{RES}

The analysis is divided into two parts: the stationary conditions and the transient conditions. In the stationary conditions, discussed in subsection \ref{STE}, only the first portion of the dataset is considered. Referring to Fig. \ref{FIG1}c), this covers a duration of $[0,1.33s]$ and consists of $n_t=4000$ snapshots. Since the PIV acquisition starts several minutes after the wind-tunnel has reached its stationary regime, this configuration can be seen as a fully developed stationary test case. In the transient conditions, discussed in subsection \ref{UNSTE}, the full set of $n_t=13500$ snapshots is considered in the decomposition.

\subsection{Analysis of the Steady State Condition}\label{STE}

\begin{figure}[h]
	\centering
	\begin{subfigure}[t]{0.65\textwidth}
		\centering
		\includegraphics[width=0.8\textwidth]{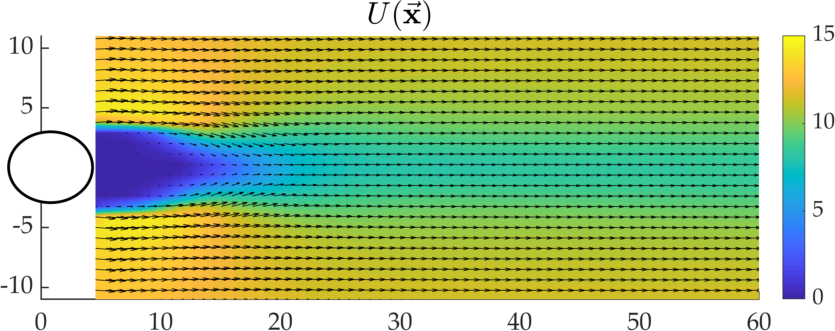}\\a)
		\vspace{2mm}
		
	\end{subfigure}
	\begin{subfigure}[t]{0.65\textwidth}
		\centering
		\includegraphics[width=0.8\textwidth]{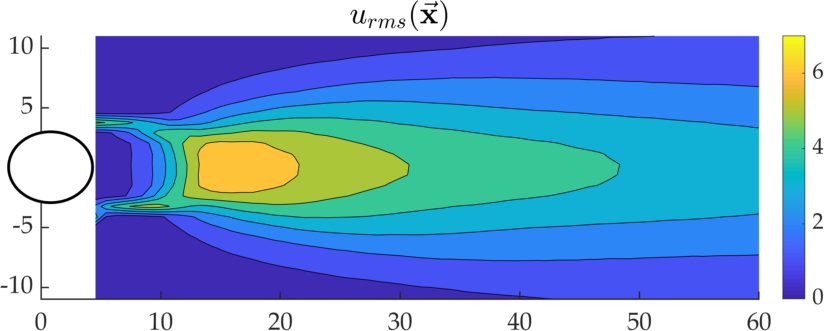}\\b)
		\label{POD_RESULTSa}
		\vspace{2mm}
	\end{subfigure}
	\caption{ Mean velocity field (a) and root mean square (b) of the investigated test case. Both  color-maps are in $m/s$ while spatial units are in mm.}
	\label{MEAN}
\end{figure}

The mean flow field $U=\langle \vec{\bm{u}} \rangle$ and the root mean square field $u_{rms}=\sqrt{\langle u'^{2}\rangle+\langle v'^{2}\rangle}$, with $(u',v')$ the mean subtracted velocity components, are shown in Figure \ref{MEAN}(a) and Figure \ref{MEAN}(b) respectively.
The root-mean-square reaches a maximum of about $u_{rms}=6.9m/s$ at approximately $16mm$ from the cylinder axis. The flow statistics being well developed, the time average is subtracted before performing all the decompositions.

\begin{figure}[h]
	\centering
	\begin{subfigure}[t]{0.6\textwidth}
		\centering
		\includegraphics[width=0.8\textwidth]{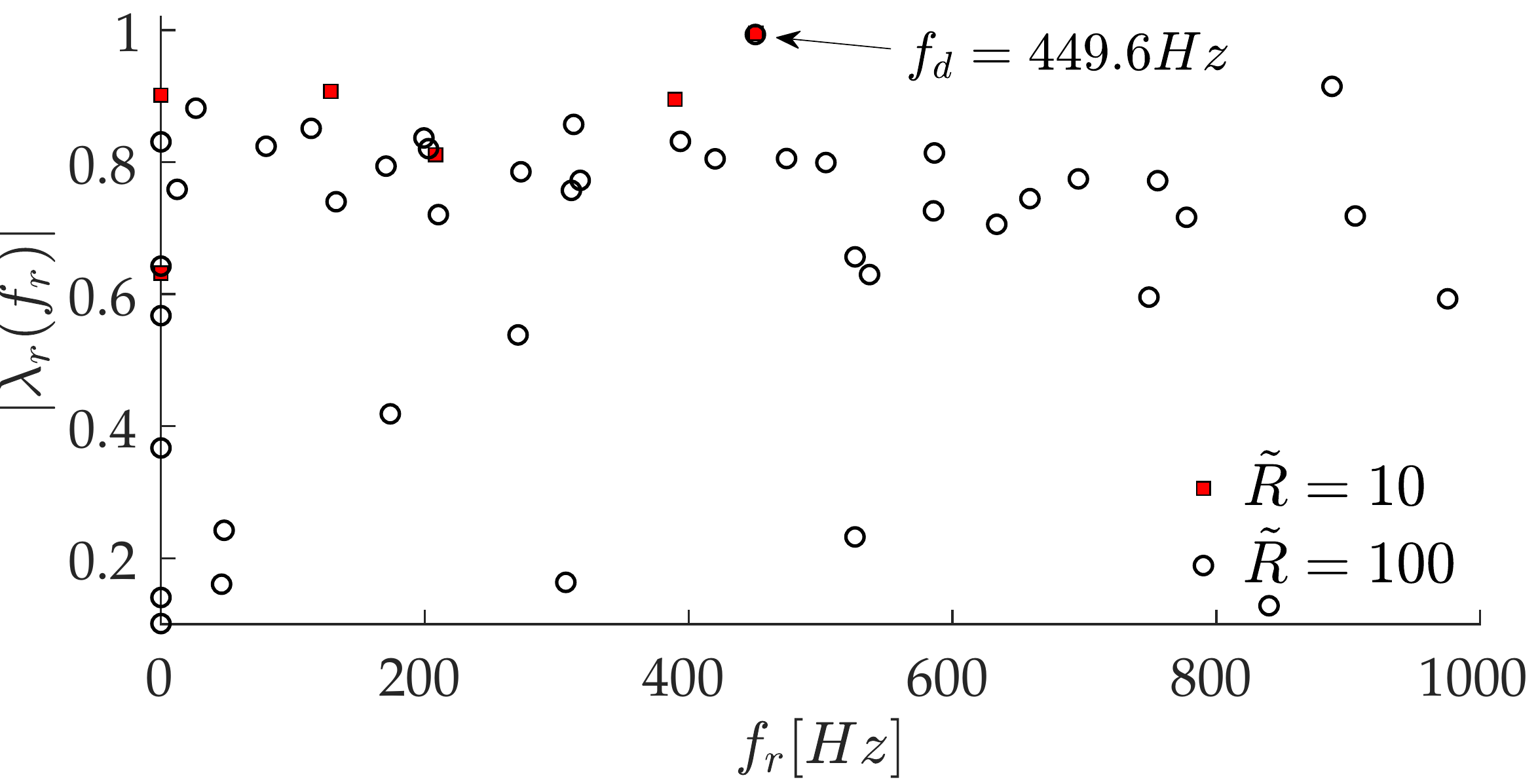}\\a)
		\vspace{2mm}
		
	\end{subfigure}
	\begin{subfigure}[t]{0.65\textwidth}
		\centering
		\includegraphics[width=0.8\textwidth]{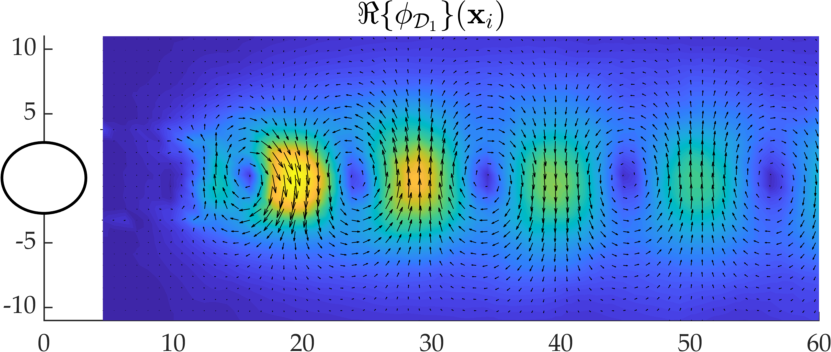}\\b)
		\label{POD_RESULTSa}
		\vspace{2mm}
	\end{subfigure}
	\begin{subfigure}[t]{0.65\textwidth}
		\centering
		\includegraphics[width=0.8\textwidth]{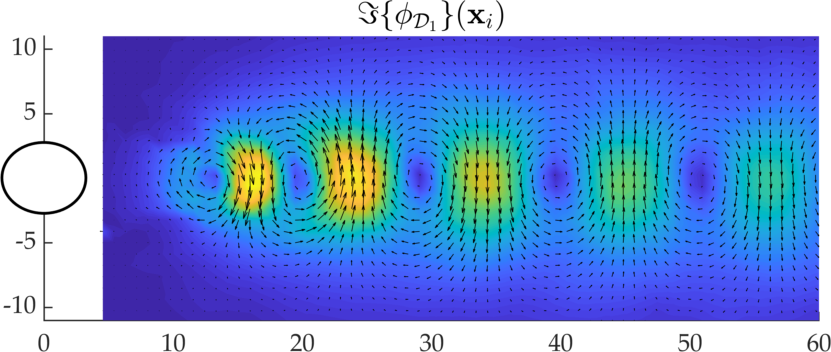}\\c)
		\label{POD_RESULTSa}
	\end{subfigure}
	\caption{(a) Eigenvalue spectra from the DMD analysis in the stationary conditions with $U_{\infty}=12.1 m/s$. The real and the imaginary parts of the spatial structure associated to the dominant ($|\lambda_r|\approx 1$) mode, with frequency $f_d=449.6 Hz$, are shown in (b) and (c) respectively. Spatial units are in mm.}
	\label{RES_DMD}
\end{figure}

The key results from the DMD/OPD are collected in Figure \ref{RES_DMD}. Figure \ref{RES_DMD}a)
shows the eigenvalue spectrum from the DMD/OPD, while \ref{RES_DMD}b) and \ref{RES_DMD}c) show the real and the imaginary part of the spatial structure associated with the dominant DMD/OPD mode. As stated earlier, no appreciable difference is observed between the propagators in \eqref{prop1} and \eqref{prop2}, and hence only the results from the first are shown.

The plot of the eigenvalue spectra is shown in terms of modulus of the reduced propagator's eigenvalue $\tilde{S}$ versus the associated frequency (only $f_n\geq 0$ are shown). The spectra is constructed for two choices of reduced POD basis, namely including $\tilde{R}=10$ (red squares) or $\tilde{R}=100$ POD modes (white circles) in \eqref{prop1} out of the $n_t=4000$ available modes.

For both reduced POD bases, one eigenvalue with modulus $|\lambda_r|\approx 1$ is visible in the plots: this is linked to the dominant frequency in the flow produced by the vortex shedding. This eigenvalue, however, has a modulus slightly below unity ($|\lambda_r|=0.9935$), possibly due to round-off errors, measurement noise, or insufficient sampling. As a result, this mode (as well as all the others) has a temporal structure that is exponentially decaying: the half-life of the dominant mode is of the order of 100 steps, i.e., about $3.3 ms$, while all the other vanishes faster.

The energy associated with the DMD modes $\sigma^{2}_{\mathcal{D}r}$ is hence of little interest and not shown. As discussed in Section \ref{DMD_OPD}, these are the correlation between the initial snapshot of the data and the computed spatial basis, but the vanishing nature of the modes and the normalization of temporal and spatial structures make them unrepresentative of their relative importance.

\begin{figure}[h]
	\centering
	\begin{subfigure}[t]{0.6\textwidth}
		\centering
		\includegraphics[width=0.8\textwidth]{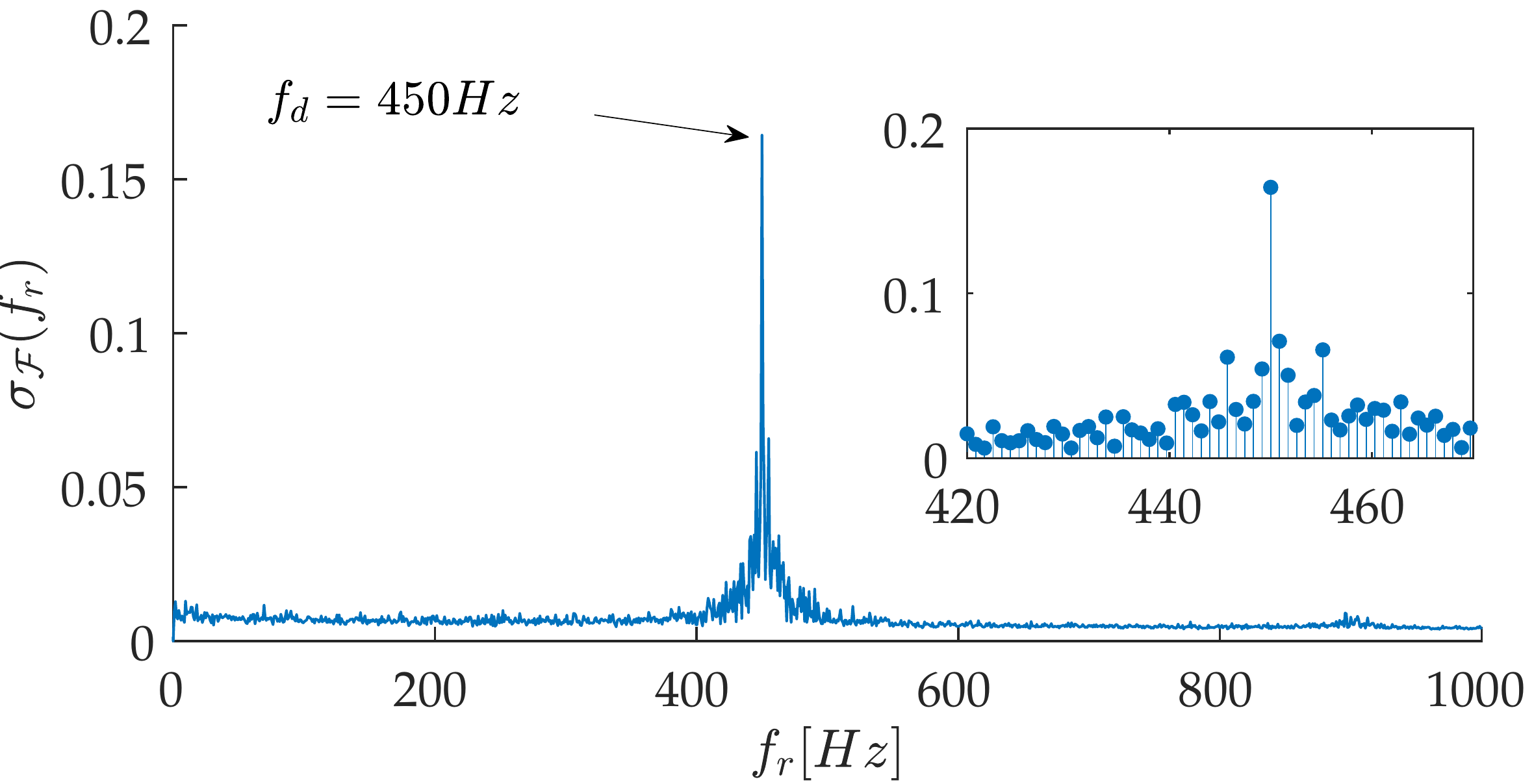}\\a)
		\label{Structures_DFT}
		\vspace{1mm}
	\end{subfigure}
	\begin{subfigure}[t]{0.65\textwidth}
		\centering
		\includegraphics[width=0.8\textwidth]{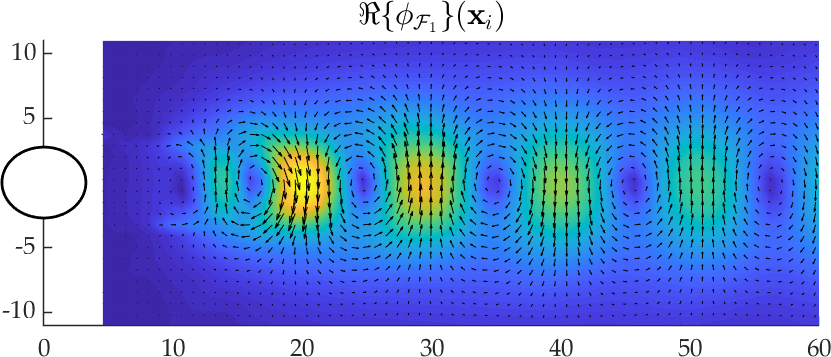}\\b)
		\label{POD_RESULTSa}
		\vspace{1mm}
	\end{subfigure}
	\begin{subfigure}[t]{0.65\textwidth}
		\centering
		\includegraphics[width=0.8\textwidth]{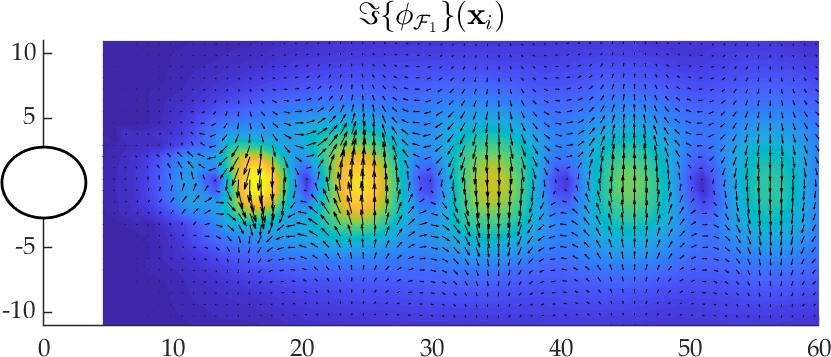}\\c)
		\label{POD_RESULTSa}
		\vspace{1mm}
	\end{subfigure}
	\caption{Spectra of the DFT amplitudes ($diag(\Sigma_{\mathcal{D}})$) the stationary test case with $U_{\infty}=12.1m/s$. The dominant frequency of the vortex shedding is clearly visible. The real and the imaginary parts of the associated spatial structure are shown in b) and c) respectively.}
	\label{ResDFT}
\end{figure}

The major limitation of the DMD is thus convergence: regardless of the size of the reduced POD basis, the vanishing of the modes make the error in \eqref{ERROR} of the order of $95 \%$ even if all the modes of the decomposition are included. 

The spatial structure associated with the dominant DMD modes in Figure \ref{RES_DMD}b)-c) features a periodic pattern with a stream-wise wavelength of the order of the cylinder diameter. The first large vortical structure of the wake is centered at a distance of $\approx 16mm$ from the cylinder. Real and imaginary parts are phase-shifted in space, due to their harmonic `traveling wave' nature.

In order to identify the structures associated with harmonic (i.e., not decaying) modes for these stationary test cases, the results of a DFT is also considered in Figure \ref{ResDFT}. Figure \ref{ResDFT}a) shows the energy contributions of the DFT modes as a function of their associated frequency (only $f_n>0$ is shown) while Figures \ref{ResDFT}b)-c) show the real and the imaginary parts of the dominant mode. The energy spectrum is narrow and centered around the dominant frequency of the vortex shedding, labeled in the plot, and in agreement with the one identified by the DMD.

Although this mode is purely harmonic (with no decay), the real part of the spatial structures is similar to the DMD ones. The imaginary part, on the other hand, is not: while the classic overall spatial shift is also visible, the DFT mode displays a saddle point in place of the vortex core of the corresponding imaginary part in the DMD mode. Since the DMD and the DFT modes evolve at nearly the same frequency, this difference can only be due to their different duration: the saddle pattern appears as more representative of the flow evolution in the entire dataset while the structure identified by the DMD mode vanishes within $1/5$ of the observation time. The flow topology of the saddle pattern is described by \cite{Hussain1987,Chen2017} as the result of the interaction between phase-shifted counter-rotating vortices produced in the von Karman street (see Fig. 15 in \cite{Chen2017}). This is also the region of larger strain and thus turbulence production; not surprisingly, the saddle point is located close to the region of larger root-mean-square velocity, as shown in Figure \ref{MEAN}b).

\begin{figure}[h!]
	\centering
	\begin{subfigure}[t]{0.65\textwidth}
		\centering
		\includegraphics[width=0.75\textwidth]{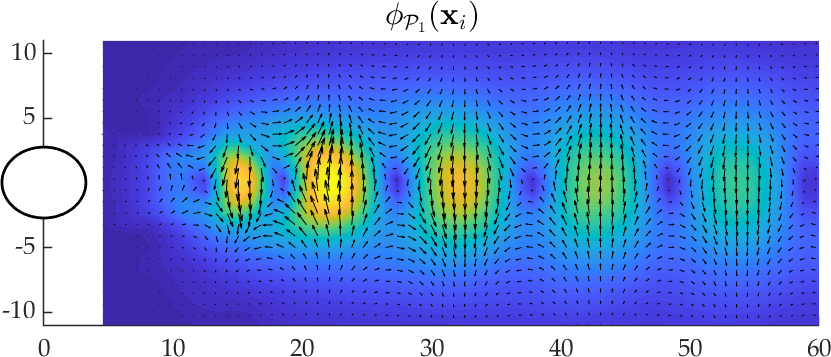}\\a)
		\label{POD_RESULTSa}
		\vspace{1mm}
	\end{subfigure}\\
	\begin{subfigure}[t]{0.65\textwidth}
		\centering
		\includegraphics[width=0.75\textwidth]{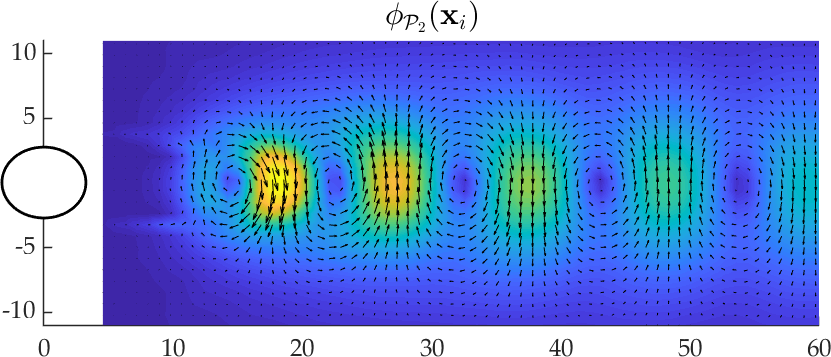}\\b)
		\vspace{1mm}
		\label{POD_RESULTSb}
	\end{subfigure}\\
	\begin{subfigure}[t]{0.58\textwidth}
		\centering
		\includegraphics[width=0.75\textwidth]{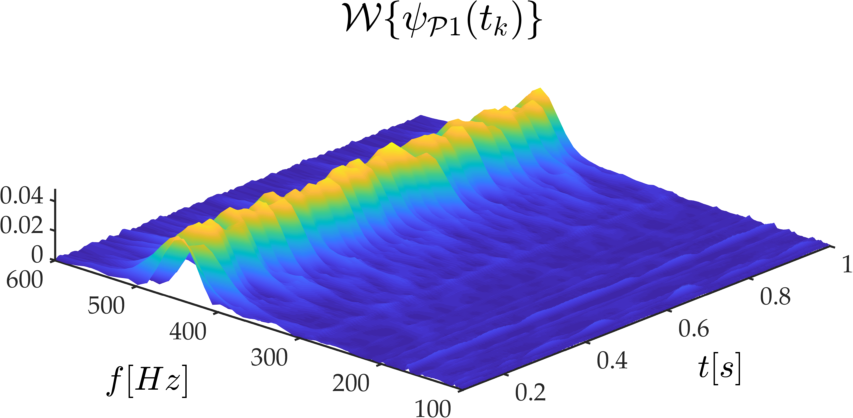}\\c)
		\vspace{1mm}
		\label{POD_RESULTSc}
	\end{subfigure}\\
	
	\begin{subfigure}[t]{0.58\textwidth}
		\centering 
		\includegraphics[width=0.75\textwidth]{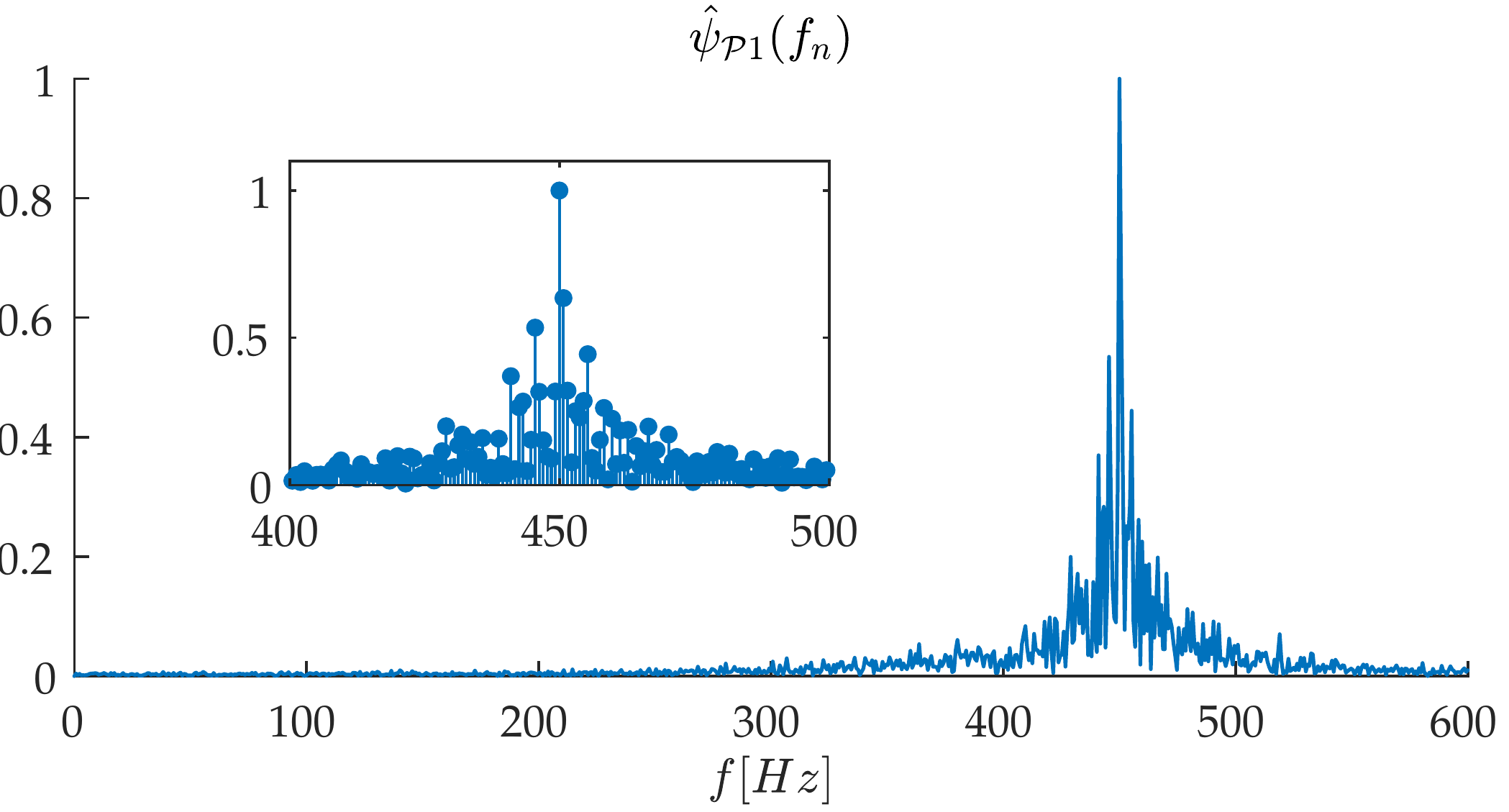}\\d)
		\label{POD_RESULTSd}
	\end{subfigure} 
	\begin{subfigure}[t]{0.35\textwidth}
		\centering
		\includegraphics[width=0.75\textwidth]{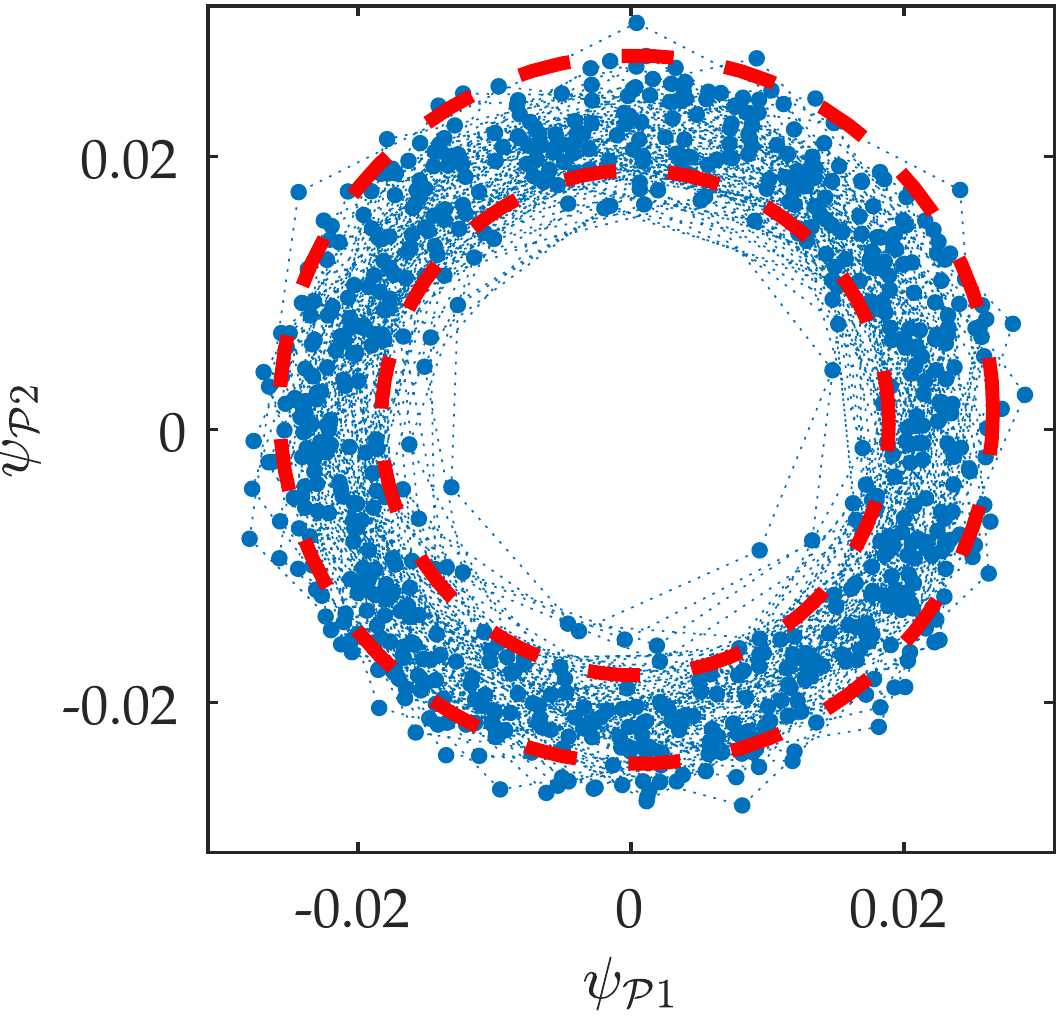}\\e)
		\label{POD_RESULTSe}
	\end{subfigure}%
	\caption{Results from the POD analysis of the stationary test case. The spatial structures of the dominant POD modes are shown in (a) and (b). Figure (c) shows amplitude of the CWT of the first temporal structure $\psi_{\mathcal{P}}$, revealing the time-frequency evolution of these modes. Figure (d) shows their DFT. These modes have similar evolution, up to a $\pi/2$ phase difference, as shown by the phase portrait (Lissajous curve) in Figure (e). }
	\label{POD_RESULTS}
\end{figure}

To conclude the analysis of the stationary test case using state-of-the-art data-driven decomposition, the results for the POD, collected in Figure \ref{POD_RESULTS}, are now considered. Figures \ref{POD_RESULTS}a,b) show the spatial structure of the two dominant POD modes. These two modes are strongly linked, with nearly identical temporal evolution and frequency content. The time-frequency evolution of the first, using Continuous Wavelet Transform (CWT), is shown in \ref{POD_RESULTS}c); the CWT of the second, being almost identical, is not shown. The CWT is performed using \textsc{Matlab}'s function \textsc{cwt} with a complex Morlet wavelet with bandwidth parameter $f_b=5$ and center frequency $f_c=1$ \citep{MATLAB} as a compromise between frequency and time resolution. The DFT of the associated temporal structures is shown in \ref{POD_RESULTS}d), normalized with respect to the dominant frequency. Figure \ref{POD_RESULTS}e) presents their phase portrait (Lissajous curves) produced by plotting one temporal structure versus the other. 

Since the POD modes are real, the traveling wave pattern is described by two modes in quadrature, which is clearly highlighted by the circular pattern produced by the Lissajous curve. The regularity of these patterns shows that the phase shift between the modes is overall time-invariant, similarly to what occurs in a DFT mode. As a result, the spatial structures of these two modes appear qualitatively similar to the real and the imaginary counter parts of the dominant DFT mode. 

Nevertheless, the Lissajous curve shows a strong amplitude modulation, as the radius of the system trajectory varies in time between two extreme values, indicated in the figure with dashed red lines. Since the frequency spectra of these modes appear time-invariant and narrow in the frequency domain, as shown by the CWT in \ref{POD_RESULTS}c), this modulation can only be explained by two mechanisms: 1) the contribution due to phenomena having broad frequency spectrum (e.g. noise of turbulence) 2) the competition between harmonic modes having frequency close to the dominant shedding frequency.

The distinction between these two mechanisms cannot be revealed either by the DFT, because its harmonic modes do not admit modulation in time, neither by the POD, because its modes have too broad frequency content. In particular, it is interesting to observe that the DFT spectra of the temporal structures of the POD, shown in Figure \ref{POD_RESULTS}d), is very similar to the DFT spectra of the entire dataset in Figure \ref{ResDFT}a). This is a direct consequence of the energy optimality, which allows these two POD modes to capture a significant amount of the dataset's energy, at the cost of mixing in the same modes a large range of scales. As shown by the stem plot zooming around the dominant peak, these modes have both contributions from frequencies close to the dominant one and larger tails of smaller amplitudes that extends over a frequency range of about $300Hz$. It is worth highlighting that the observation time covers about $5330$ periods of the dominant frequency, and hence these secondary peaks are unlikely due to windowing phenomena (i.e., spectral leakage, see \cite{Harris1978}).


\begin{figure}[h!]
	\centering
	\includegraphics[width=0.3\textwidth]{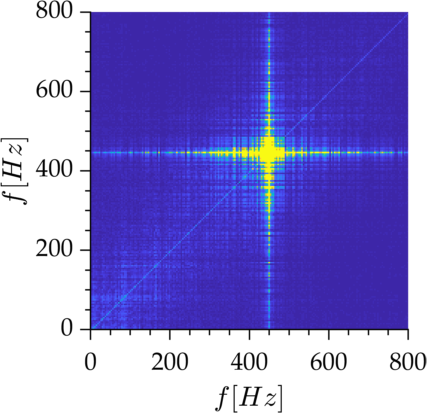}\hspace{2mm}
	\includegraphics[width=0.6\textwidth]{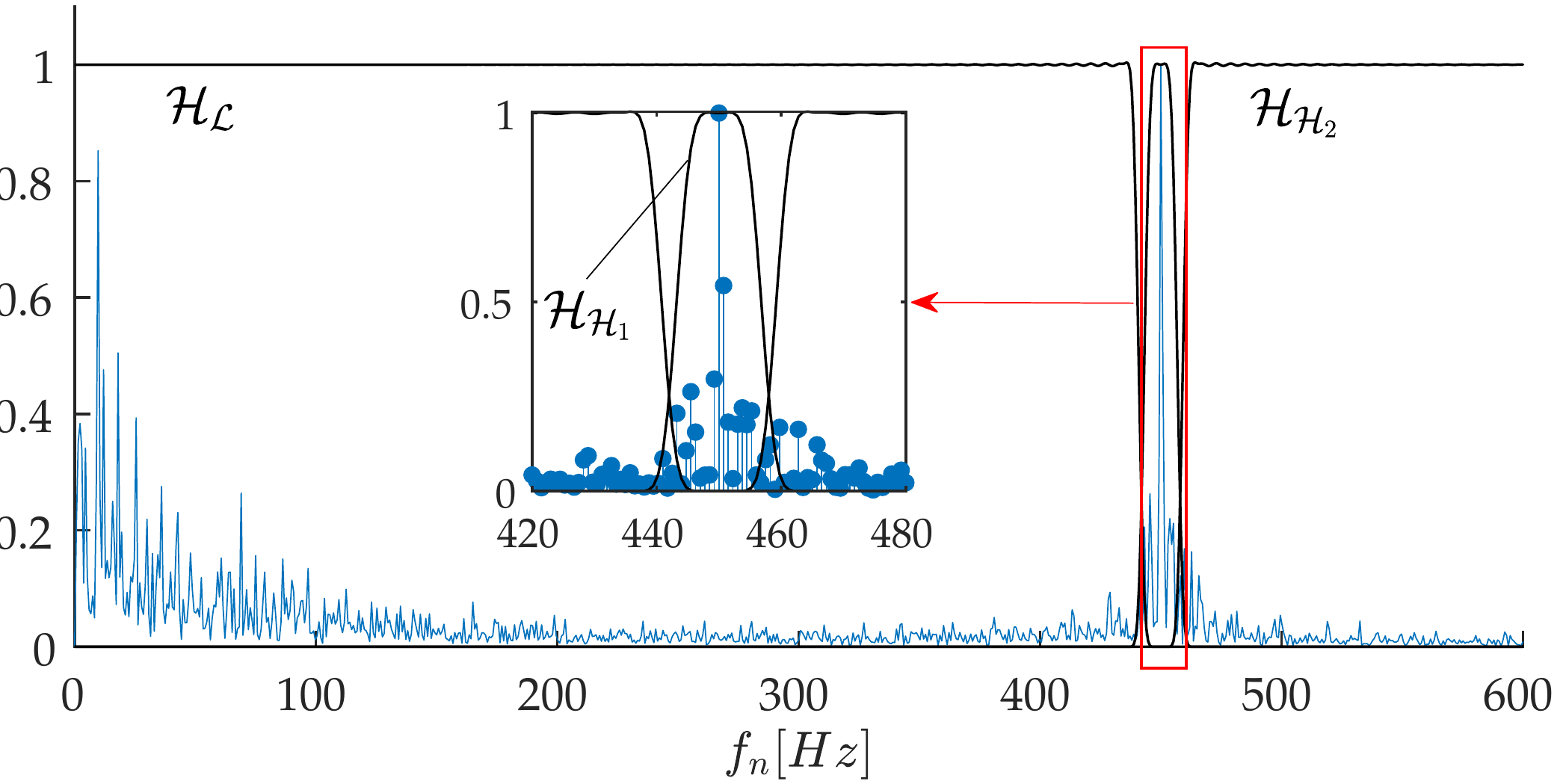} \\
	\hspace{2mm}a) \hspace{55mm} b)\hspace{22mm}
	\caption{(a) Contour of the modulus of the Fourier transform of the correlation matrix $\widehat{K}$ and (b) auto-correlation spectra (i.e. diagonal of $\widehat{\bm{K}}$) with the modulus of the transfer functions of the selected scales ($diag(\Sigma_{\mathcal{D}})$).}
	\label{MRA_STATS}
\end{figure}

The results from the mPOD are now discussed. For the purpose of this work and to illustrate the flexibility of this decomposition, it is interesting to obtain modes that have cleaner spectra than the POD ones in order to analyze the origin of the shedding modulation previously described. On the other hand, it is also of interest to achieve this without significantly compromising the convergence of the POD. To this end, only three scales are selected, starting from the one that should isolate the vortex shedding. The selection of the frequency range of each scale is illustrated in Figure \ref{MRA_STATS} along with the auto-correlation spectra of the dataset (shown in \ref{MRA_STATS}b), that is the diagonal of the 2D Fourier transform of the correlation matrix $\widehat{K}$ (shown in \ref{MRA_STATS}a).

The correlation spectra show an important frequency content at a range $<100 Hz$; this is not visible in the first two dominant POD modes, but it is strongly present in the third and fourth POD modes (not shown). The scale centered around the vortex shedding is identified by a band-pass filter $\mathcal{H}_1$ with pass-band $f\in[445,455]Hz$, while the other two scales take the remaining larger scales (filter $|\mathcal{H}_{\mathcal{L}}|=1$ in $f\approx [0,445]Hz$) and finer scales (filter $|\mathcal{H}_{2}|=1$ in $f\approx [455,1500]Hz$). It is essential to observe that these spectral constraints only impose that a mode having frequency content in one scale does not have frequency content in others.  Each scale is equipped with its POD, and the final mPOD basis is constructed by sorting these POD from various scales by energy contribution (associated eigenvalue).

\begin{figure}[h!]
	\centering
	\begin{subfigure}[t]{0.6\textwidth}
		\centering
		\includegraphics[width=0.75\textwidth]{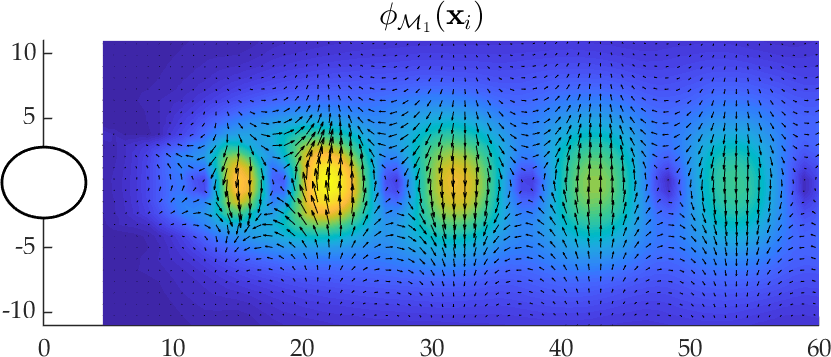}\\a)
		\label{POD_RESULTSa}
		\vspace{1mm}
	\end{subfigure}\\
	\begin{subfigure}[t]{0.6\textwidth}
		\centering
		\includegraphics[width=0.75\textwidth]{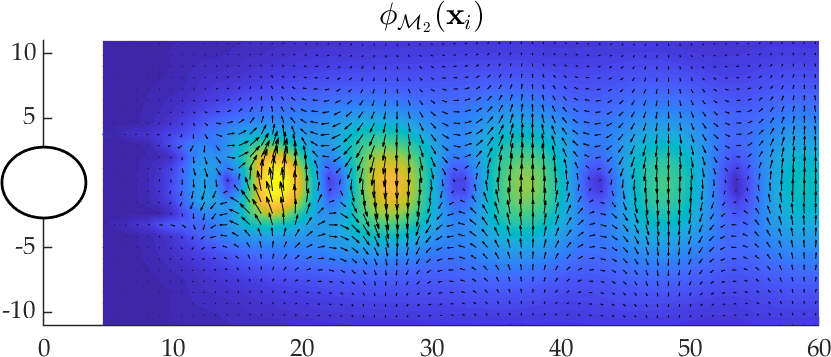}\\b)
		\vspace{1mm}
		\label{POD_RESULTSb}
	\end{subfigure}\\
	\begin{subfigure}[t]{0.58\textwidth}
		\centering
		\includegraphics[width=0.75\textwidth]{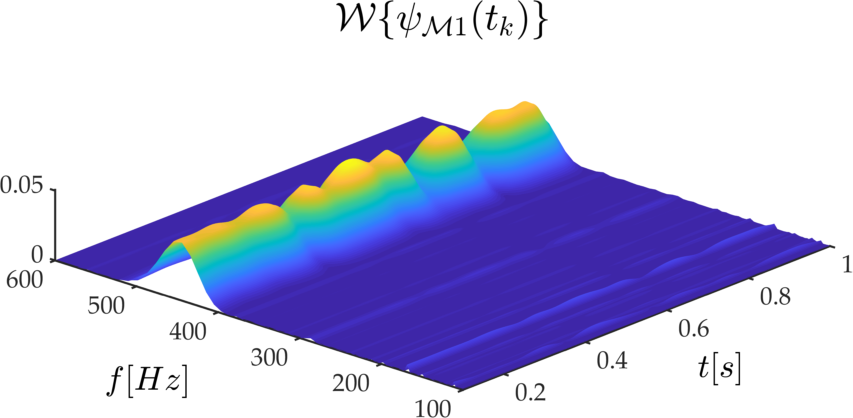}\\c)
		\vspace{1mm}
		\label{POD_RESULTSc}
	\end{subfigure}\\
	
	\begin{subfigure}[t]{0.58\textwidth}
		\centering 
		\includegraphics[width=0.75\textwidth]{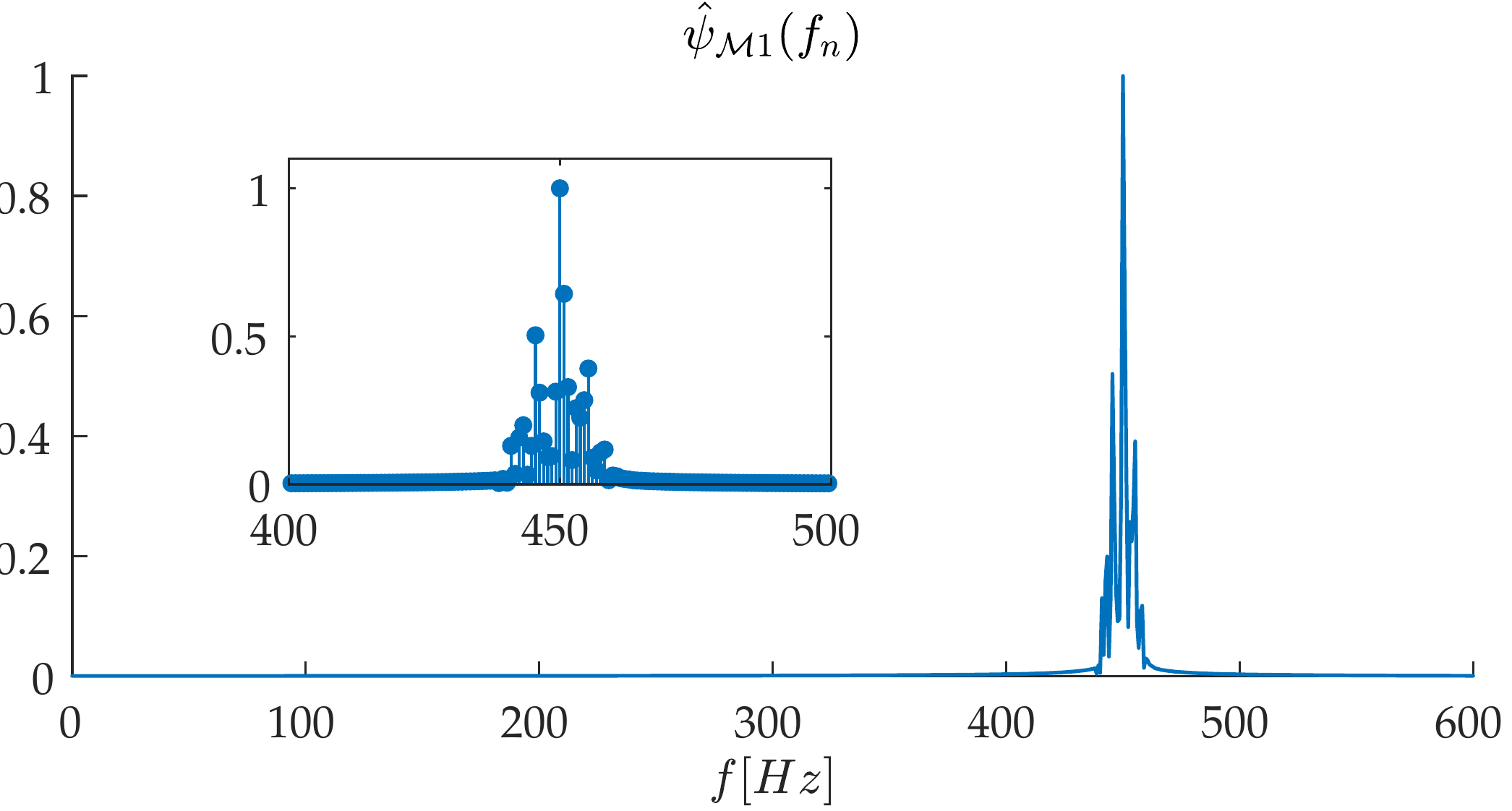}\\d)
		\label{POD_RESULTSd}
	\end{subfigure} 
	\begin{subfigure}[t]{0.35\textwidth}
		\centering
		\includegraphics[width=0.75\textwidth]{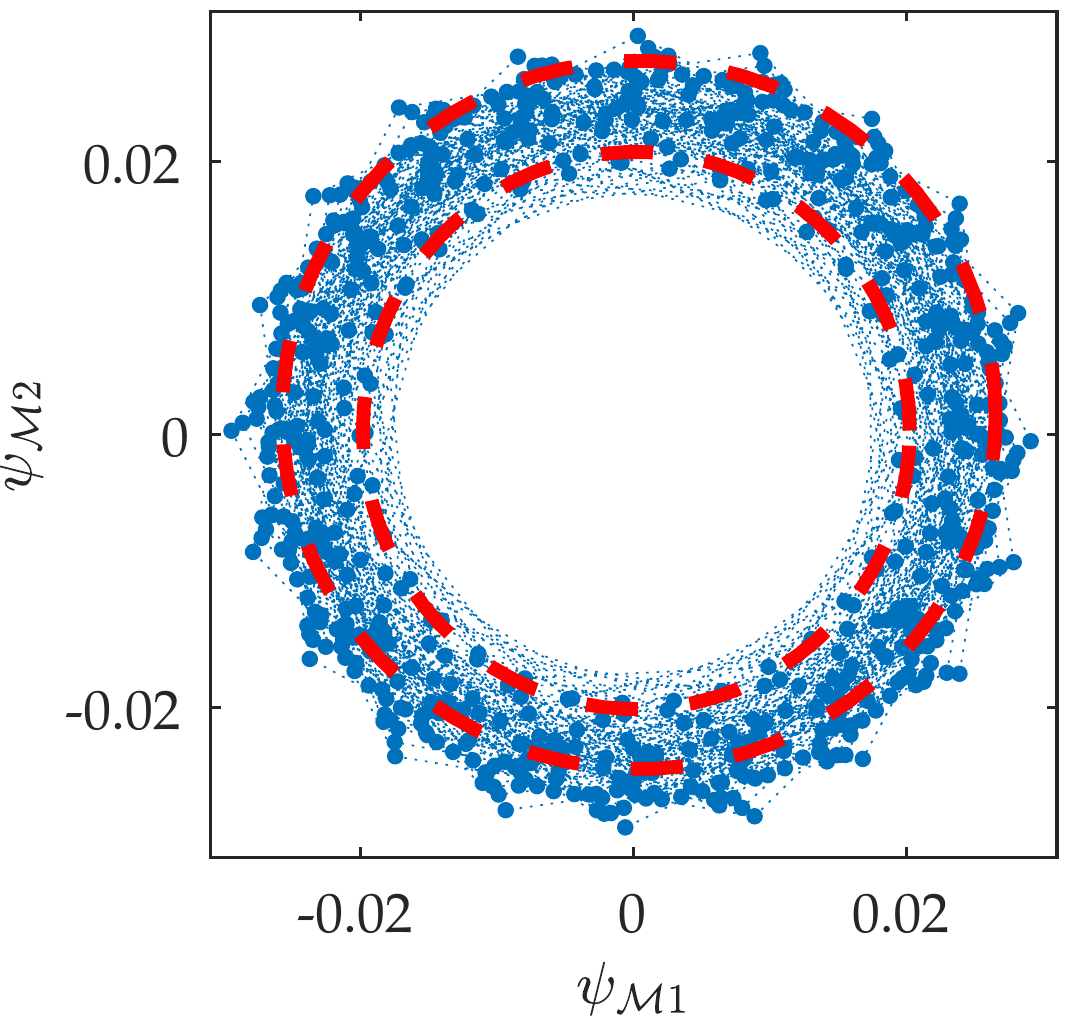}\\e)
		\label{POD_RESULTSe}
	\end{subfigure}%
	\caption{Same as Figure \ref{POD_RESULTS}, but for the first two mPOD modes.}
	\label{fullMPOD_STAT1}
\end{figure}

All the filters are constructed using a windowing method with a Hamming window and kernels of size $n_I=2000$. The filtering is performed on the temporal correlation matrix using Matlab's function \textsc{imfilter} with replicating boundary conditions (`\textsc{replicate}' option). All the Matlab codes for the mPOD implementation are available online and described by \cite{Ninni}.

\begin{figure}[h!]
	\centering
	\begin{subfigure}[t]{0.6\textwidth}
		\centering
		\includegraphics[width=0.75\textwidth]{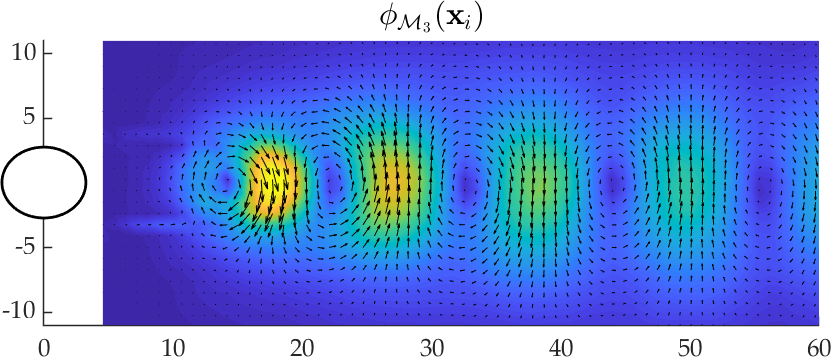}\\a)
		\label{POD_RESULTSa}
		\vspace{1mm}
	\end{subfigure}\\
	\begin{subfigure}[t]{0.6\textwidth}
		\centering
		\includegraphics[width=0.75\textwidth]{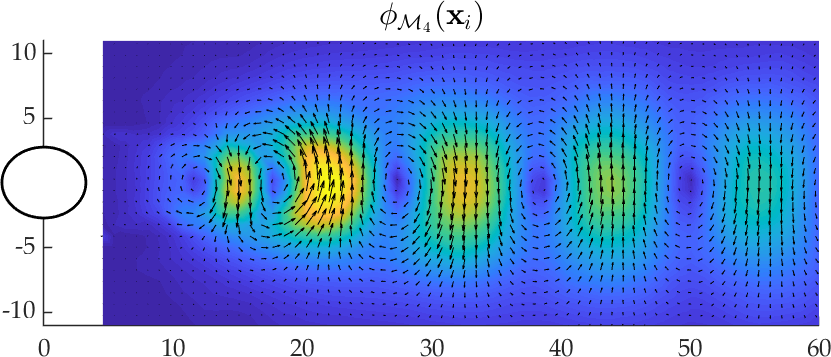}\\b)
		\vspace{1mm}
		\label{POD_RESULTSb}
	\end{subfigure}\\
	\begin{subfigure}[t]{0.55\textwidth}
		\centering
		\includegraphics[width=0.75\textwidth]{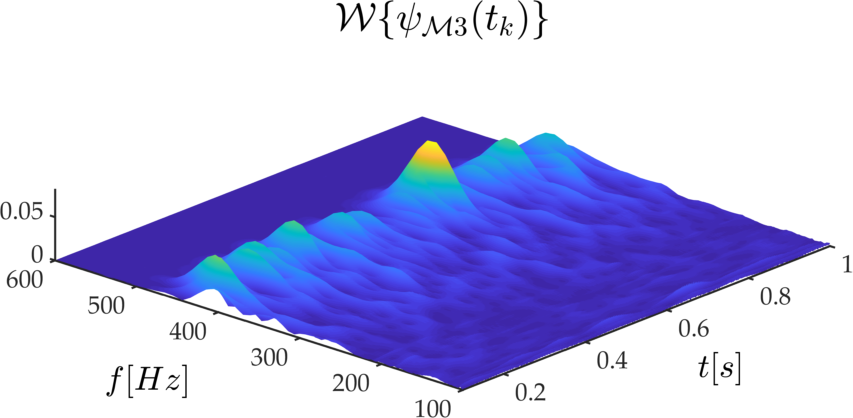}\\c)
		\vspace{1mm}
		\label{POD_RESULTSc}
	\end{subfigure}\\
	
	\begin{subfigure}[t]{0.58\textwidth}
		\centering 
		\includegraphics[width=0.75\textwidth]{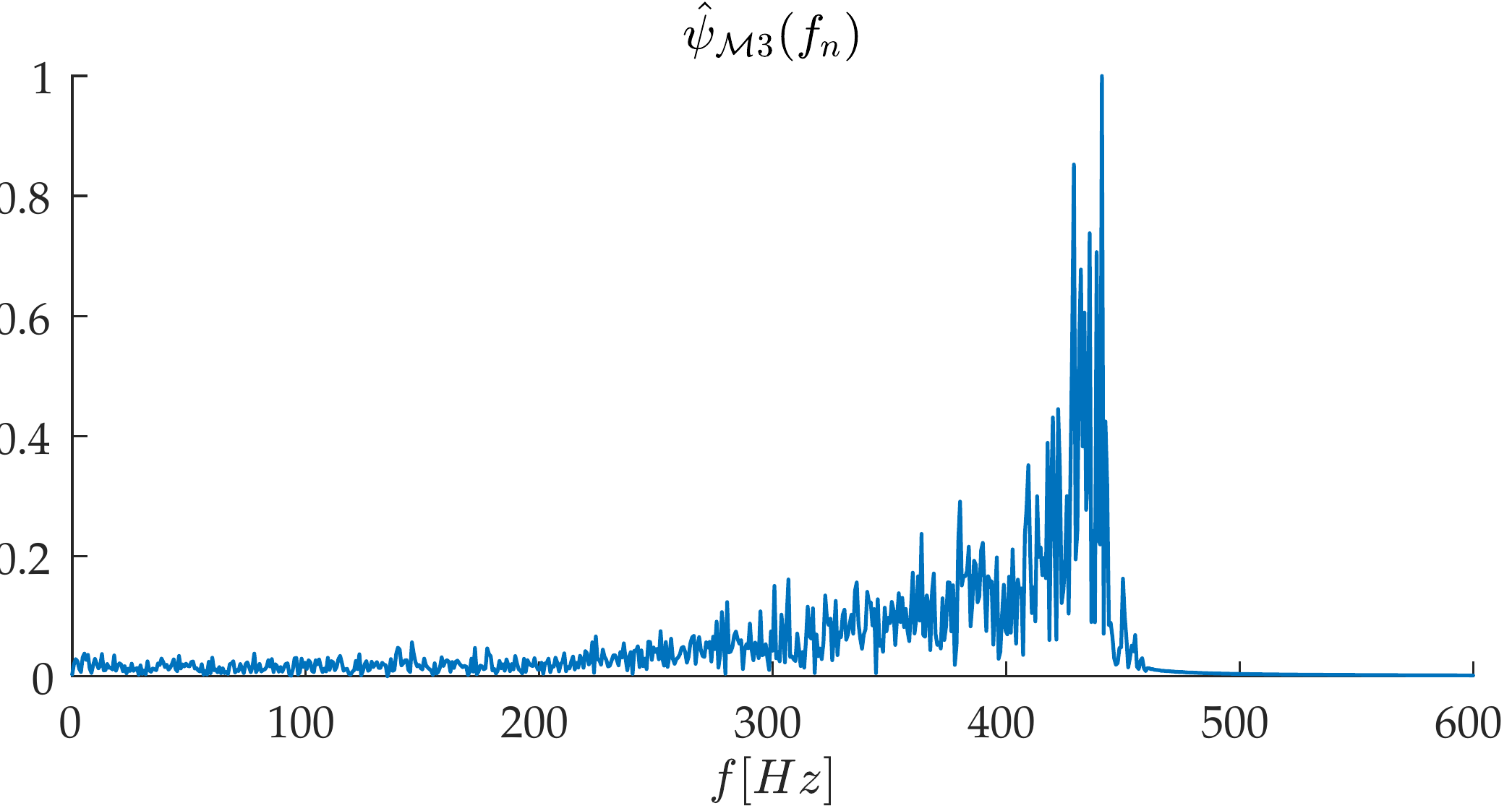}\\d)
		\label{POD_RESULTSd}
	\end{subfigure} 
	\begin{subfigure}[t]{0.35\textwidth}
		\centering
		\includegraphics[width=0.75\textwidth]{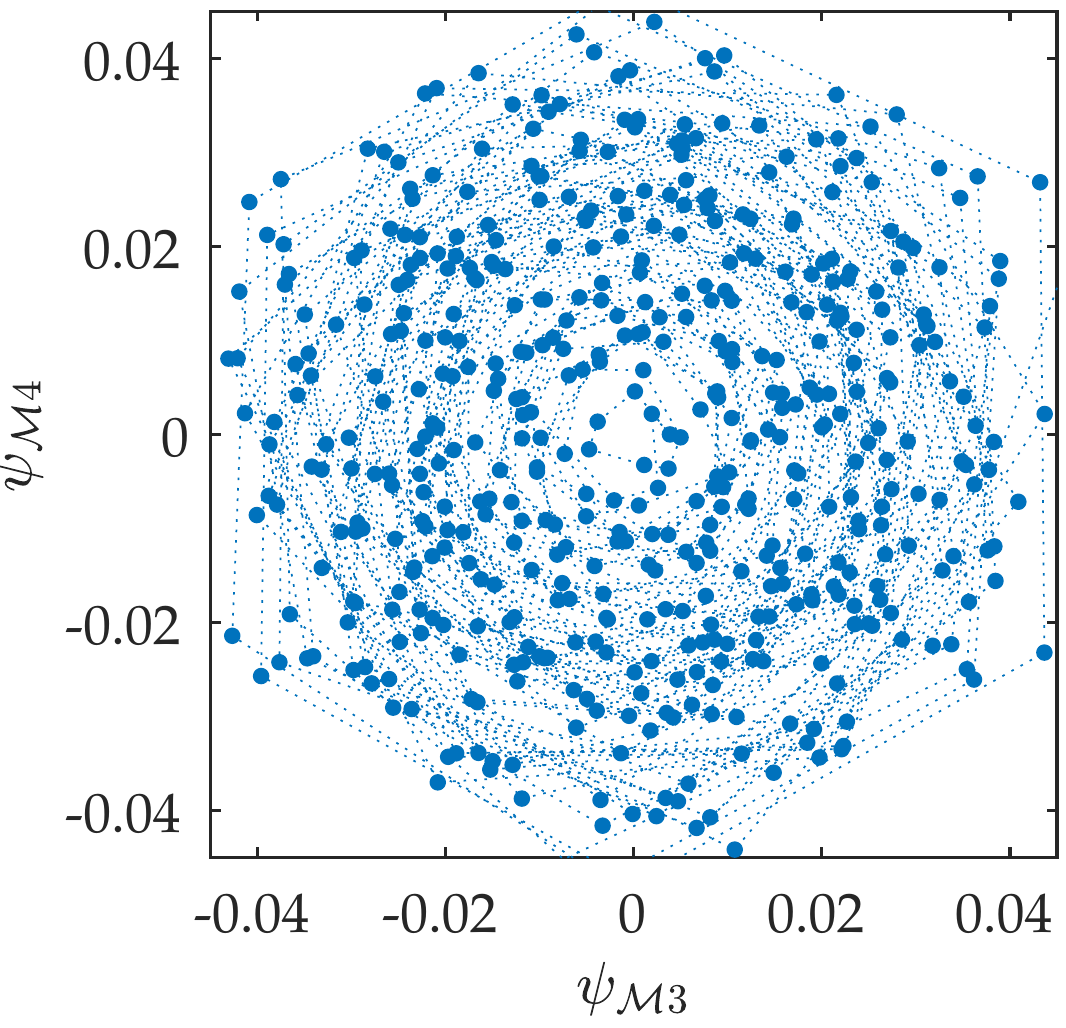}\\e)
		\label{POD_RESULTSe}
	\end{subfigure}%
	\caption{Same as Figure \ref{POD_RESULTS}, but for the third and fourth mPOD modes.}
	\label{fullMPOD_STAT2}
\end{figure}

The first two mPOD modes are shown in figure \ref{fullMPOD_STAT1}; the third and fourth are shown in figure \ref{fullMPOD_STAT2}. Both pairs share the same frequency spectra as for the POD pairs and the figures follow the same structure of figure \ref{POD_RESULTS}.

The first mPOD pair of modes arise from the scale $\mathcal{H}_1$, that is the one centered around the dominant vortex shedding frequency. The spatial structures of both modes feature the saddle region in the wake flow, and their CWT shows the modulation of their evolution in time.
Because of their narrow frequency bandwidth, this modulation can only be linked to the interaction of different harmonic modes close to the shedding, which results in a beating phenomenon. This phenomenon is well described in standard textbooks on mechanical vibrations (e.g., \cite{Rao}) and results from the interference of two harmonics motion having a similar frequency. The phase portrait of these modes (Figure \ref{fullMPOD_STAT1}e) shows that the beating accounts for a significant portion of the modulation observed in the POD modes.

The second mPOD pair of modes in Figure \ref{fullMPOD_STAT2} arise from the large scale $\mathcal{L}$ and is characterized by a frequency content which extends over the entire allowed spectra. A finer frequency selection could be used to reduce the frequency bandwidth of these modes, but for the purposes of this work it suffices to observe that these are not stationary: comparing the CWT transform of these modes in Figure \ref{fullMPOD_STAT2}c) with the one of the first two in Figure \ref{fullMPOD_STAT1} shows that their contributions are complementary-- the mode pair $3-4$ is most active when the mode pair $1-2$ is not. This intermittency results in the phase portraits that continuously spirals in and out (see Fig. \ref{fullMPOD_STAT2}e).

The spatial structures of these modes show the dominant vortical structure also observed in the harmonic decomposition. The mPOD describes, therefore, the dataset as a rather intermittent interplay of modes: the vortex shedding appears to slightly change its frequency content in time resulting in a temporal modulation and beating of the shedding amplitude mostly occurring at the dominant frequency detected by DFT and DMD. This intermittent effect is possibly linked to the three-dimensional nature of the flow and particularly to the interaction between the quasi two dimensional Karman street with span-wise structures, that are characterized by similar frequency content \citep{Zhang2000,Zhou2003}. 

Modes belonging to the highest frequency portion $\mathcal{H}_2$ have much lower energy and are not among the first ten mPOD modes-- these are therefore not shown.

\begin{figure}[h!]
	\centering
	\begin{subfigure}[t]{0.75\textwidth}
		\centering 
		\includegraphics[width=0.8\textwidth]{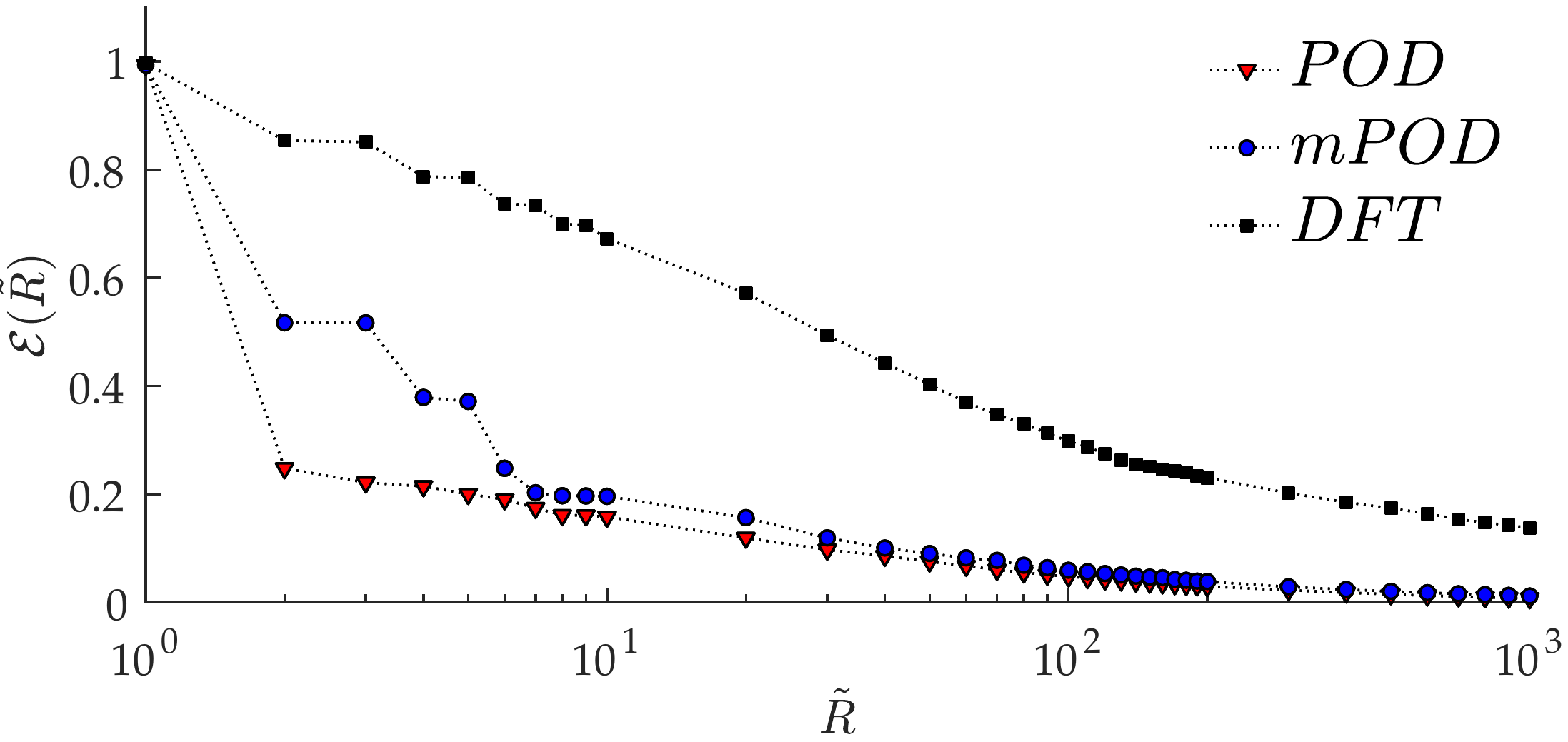}\\a)
		\label{conv1}
	\end{subfigure} 
	\caption{Decomposition error $\mathcal{E}$ in \eqref{ERROR} as a function of the number of modes included in the summation \eqref{EQ1} for the POD, the mPOD and the DFT of the stationary dataset at $U_\infty=12.1m/s$. }
	\label{DECOconv}
\end{figure}

Finally, to conclude the analysis of the stationary test case, the decomposition convergence is discussed. Figure \ref{DECOconv} shows the $L_2$ error over the full dataset (in this case, the first $n_t=4000$ snapshots) as a function of the number of modes used in the approximation. The convergence results of the DMD/OPD are not shown since this decomposition does not converge to the dataset. While the spectral constraints in the mPOD do not allow to reach the optimal convergence of the POD, the two decompositions reach comparable convergence for $\tilde{R}>8$. In particular, it appears that the information contained in the first five POD modes (producing $\mathcal{E}=0.204$) is distributed to the first six mPOD modes (producing $\mathcal{E}=0.200$). This result shows that the mPOD allows for more degrees of freedom than the POD in the modal analysis while maintaining comparable convergence, contrary to harmonic methods. For comparison purposes, the plot also shows the convergence of the DFT in time, which is considerably poorer. This represents the limit at which the mPOD tend if the spectral constraints in the MRA are increased to the extreme of having only one harmonic per scale.

\subsection{Analysis of the Transient Conditions}\label{UNSTE}

The full set of velocity fields is now considered, with free stream velocity evolving, as shown in Figure \ref{FIG1}c. Since the flow is in transient conditions, the time-averaged flow is of little significance, and it is therefore not removed from the dataset before performing POD and mPOD. For the DFT, the mean removal is only a matter of plotting purposes (since the mode at $f=0$ is available on the basis), while for the DMD/OPD, the time-averaged flow removal turned out to have no impact on the presented results. 

\begin{figure}[h!]
	\centering
	\includegraphics[width=0.65\textwidth]{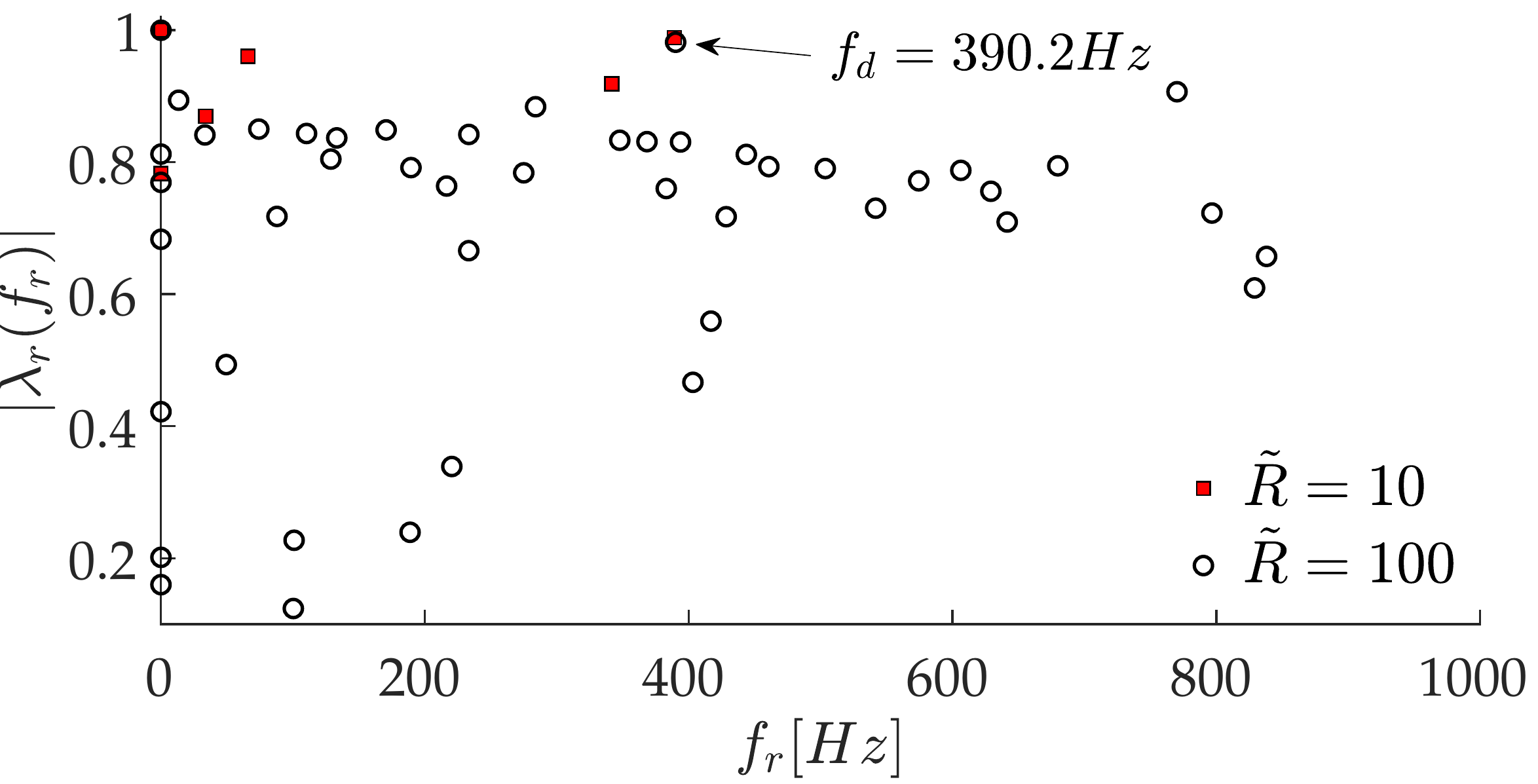}
	\caption{Eigenvalue spectra from the DMD analysis in of the full dataset (transient conditions). The computed dominant frequency is somewhere in between the two expected ones. }
	\label{DMD_SP2}
\end{figure}

The eigenvalue spectra of the DMD/OPD with the same reduced propagator described in the previous case is shown in Figure \ref{DMD_SP2}. A dominant frequency of $f_d=390.2 Hz$ is observed. This is somewhere in between the one linked to the vortex shedding in the first ($f_d\approx 450 Hz$) and the second ($f_d\approx 303 Hz$) stationary conditions. As for the previous case, this decomposition does not converge to the dataset. The associated spatial structures are, therefore, not shown.

\begin{figure}[h!]
	\centering
	\begin{subfigure}[t]{0.6\textwidth}
		\centering
		\includegraphics[width=0.75\textwidth]{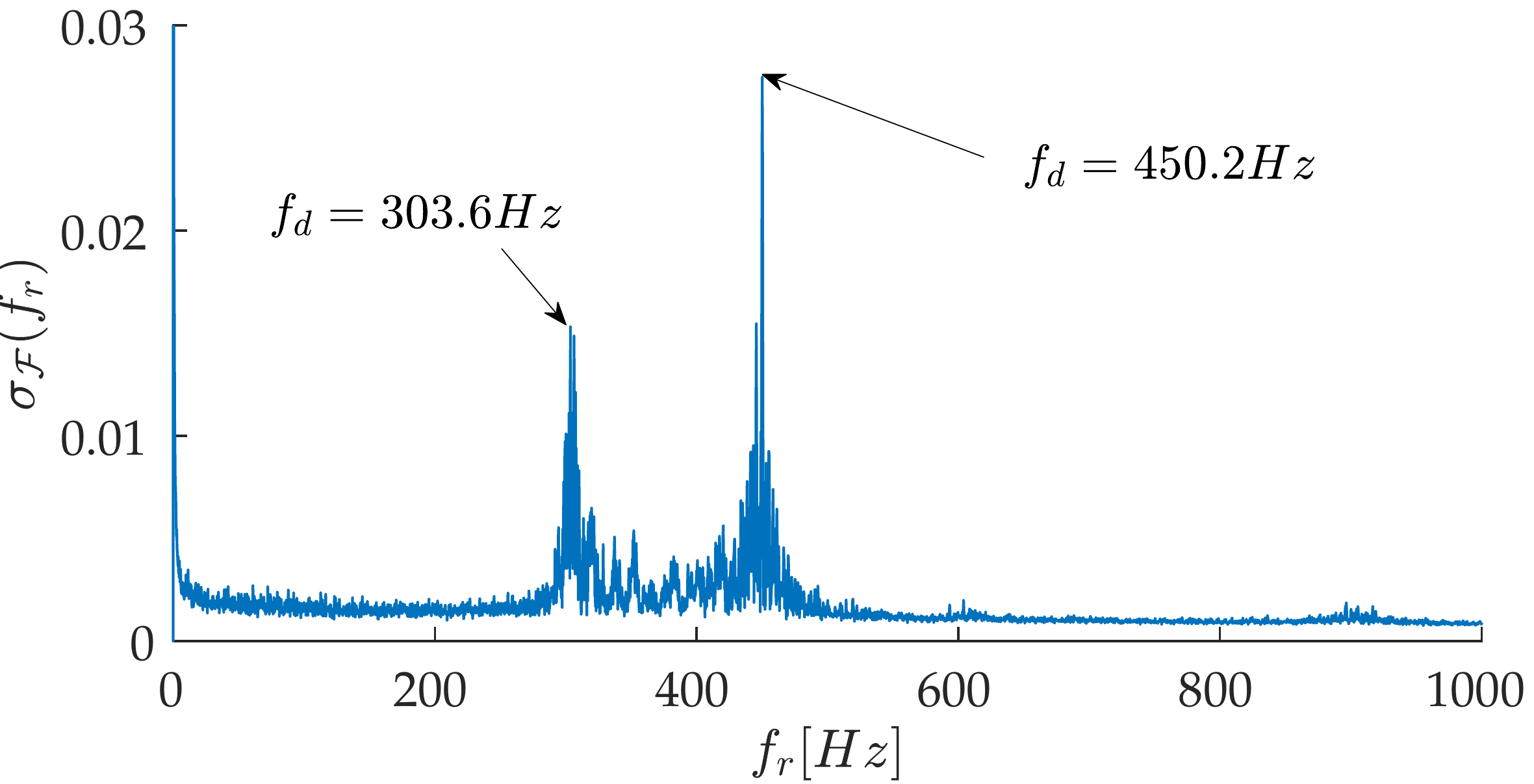}\\a)
		\vspace{1mm}
	\end{subfigure}
	\begin{subfigure}[t]{0.65\textwidth}
		\centering
		\includegraphics[width=0.75\textwidth]{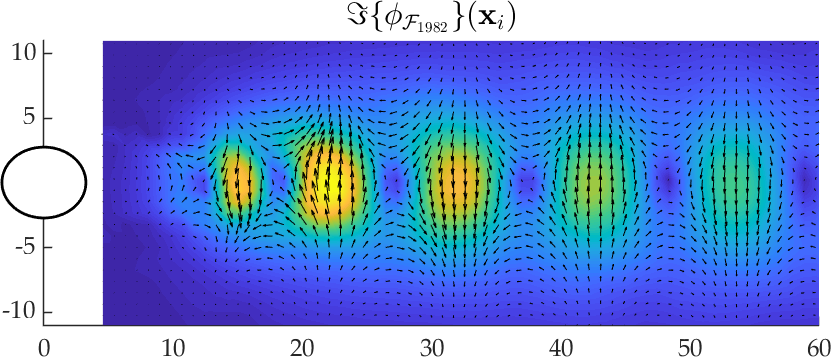}\\b)
		\vspace{1mm}
	\end{subfigure}
	\begin{subfigure}[t]{0.68\textwidth}
		\centering
		\includegraphics[width=0.75\textwidth]{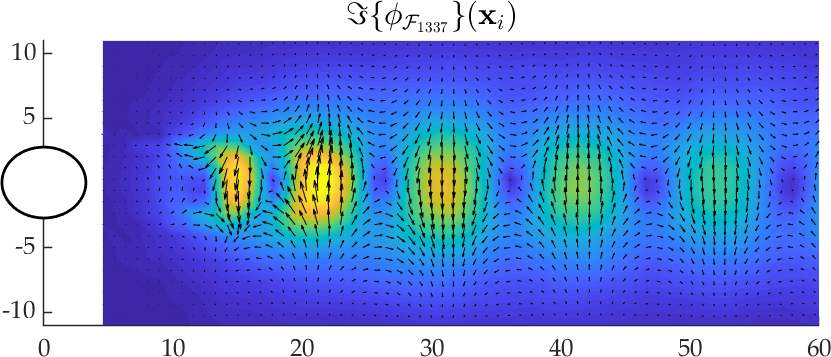}\\c)
		\vspace{1mm}
	\end{subfigure}
	\caption{(a) Spectra of the DFT amplitudes ($diag(\Sigma_{\mathcal{D}})$) for the transient test case, showing the two harmonics dominating in the two stationary regimes. The imaginary part of the spatial structures associated to the frequency of the shedding at $U_{\infty}=12.1 m/s$ (that is $f_d=450 Hz$) is shown in (b) while the one associated to the frequency of the shedding at $U_{\infty}=7.9 m/s$ (that is $f_d=303.6 Hz$) is shown in (c).}
	\label{ResDFT_UNST}
\end{figure}

The results of the DFT are shown in Figure \ref{ResDFT_UNST}. Figure \ref{ResDFT_UNST}a) shows the energy distribution as a function of the associated frequency. The frequencies of the vortex shedding in the two stationary test cases are visible and labeled. Figures \ref{ResDFT_UNST}b) and \ref{ResDFT_UNST}c) shows the real parts of the spatial structures associated with the dominant frequencies in the two shedding regimes.

These structures are similar: the first is associated with the stationary regime at $U_{\infty}=12.1 m/s$ while the second is linked to the stationary regime at $U_{\infty}=7.9 m/s$. The saddle pattern observed in the previous stationary test case is present in both, with a slight shift linked to the different free stream velocity in which they become more representative. On the other hand, since these modes are purely harmonic, it is not possible to infer any temporal information regarding their occurrence. Moreover, since these modes are obtained by projecting the entire dataset onto harmonics that are mostly present in different time intervals, a distortion is produced in the spatial structures if compared to the ones obtained in an entirely stationary test case. This can be observed by comparing the imaginary part of the structure at Figure \ref{ResDFT_UNST}b) with the one in Figure \ref{ResDFT}c) where only the stationary case was considered.

\begin{figure}[h!]
	\centering
	\begin{subfigure}[t]{0.65\textwidth}
		\centering
		\includegraphics[width=0.75\textwidth]{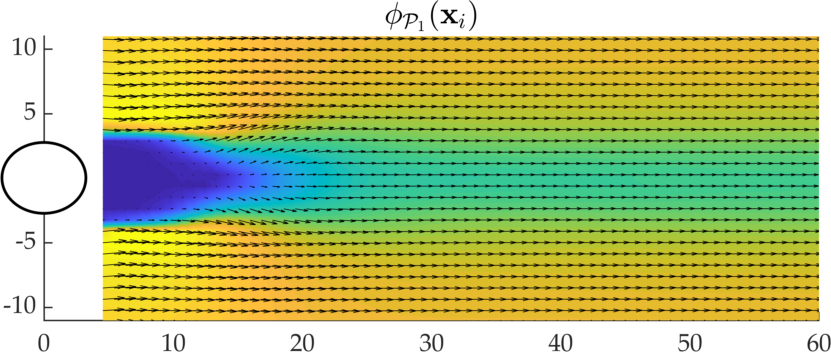}\\a)
		\vspace{1mm}
	\end{subfigure}\\
	\begin{subfigure}[t]{0.55\textwidth}
		\centering
		\includegraphics[width=0.75\textwidth]{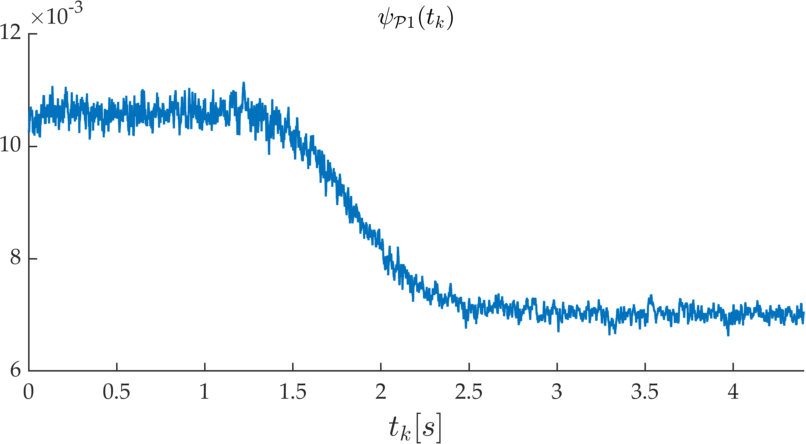}\\b)
		\vspace{1mm}
	\end{subfigure}\\
	\caption{Spatial structure (a) and temporal structure (b) of the dominant POD mode for the transient test case.}
	\label{POD_1}
\end{figure}

The POD is now discussed. Since the temporal average is not removed, it is of interest to consider the first mode separately, which accounts for about $85.5\%$ of decomposition convergence, i.e., $\mathcal{E}(1)=15.5\%$ from \eqref{ERROR}. Its spatial and temporal structures are shown in Figure \ref{POD_1}, and describe the large scale variation of the flow. In the general framework of a triple decomposition \citep{Reynolds1972}, this mode could be seen as the base flow, i.e., the slowly varying contribution or `shift mode'  \citep{NOACK2003,Tadmor2010,Bourgeois2013}. Nevertheless, since the POD sets no constraints on the frequency content of its modes, the temporal evolution exhibits additional higher frequencies.

\begin{figure}[h!]
	\centering
	\begin{subfigure}[t]{0.65\textwidth}
		\centering
		\includegraphics[width=0.75\textwidth]{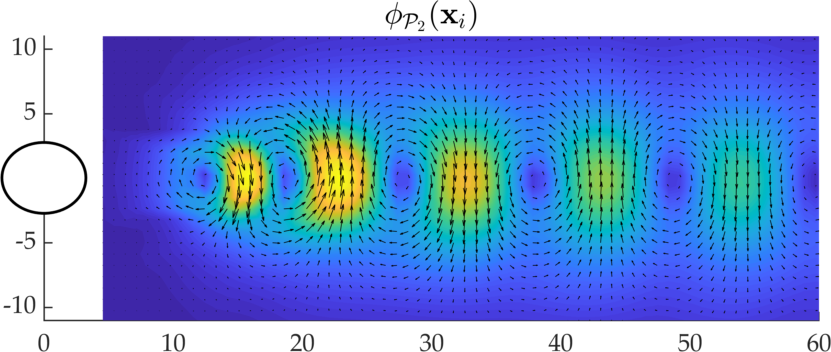}\\a)
		\vspace{1mm}
	\end{subfigure}\\
	\begin{subfigure}[t]{0.65\textwidth}
		\centering
		\includegraphics[width=0.75\textwidth]{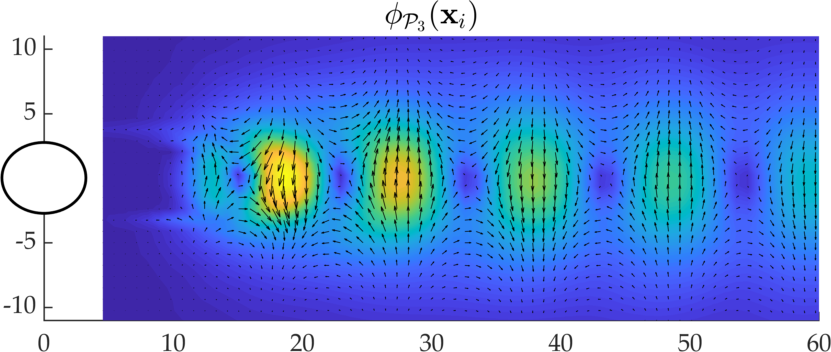}\\b)
		\vspace{1mm}
	\end{subfigure}\\
	\begin{subfigure}[t]{0.58\textwidth}
		\centering
		\includegraphics[width=0.75\textwidth]{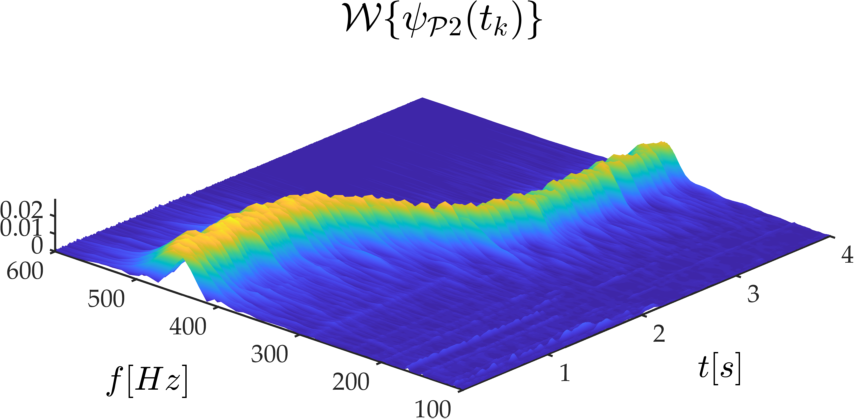}\\c)
		\vspace{1mm}
	\end{subfigure}\\
	\begin{subfigure}[t]{0.58\textwidth}
		\centering 
		\includegraphics[width=0.75\textwidth]{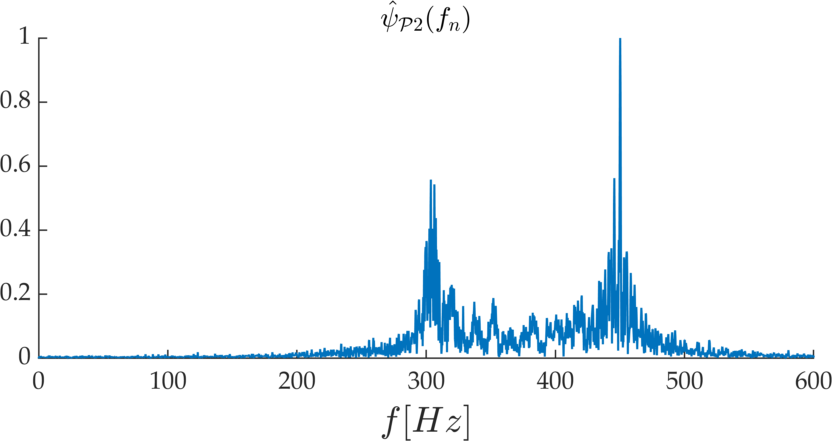}\\d)
	\end{subfigure} 
	\begin{subfigure}[t]{0.35\textwidth}
		\centering
		\includegraphics[width=0.75\textwidth]{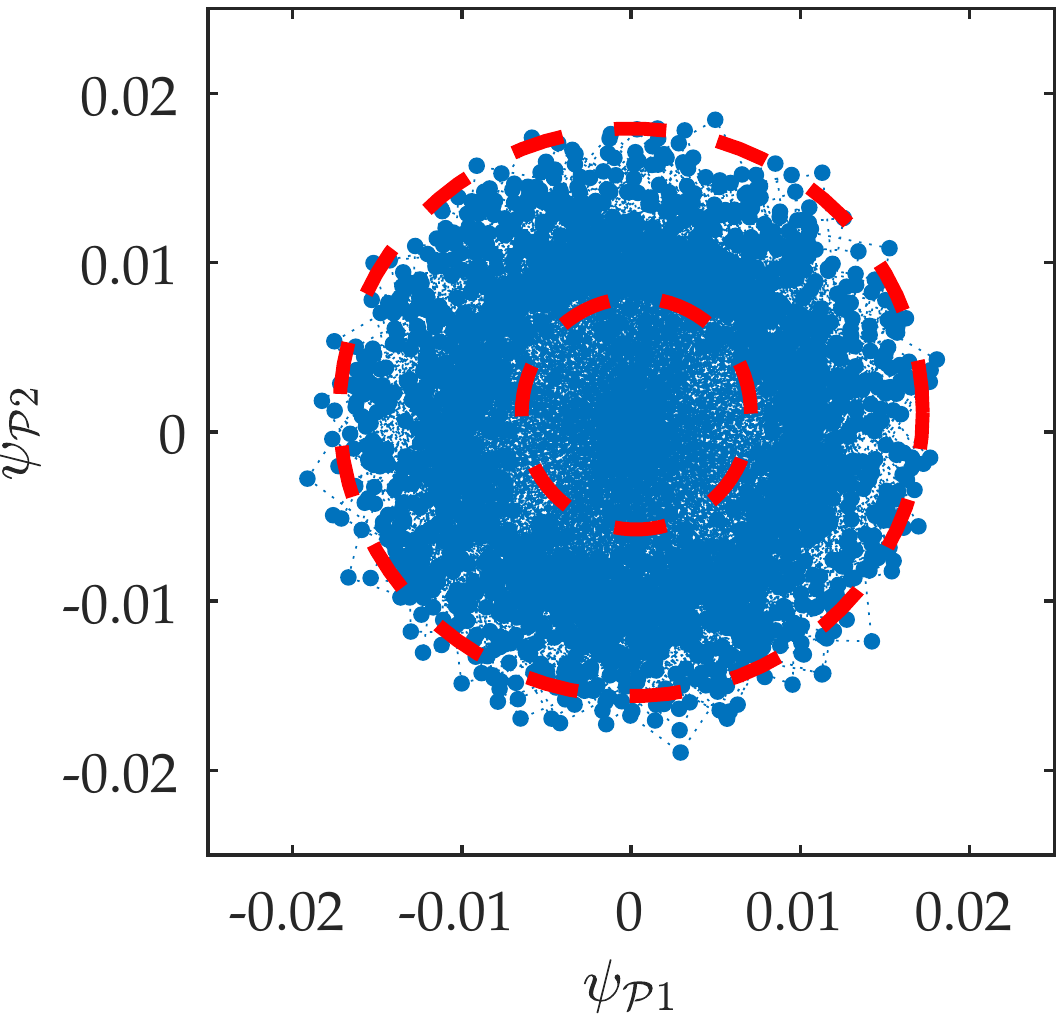}\\e)
	\end{subfigure}%
	\caption{Second and third POD modes for the transient case. Figure structured as Figure \ref{POD_RESULTS}. }
	\label{POD_RESULTS_UNST}
\end{figure}

The first pair of POD modes linked to the shedding frequencies (modes 2 and 3) are shown in Figure \ref{POD_RESULTS_UNST}, following the same structure as Figure \ref{POD_RESULTS}.
These two POD modes account for an additional $11.54\%$ of the decomposition convergence, and the error produced by an approximation including the first three POD modes is $\mathcal{E}(3)=3.46\%$. These POD modes describe the evolution of the vortex shedding in the \emph{entire} dataset, i.e., including both the stationary conditions and the transient. The CWT of their temporal structure shows how the frequency changes in time and evolves, interestingly, so as to keep the Strouhal number constant also during the transitory. The DFT of their temporal structure shows the two peaks and is, again, very close to the DFT of the entire dataset shown in Figure \ref{ResDFT_UNST}. As a result, the phase portrait is difficult to interpret, and so are the associated spatial structures that are, in fact, a combination of the ones to be expected for the different phases, namely from Figure \ref{FIG1}c), the steady-state 1, the transitory, and the steady-state 2.

\begin{figure}[h!]
	\centering
	\includegraphics[width=0.3\textwidth]{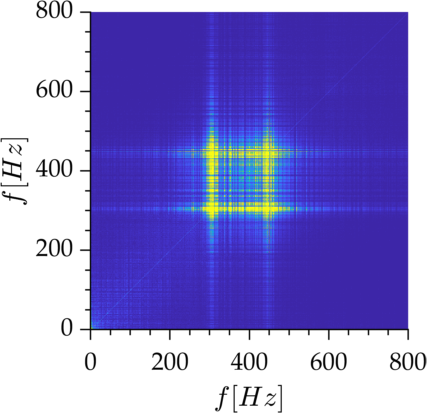}\hspace{2mm}
	\includegraphics[width=0.6\textwidth]{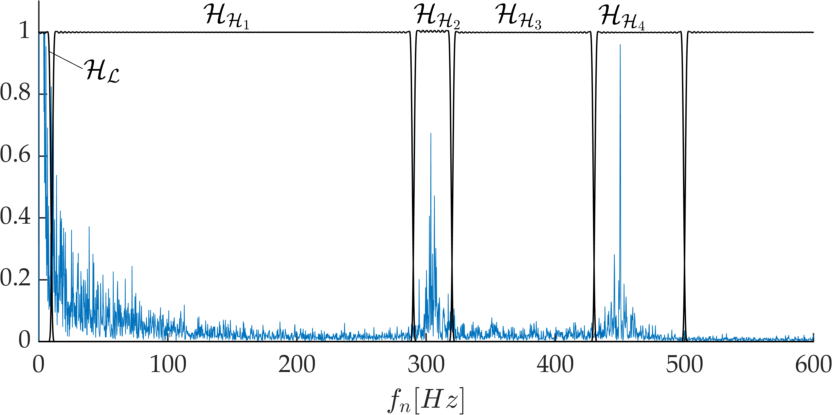} \\
	\hspace{2mm}a) \hspace{55mm} b)\hspace{22mm}
	\caption{Same as Figure \ref{MRA_STATS}, but for the transient test case. The spectrum is divided into six scales. $H_{\mathcal{L}}$ is associated with the large scale variation of the flow, $H_{\mathcal{H}2}$ and $H_{\mathcal{H}4}$ with the vortex shedding in the two steady states, while the remaining scales cover the intermediate and high frequency portions.}
	\label{MRA_STATS_UNST}
\end{figure}

\begin{figure}[h!]
	\centering
	\begin{subfigure}[t]{0.65\textwidth}
		\centering
		\includegraphics[width=0.75\textwidth]{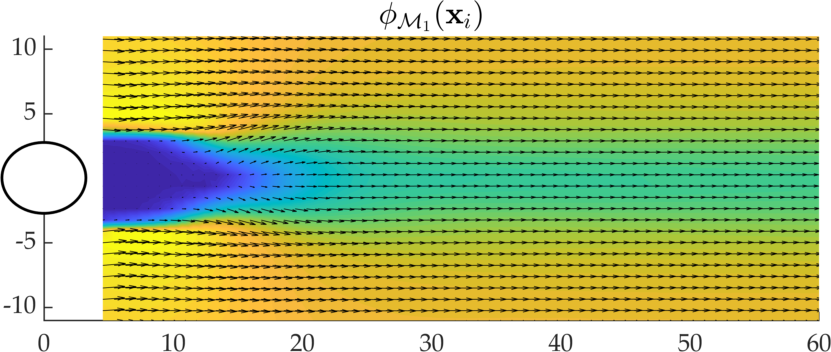}\\a)
		\vspace{1mm}
	\end{subfigure}\\
	\begin{subfigure}[t]{0.55\textwidth}
		\centering
		\includegraphics[width=0.75\textwidth]{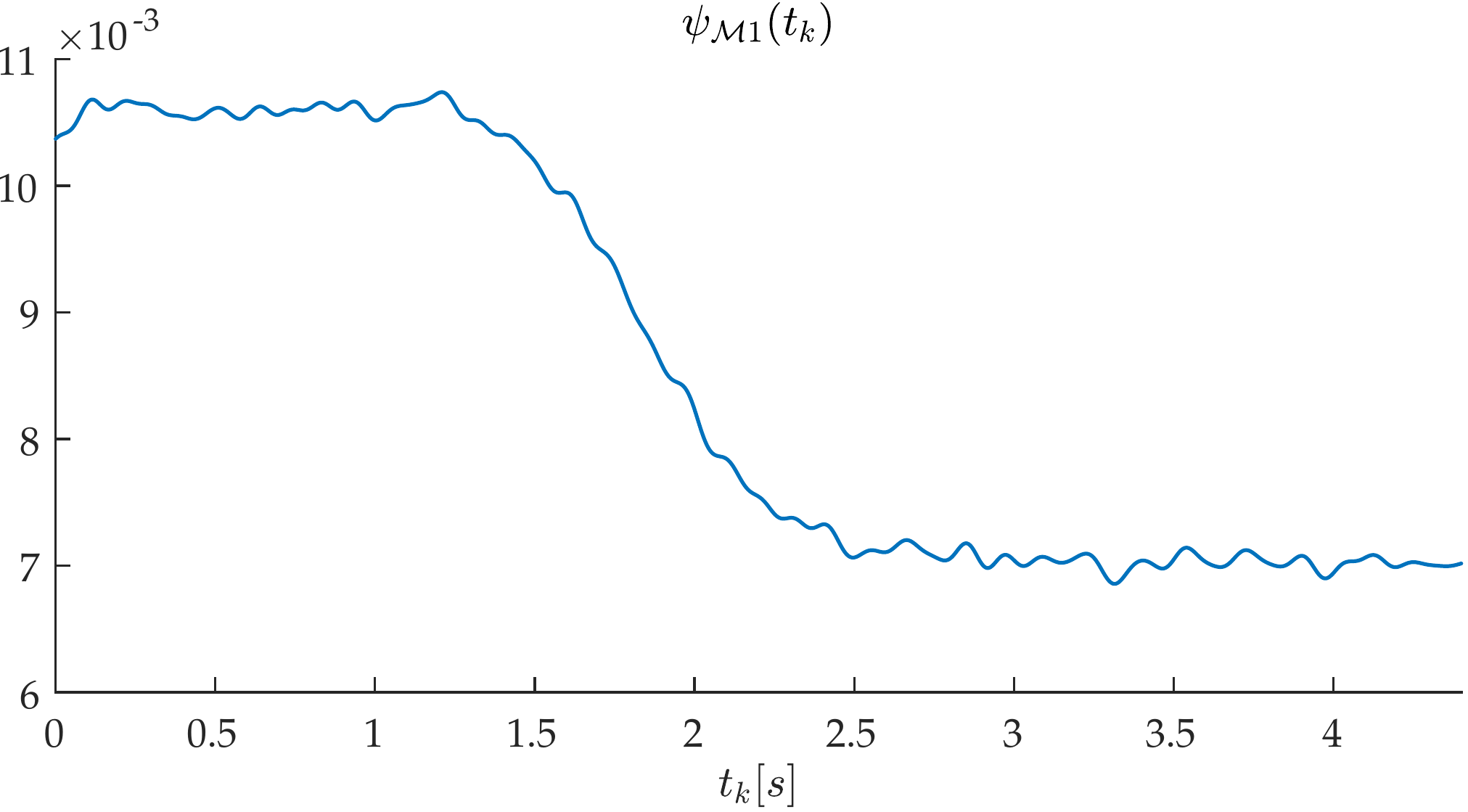}\\b)
		\vspace{1mm}
	\end{subfigure}\\
	\caption{Spatial structure (a) and temporal structure (b) of the dominant mPOD mode for the transient test case.}
	\label{mPOD_1}
\end{figure}

The results from the mPOD are now discussed. For the purpose of this test case, it is interesting to separate the modes that are associated with the two steady conditions and the ones associated with the transient. As with the previous test case, Figure \ref{MRA_STATS_UNST}a) shows the contour map of the modulus of the Fourier transformed transfer function $\widehat{K}$ while Figure \ref{MRA_STATS_UNST}b) shows its diagonal together with the filter transfer function used to identify the scales. The dataset is partitioned into six scales:

\begin{center}
  \begin{tabular}{| l | c | c | c | c | c | c | c |}
    \hline
    Scale &
    $\mathcal{H}_{\mathcal{L}}$ &
    $\mathcal{H}_{1}$ &
    $\mathcal{H}_{2}$ &
    $\mathcal{H}_{3}$ &
    $\mathcal{H}_{4}$ &
    $\mathcal{H}_{5}$ &
    \\
    \hline
    $|\mathcal{H}_{n}|\approx 1 \,\forall f \in$ & 
    $[0,10]$ & 
    $[10,290]$ & 
    $[290,320]$ & 
    $[320,430]$ & 
    $[430,470]$ & 
    $[470,1500]$ & 
    $Hz$ \\
    \hline
  \end{tabular}
\end{center}

The first scale is focused on the slow variation of the flow, identified by the low pass filter $\mathcal{H}_{\mathcal{L}}$. The third scale, $\mathcal{H}_{2}$, is centered around the first peak, while the fifth scale, $\mathcal{H}_{4}$, is centered around the second peak. Scales $\mathcal{H}_{1}$, $\mathcal{H}_{3}$ and $\mathcal{H}_{5}$ cover the remaining portions of the spectrum.

\begin{figure}[h!]
	\centering
	\begin{subfigure}[t]{0.65\textwidth}
		\centering
		\includegraphics[width=0.75\textwidth]{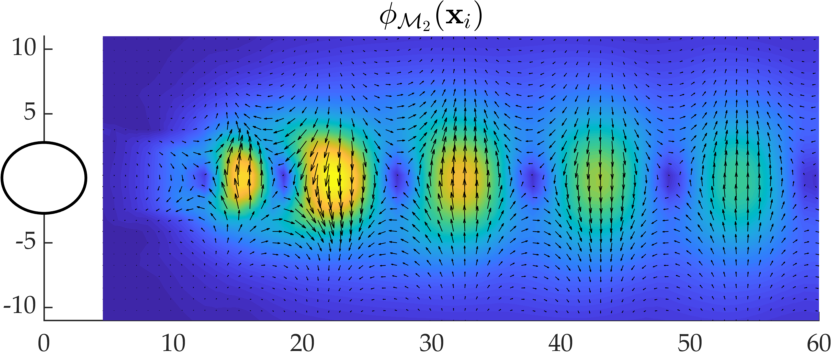}\\a)
		\label{POD_RESULTSa}
		\vspace{1mm}
	\end{subfigure}\\
	\begin{subfigure}[t]{0.65\textwidth}
		\centering
		\includegraphics[width=0.75\textwidth]{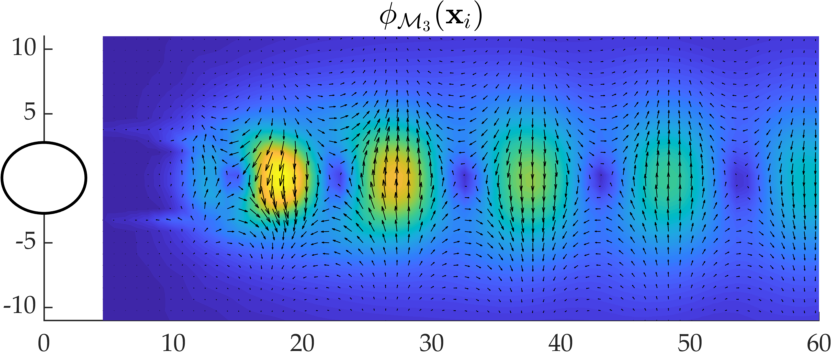}\\b)
		\vspace{1mm}
		\label{POD_RESULTSb}
	\end{subfigure}\\
	\begin{subfigure}[t]{0.55\textwidth}
		\centering
		\includegraphics[width=0.75\textwidth]{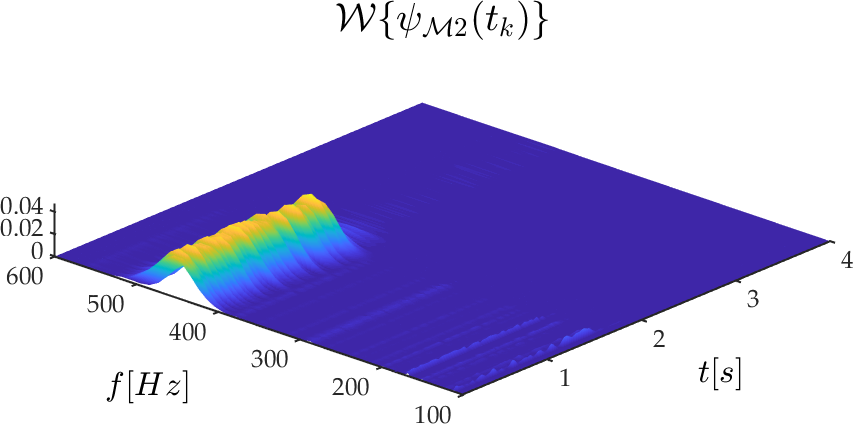}\\c)
		\vspace{1mm}
		\label{POD_RESULTSc}
	\end{subfigure}\\
	
	\begin{subfigure}[t]{0.58\textwidth}
		\centering 
		\includegraphics[width=0.75\textwidth]{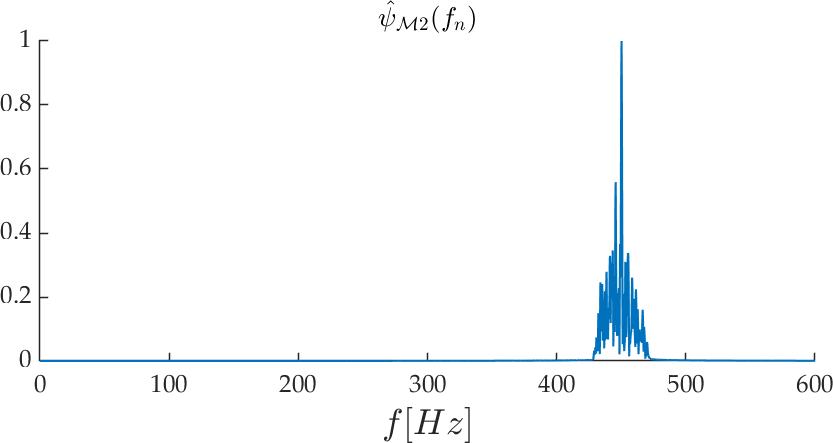}\\d)
		\label{POD_RESULTSd}
	\end{subfigure} 
	\begin{subfigure}[t]{0.35\textwidth}
		\centering
		\includegraphics[width=0.75\textwidth]{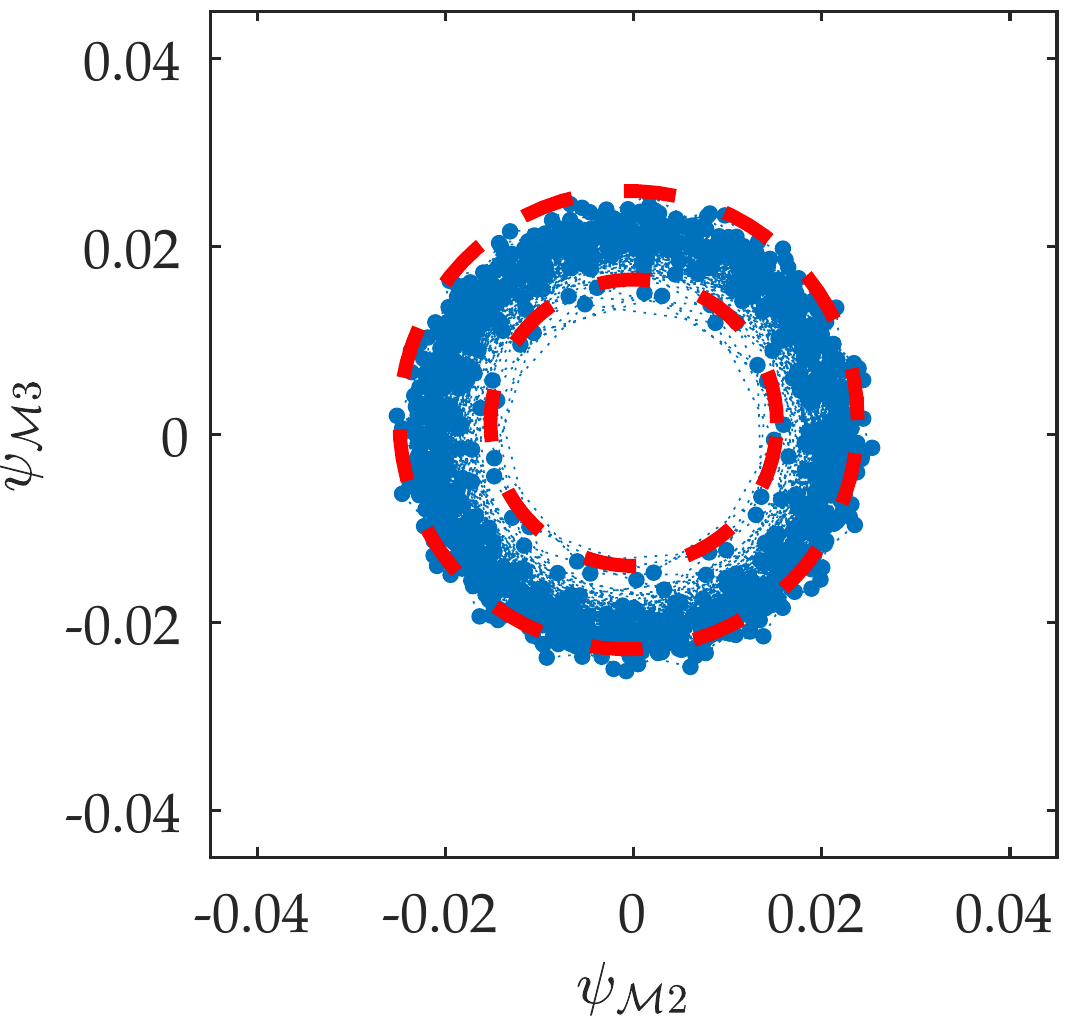}\\e)
		\label{POD_RESULTSe}
	\end{subfigure}%
	\caption{Same as Figure \ref{POD_RESULTS_UNST}, but for the second and third mPOD modes.}
	\label{fullmPOD_UNST1}
\end{figure}

\begin{figure}[h!]
	\centering
	\begin{subfigure}[t]{0.65\textwidth}
		\centering
		\includegraphics[width=0.75\textwidth]{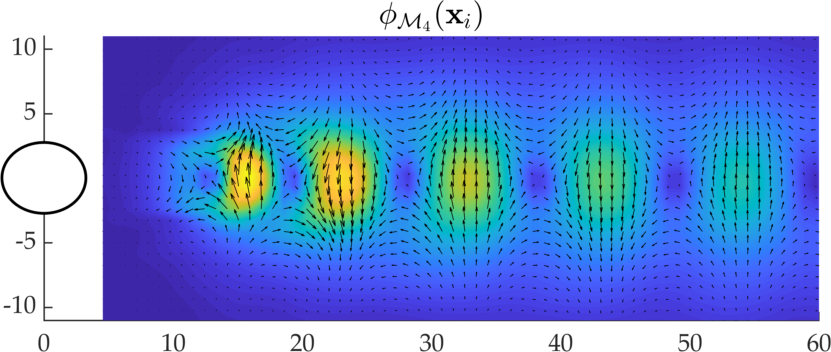}\\a)
		\label{POD_RESULTSa}
		\vspace{1mm}
	\end{subfigure}\\
	\begin{subfigure}[t]{0.65\textwidth}
		\centering
		\includegraphics[width=0.75\textwidth]{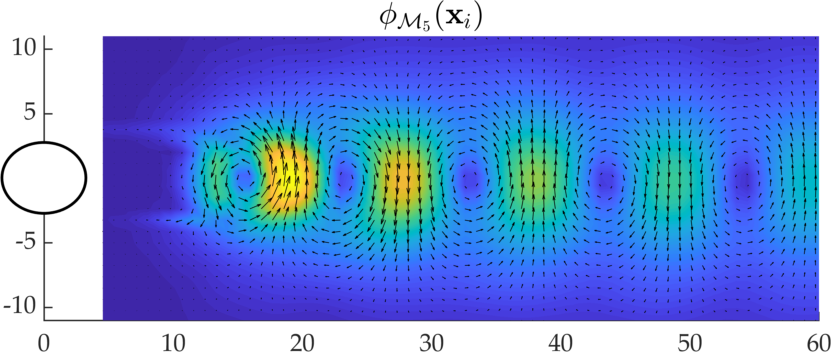}\\b)
		\vspace{1mm}
		\label{POD_RESULTSb}
	\end{subfigure}\\
	\begin{subfigure}[t]{0.55\textwidth}
		\centering
		\includegraphics[width=0.75\textwidth]{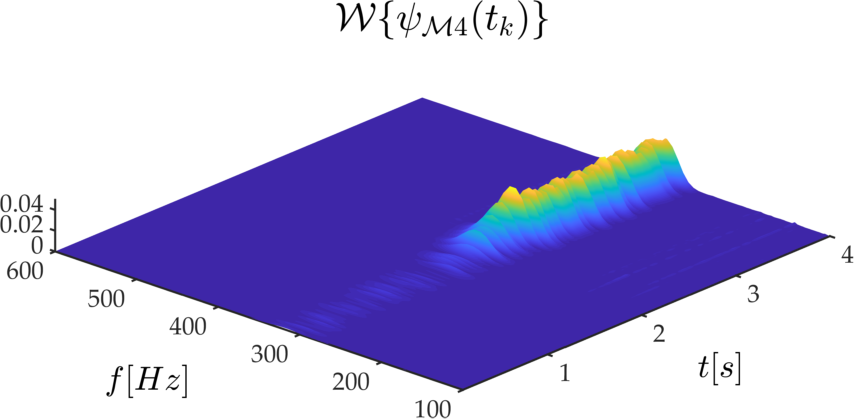}\\c)
		\vspace{1mm}
		\label{POD_RESULTSc}
	\end{subfigure}\\
	
	\begin{subfigure}[t]{0.58\textwidth}
		\centering 
		\includegraphics[width=0.75\textwidth]{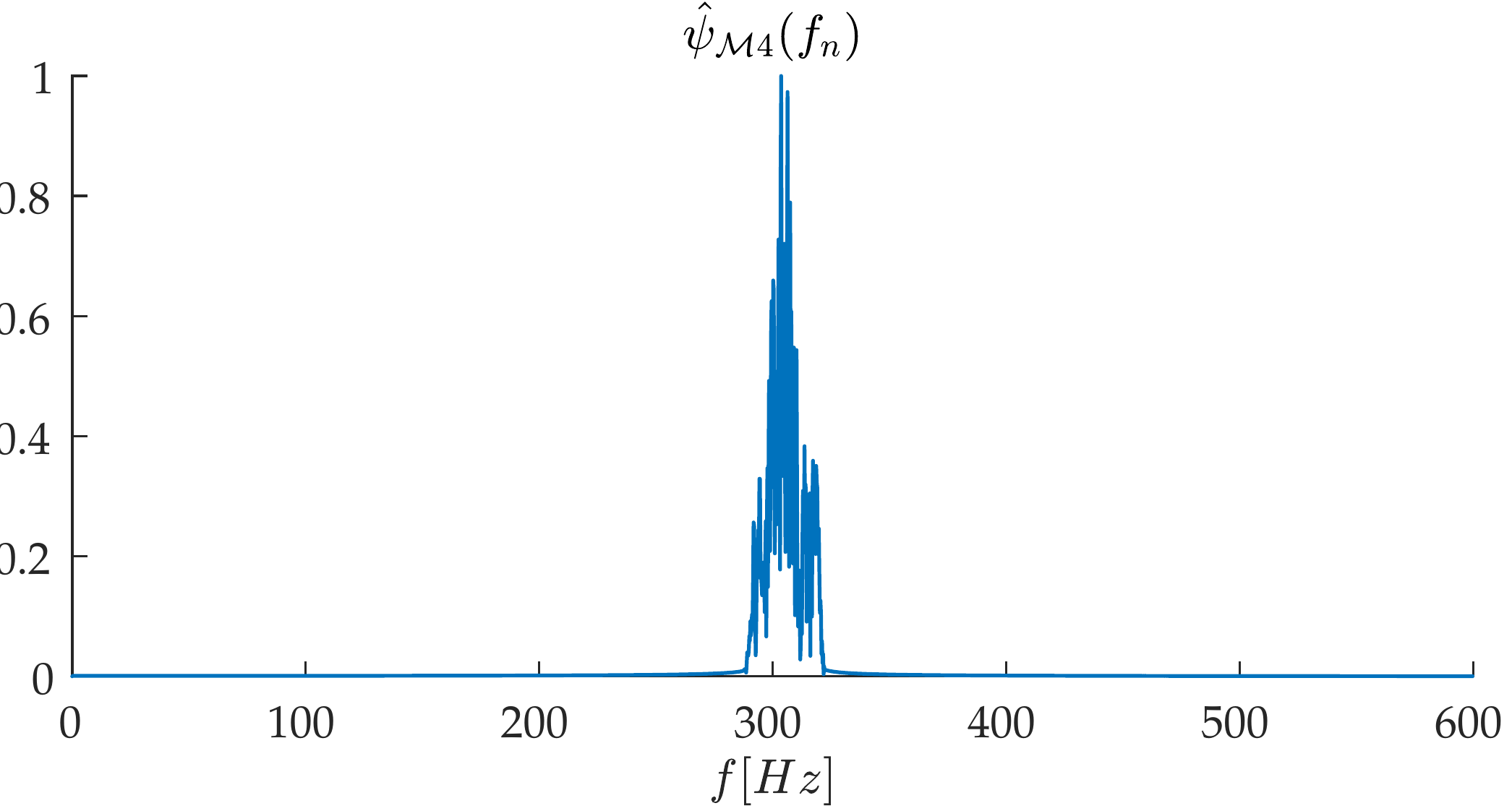}\\d)
		\label{POD_RESULTSd}
	\end{subfigure} 
	\begin{subfigure}[t]{0.35\textwidth}
		\centering
		\includegraphics[width=0.75\textwidth]{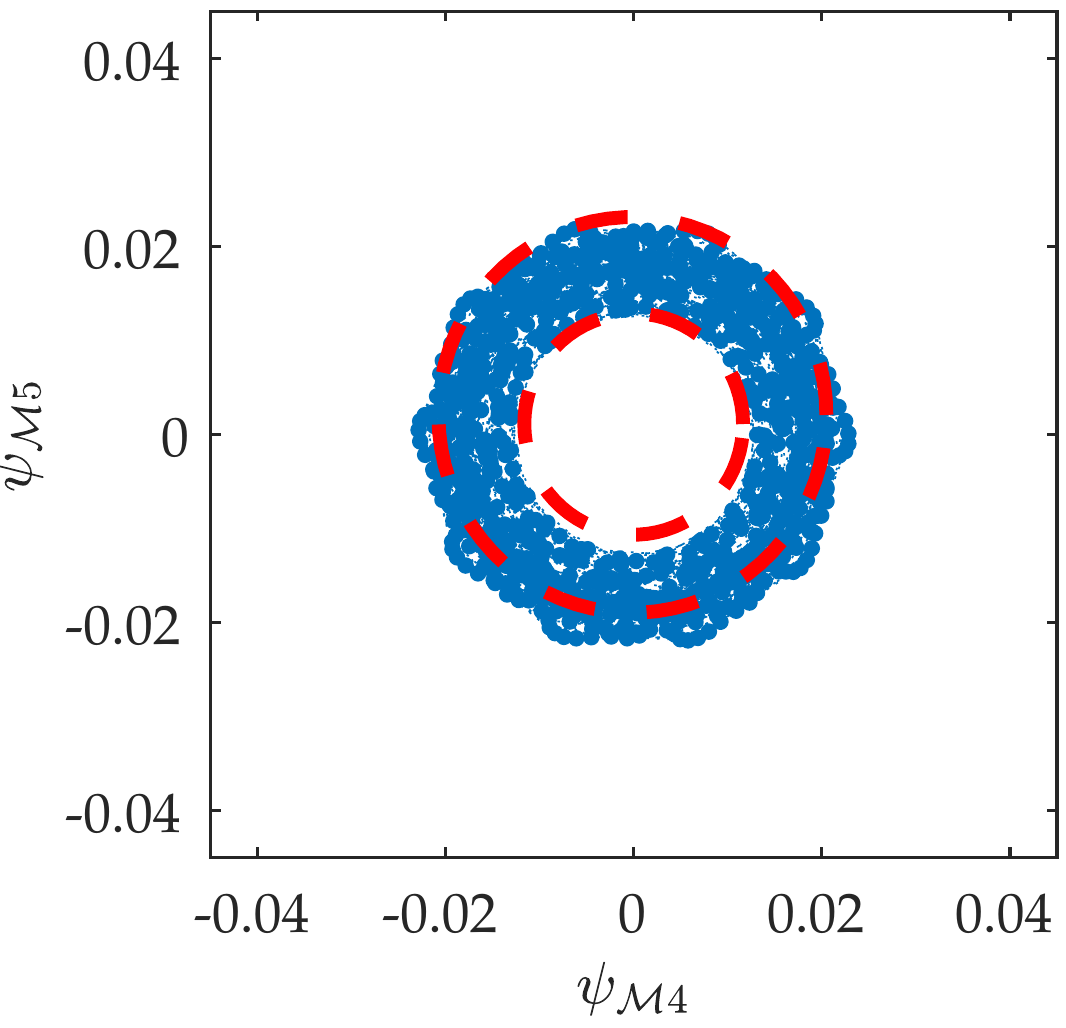}\\e)
		\label{POD_RESULTSe}
	\end{subfigure}%
	\caption{Same as Figure \ref{POD_RESULTS_UNST}, but for the fourth and fifth mPOD modes.}
	\label{fullmPOD_UNST2}
\end{figure}

\begin{figure}[h]
	\centering
	\begin{subfigure}[t]{0.65\textwidth}
		\centering
		\includegraphics[width=0.75\textwidth]{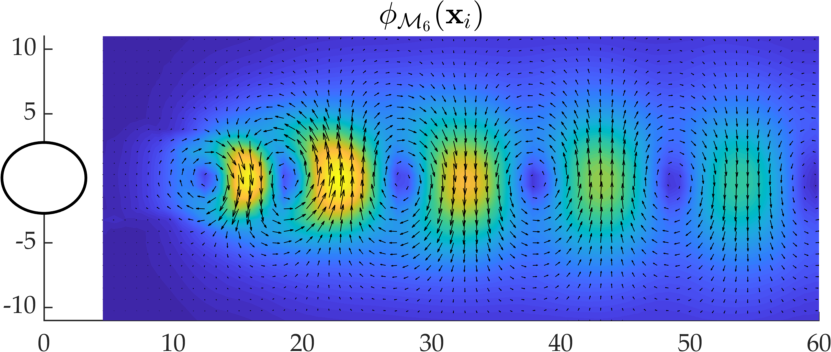}\\a)
		\label{POD_RESULTSa}
		\vspace{1mm}
	\end{subfigure}\\
	\begin{subfigure}[t]{0.65\textwidth}
		\centering
		\includegraphics[width=0.75\textwidth]{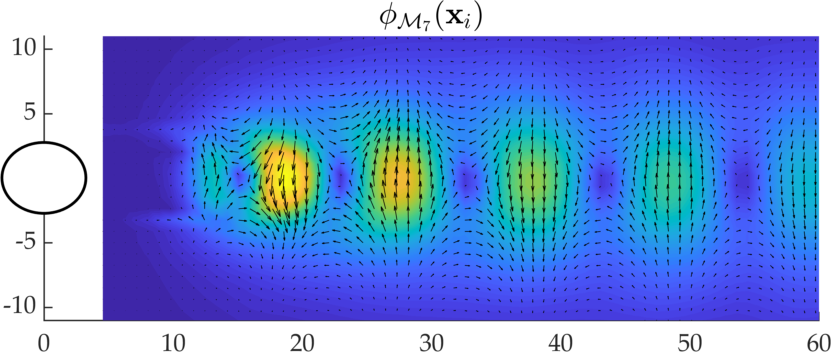}\\b)
		\vspace{1mm}
		\label{POD_RESULTSb}
	\end{subfigure}\\
	\begin{subfigure}[t]{0.55\textwidth}
		\centering
		\includegraphics[width=0.75\textwidth]{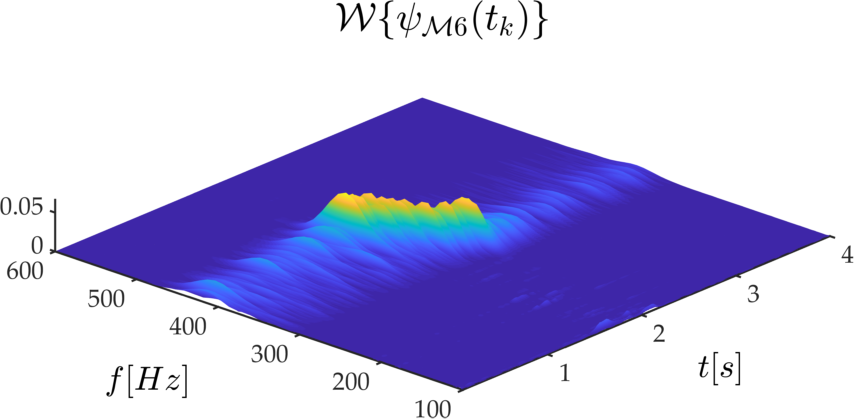}\\c)
		\vspace{1mm}
		\label{POD_RESULTSc}
	\end{subfigure}\\
	
	\begin{subfigure}[t]{0.58\textwidth}
		\centering 
		\includegraphics[width=0.75\textwidth]{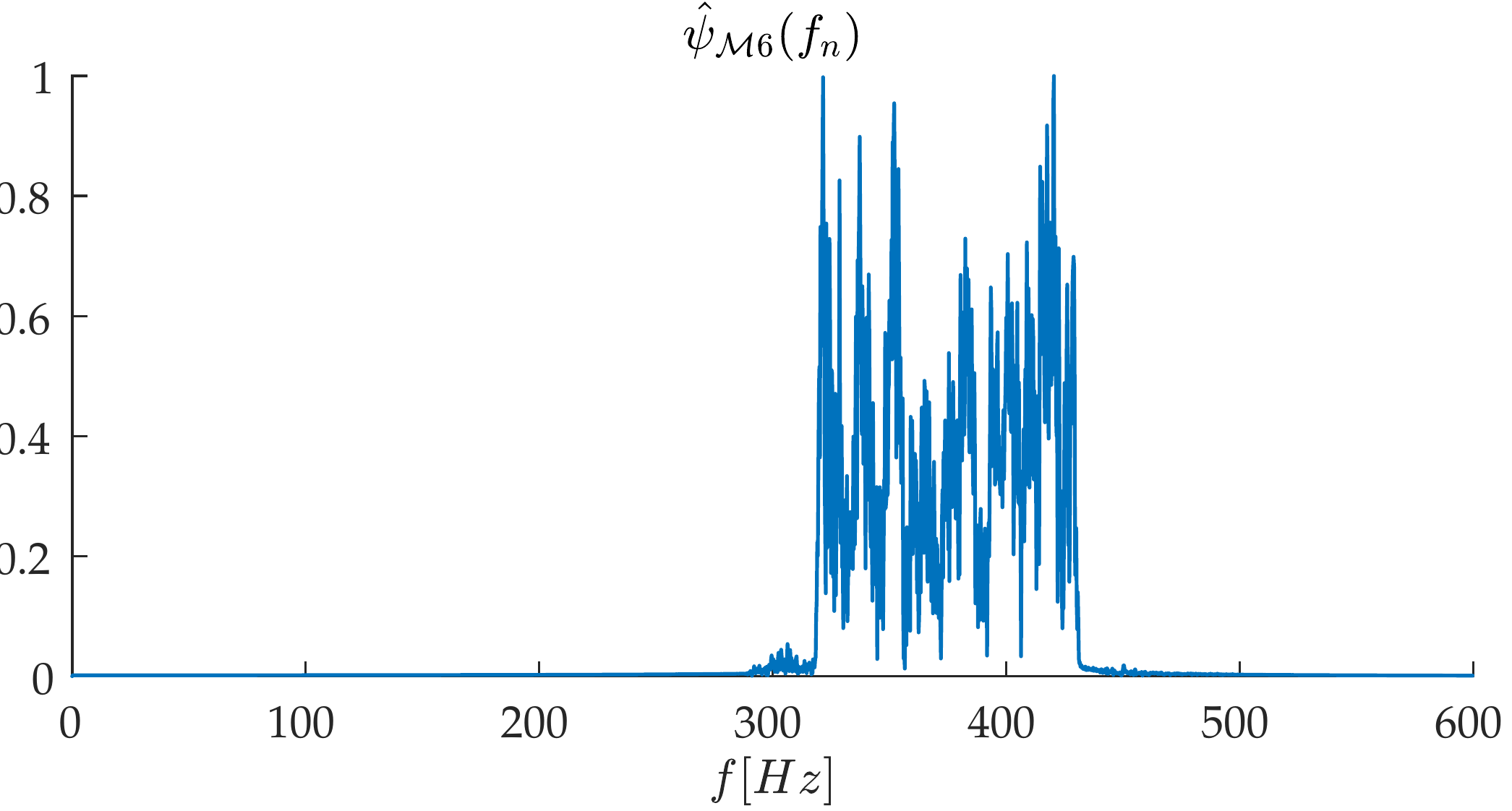}\\d)
		\label{POD_RESULTSd}
	\end{subfigure} 
	\begin{subfigure}[t]{0.35\textwidth}
		\centering
		\includegraphics[width=0.75\textwidth]{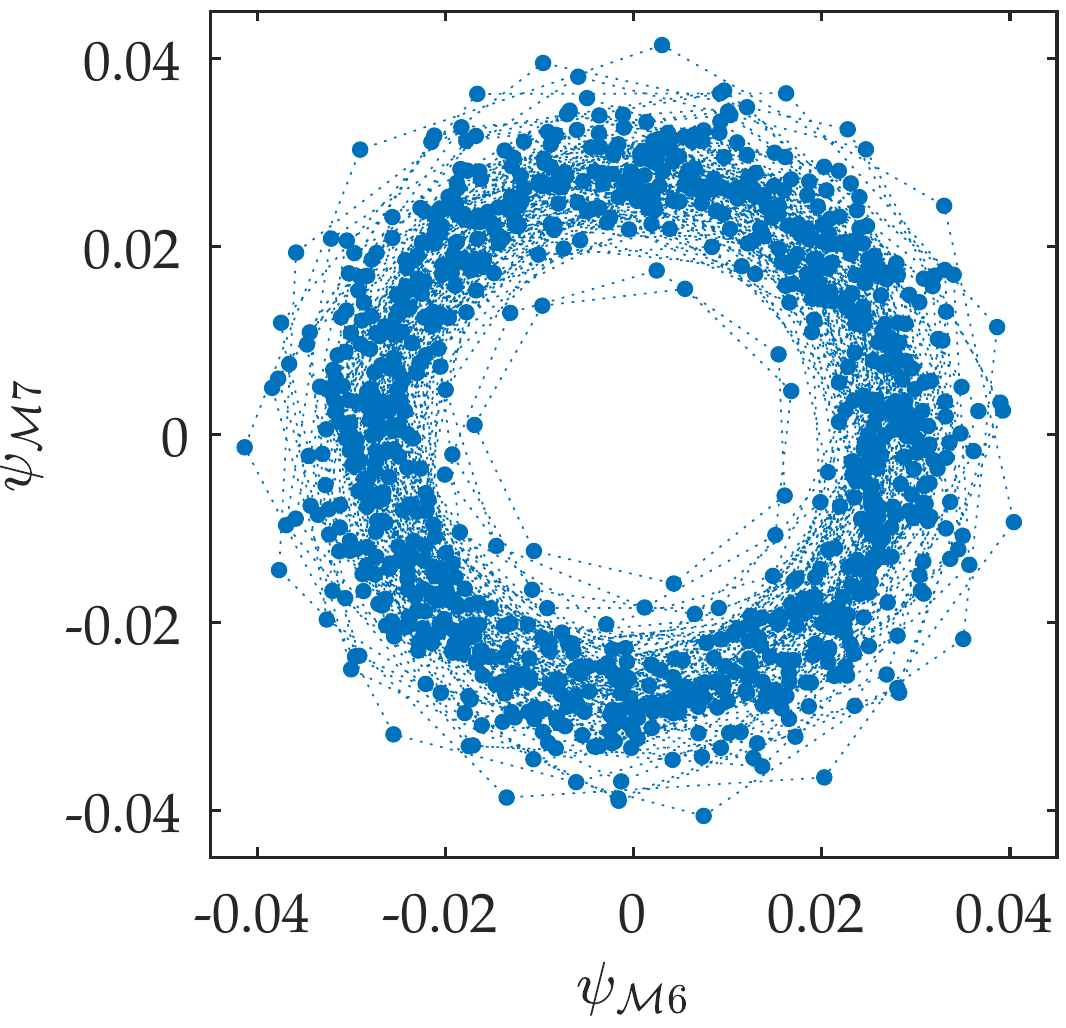}\\e)
		\label{POD_RESULTSe}
	\end{subfigure}%
	\caption{Same as Figure \ref{POD_RESULTS_UNST}, but for the sixth and seventh mPOD modes.}
	\label{fullmPOD_UNST3}
\end{figure}

The spatial structure and the temporal evolution of the first mPOD modes are shown in Figure \ref{mPOD_1}. This mode is analogous to the first POD mode in Figure \ref{POD_1}, with a smoother temporal evolution as its frequency content is bounded to $f<10Hz$. The remaining mPOD modes, obtained in pairs of equal energy, are shown in Figures \ref{fullmPOD_UNST1}, \ref{fullmPOD_UNST2}, and \ref{fullmPOD_UNST3}, keeping the usual structure of the figure as in the previous subsection.

The first pair of modes in Figure \ref{fullmPOD_UNST1} is associated to the first stationary condition, with no significant contribution appearing outside the spectral bandwidth that characterizes the first vortex shedding. Interestingly, this mode has also no appreciable contribution for $t>1.5s$, as shown by the CWT of its temporal structure in \ref{fullmPOD_UNST1}c): while the filters bound the spectral constraints of the mode, the diagonalization of the correlation matrix at each of the corresponding scales also allows for temporal localization. The associated spatial structures resemble more closely the ones observed in the stationary test case investigated in the previous section (see Figure \ref{fullMPOD_STAT1}a). Similarly, the second pair of modes in  Figure \ref{fullmPOD_UNST2} captures the vortex shedding in the second steady-state condition, while Figure \ref{fullmPOD_UNST3} allows analyzing the modes associated with the flow transient.

\begin{figure}[h!]
	\centering
	\begin{subfigure}[t]{0.7\textwidth}
		\centering 
		\includegraphics[width=0.8\textwidth]{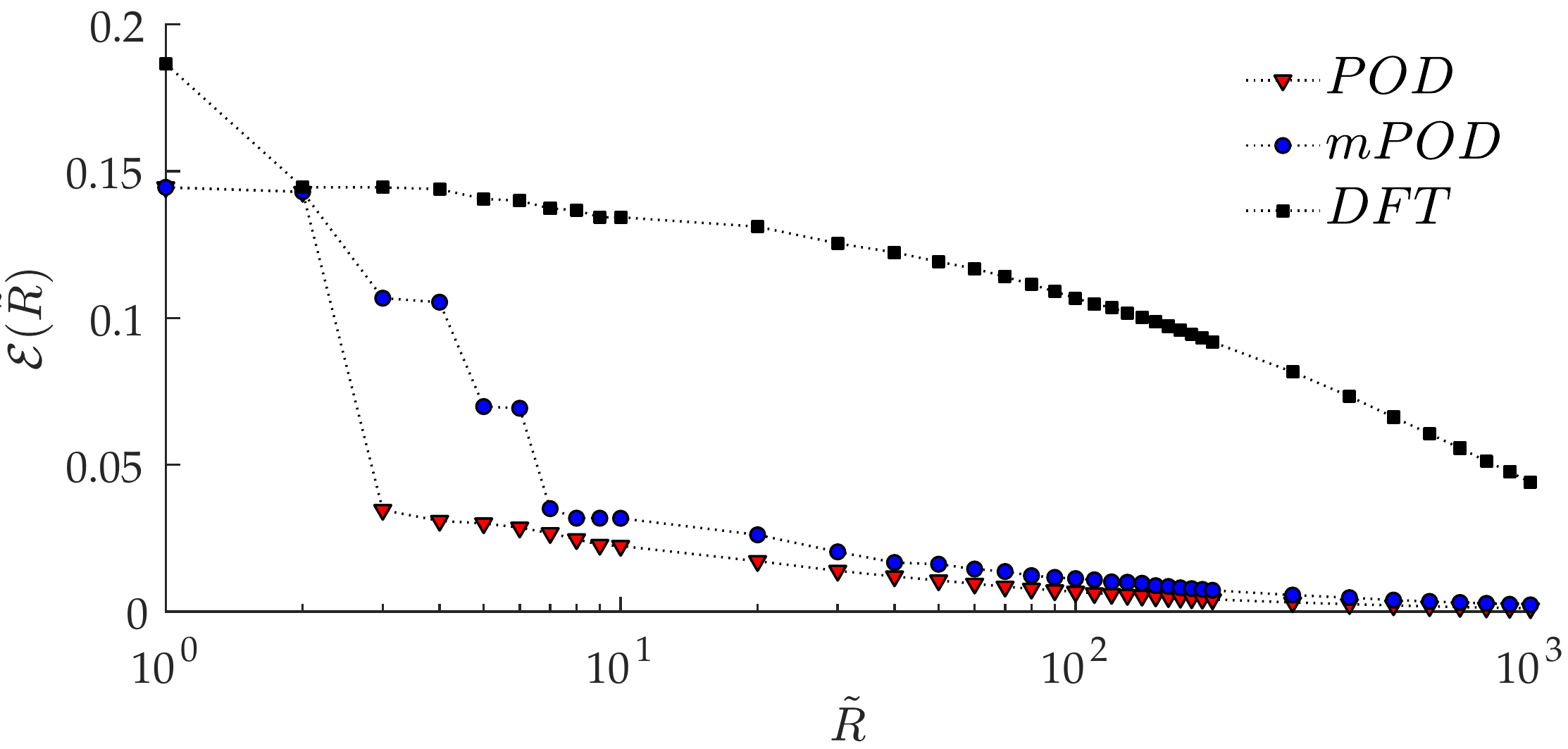}\\a)
		\label{conv1}
	\end{subfigure} 
	\caption{Decomposition error $\mathcal{E}$ in \eqref{ERROR} as a function of the number of modes included in the summation \eqref{EQ1} for the POD, the mPOD and the DFT of the transient test case. }
	\label{DECOconvUNST}
\end{figure}

To conclude, Figure \ref{DECOconvUNST} shows the decomposition convergence for the POD, the DFT, and the mPOD for the transient test case. Recalling that for this dataset the temporal average has not been removed,  the decomposition convergence starts in the DFT from $\mathcal{E}_{\mathcal{F}}(1)=18.7\%$-- that is the temporal average accounts for $81.3\%$ of the convergence. Since the spectrum has been here partitioned more than in the previous case, the loss of decomposition convergence of the mPOD over the POD is here slightly more pronounced. The information in the first three POD modes is now distributed over the first seven mPOD modes. Nevertheless, the mPOD allows for much better convergence than the DFT while enabling localization both in the frequency and in the time domain and hence-- through a projection of the data-- in space. 

\clearpage

\section{Conclusions and Outlook}\label{CON}

A brief review of data-driven decompositions was presented, including the recently developed Multiscale Proper Orthogonal Decomposition (mPOD). This decomposition blends Multiresolution Analysis (MRA) and POD, generalizing POD and DMD in a single formulation, which combines the energy optimality of the first and the spectral purity of the second. The Multiscale Proper Orthogonal Decomposition (mPOD) has been tested on Time-Resolved PIV measurements of the flow past a cylinder in stationary and in transitory conditions. 

In stationary conditions, the dataset consists of $4000$ snapshots sampled at a frequency of $3kHz$ over an observation time of $1.33s$. The free stream velocity is kept constant at $U_{\infty}=12.1m/s$ producing the canonical von Karman vortex shedding at a frequency of $f_d\approx 450 Hz$. While the DMD algorithm correctly identifies the dominant frequency of the vortex shedding, the standard implementation does not converge to the dataset as all the computed modes vanish within a small fraction of the observation time. The DFT, on the other hand, allows detecting the dominant harmonic modes while still ensuring convergence, which is nevertheless rather poor and requires about $1000$ modes to reach a decomposition error of approximately $\mathcal{E}\approx 20\%$.
On the other extreme, such decomposition error is reached by the POD using only three modes. These capture the overall spatial structures of the vortex shedding and reveal an amplitude modulation whose energy content is scattered across the whole range of discrete frequencies. The mPOD proved capable of separating the two sources of such modulation, namely the contribution of harmonic phenomena evolving at a similar frequency, most probably linked to three-dimensional effects on the wake flows, from those linked to other contributions with a broader spectrum. This result is achieved while ensuring a decomposition convergence comparable to the POD, with $7$ modes sufficient to have a decomposition error of $\mathcal{E}\approx 20\%$. 

In the transient conditions, the dataset consists of $13500$ snapshots sampled over an observation time of $4.5s$. The free stream velocity is varied from $U_{\infty}=12.1m/s$ to $U_{\infty}=7.9m/s$, producing a change of frequency in the vortex shedding from $f_d\approx 450 Hz$ to $f_d\approx 300 Hz$. Because of the nonlinearity of such transitory, the DMD is unable to detect the relevant frequencies and to converge to the dataset. While the DFT is able to identify the shedding frequencies and to converge to the data (yet requiring $1230$ modes to achieve $\mathcal{E}\approx 20\%$), no time localization is possible due to the infinite duration of its basis.

The optimal convergence of the POD results in the opposite problem, with only three modes describing the entire dataset and producing $\mathcal{E}_{\mathcal{P}}\approx 3.4\%$. This is achieved at the cost of assigning to a single mode (and its phase-shifted counterpart) the whole evolution of the vortex shedding. The corresponding spatial structures are consequently a mix of features from these different phases of the flow evolution, which becomes indistinguishable. The spectral constraints of the mPOD allow for distinguishing the various steps in the evolution of the dataset both in the time and the frequency domains, assigning different modes accordingly. Thanks to the MRA architecture, this is achieved with a minor loss in the decomposition convergence with respect to the optimal POD, with only $7mPOD$ modes required to achieve $\mathcal{E}\approx 4\%$.

In conclusion, the results prove the enhanced feature detection capabilities of the mPOD on TR-PIV datasets and the significant enhancement advantages of combining energy optimality and spectral constraints. Future works are now aimed at testing different forms of MRA on the resulting mPOD, in implementing methods for the objective definition of the MRA scales and in developing a faster algorithm that hinges on the sparsity of the correlation matrices of each scale in the frequency domain.

\section*{Acknowledgements} 
The authors gratefully acknowledge an insightful discussion on the cylinder wake flow with Prof. Bernd Noack.

	\bibliography{Main}

\begin{thebibliography}{72}
\providecommand{\natexlab}[1]{#1}
\providecommand{\url}[1]{\texttt{#1}}
\expandafter\ifx\csname urlstyle\endcsname\relax
  \providecommand{\doi}[1]{doi: #1}\else
  \providecommand{\doi}{doi: \begingroup \urlstyle{rm}\Url}\fi

\bibitem[Amor et~al.(2019)Amor, P{\'{e}}rez, Schlatter, Vinuesa, and
  Clainche]{Amor2019}
Christian Amor, Jos{\'{e}}~M. P{\'{e}}rez, Philipp Schlatter, Ricardo Vinuesa,
  and Soledad~Le Clainche.
\newblock Soft computing techniques to analyze the turbulent wake of a
  wall-mounted square cylinder.
\newblock In \emph{Advances in Intelligent Systems and Computing}, pages
  577--586. Springer International Publishing, may 2019.
\newblock \doi{10.1007/978-3-030-20055-8_55}.

\bibitem[Aubry et~al.(1991)Aubry, Guyonnet, and Lima]{Aubry1991}
Nadine Aubry, R\'{e}gis Guyonnet, and Ricardo Lima.
\newblock Spatiotemporal analysis of complex signals: Theory and applications.
\newblock \emph{Journal of Statistical Physics}, 64\penalty0 (3-4):\penalty0
  683--739, aug 1991.
\newblock \doi{10.1007/bf01048312}.

\bibitem[Benner et~al.(2015)Benner, Gugercin, and Willcox]{Benner2015}
Peter Benner, Serkan Gugercin, and Karen Willcox.
\newblock A survey of projection-based model reduction methods for parametric
  dynamical systems.
\newblock \emph{{SIAM} Review}, 57\penalty0 (4):\penalty0 483--531, jan 2015.
\newblock \doi{10.1137/130932715}.

\bibitem[Berger et~al.(2014)Berger, Shea, Berry, Noack, Gogineni, and
  Glauser]{Berger_2014}
Zachary~P. Berger, Patrick~R. Shea, Matthew~G. Berry, Bernd~R. Noack, Sivaram
  Gogineni, and Mark~N. Glauser.
\newblock Active flow control for high speed jets with large window {PIV}.
\newblock \emph{Flow, Turbulence and Combustion}, 94\penalty0 (1):\penalty0
  97--123, nov 2014.
\newblock \doi{10.1007/s10494-014-9580-2}.

\bibitem[Bergmann and Cordier(2008)]{Bergmann2008}
M.~Bergmann and L.~Cordier.
\newblock Optimal control of the cylinder wake in the laminar regime by
  trust-region methods and {POD} reduced-order models.
\newblock \emph{Journal of Computational Physics}, 227\penalty0 (16):\penalty0
  7813--7840, aug 2008.
\newblock \doi{10.1016/j.jcp.2008.04.034}.

\bibitem[Berkooz et~al.(1993)Berkooz, Holmes, and Lumley]{Berkooz1993}
G~Berkooz, P~Holmes, and J~L Lumley.
\newblock The proper orthogonal decomposition in the analysis of turbulent
  flows.
\newblock \emph{Annual Review of Fluid Mechanics}, 25\penalty0 (1):\penalty0
  539--575, jan 1993.
\newblock \doi{10.1146/annurev.fl.25.010193.002543}.

\bibitem[Bishop(2011)]{Bishop2011}
Christopher~M. Bishop.
\newblock \emph{Pattern Recognition and Machine Learning}.
\newblock Springer New York, 2011.
\newblock ISBN 0387310738.

\bibitem[Bourgeois et~al.(2013)Bourgeois, Noack, and Martinuzzi]{Bourgeois2013}
J.~A. Bourgeois, B.~R. Noack, and R.~J. Martinuzzi.
\newblock Generalized phase average with applications to sensor-based flow
  estimation of the wall-mounted square cylinder wake.
\newblock \emph{Journal of Fluid Mechanics}, 736:\penalty0 316--350, nov 2013.
\newblock \doi{10.1017/jfm.2013.494}.

\bibitem[Brunton and Noack(2015)]{Brunton2015}
Steven~L. Brunton and Bernd~R. Noack.
\newblock Closed-loop turbulence control: Progress and challenges.
\newblock \emph{Applied Mechanics Reviews}, 67\penalty0 (5):\penalty0 050801,
  aug 2015.
\newblock \doi{10.1115/1.4031175}.

\bibitem[Cammilleri et~al.(2013)Cammilleri, Gueniat, Carlier, Pastur, Memin,
  Lusseyran, and Artana]{Camilleri}
A.~Cammilleri, F.~Gueniat, J.~Carlier, L.~Pastur, E.~Memin, F.~Lusseyran, and
  G.~Artana.
\newblock {POD}-spectral decomposition for fluid flow analysis and model
  reduction.
\newblock \emph{Theor. Comput. Fluid Dyn.}, 27\penalty0 (6):\penalty0 787--815,
  feb 2013.
\newblock \doi{10.1007/s00162-013-0293-2}.

\bibitem[Chen et~al.(2017)Chen, Zhou, Antonia, and Zhou]{Chen2017}
J.~G. Chen, Y.~Zhou, R.~A. Antonia, and T.~M. Zhou.
\newblock Characteristics of the turbulent energy dissipation rate in a
  cylinder wake.
\newblock \emph{Journal of Fluid Mechanics}, 835:\penalty0 271--300, nov 2017.
\newblock \doi{10.1017/jfm.2017.765}.

\bibitem[Chen et~al.(2012)Chen, Tu, and Rowley]{Chen2012}
Kevin~K. Chen, Jonathan~H. Tu, and Clarence~W. Rowley.
\newblock Variants of dynamic mode decomposition: Boundary condition, koopman,
  and fourier analyses.
\newblock \emph{Journal of Nonlinear Science}, 22\penalty0 (6):\penalty0
  887--915, apr 2012.
\newblock \doi{10.1007/s00332-012-9130-9}.

\bibitem[Citriniti and George(2000)]{CITRINITI2000}
J.~H. Citriniti and W.~K. George.
\newblock Reconstruction of the global velocity field in the axisymmetric
  mixing layer utilizing the proper orthogonal decomposition.
\newblock \emph{Journal of Fluid Mechanics}, 418:\penalty0 137--166, sep 2000.
\newblock \doi{10.1017/s0022112000001087}.

\bibitem[Ergin et~al.(2014)Ergin, Watz, Erglis, and Cebers]{Bo2}
F.~G. Ergin, B.B. Watz, K.~Erglis, and A.~Cebers.
\newblock Modal analysis of magnetic microconvection.
\newblock \emph{Magnetohydrodynamics}, 50\penalty0 (4):\penalty0 339--352,
  2014.

\bibitem[Erik et~al.(2007)Erik, Dalibor, and M.]{Meyer}
Meyer~Knud Erik, Cavar Dalibor, and Pedersen~Jakob M.
\newblock Pod as tool for comparison of piv and les data.
\newblock In \emph{7th International Symposium on Particle Image Velocimetry},
  Rome, Italy, September 2007.

\bibitem[Ghil(2002)]{Ghil2002}
M.~Ghil.
\newblock Advanced spectral methods for climatic time series.
\newblock \emph{Reviews of Geophysics}, 40\penalty0 (1), 2002.
\newblock \doi{10.1029/2000rg000092}.

\bibitem[Gronskis et~al.(2009)Gronskis, Adamo, Cammilleri, and
  Artana]{Gronskis2009}
A~Gronskis, J~D Adamo, A~Cammilleri, and G~Artana.
\newblock Reduced order models for wake control with a spinning cylinder.
\newblock \emph{Journal of Physics: Conference Series}, 166:\penalty0 012016,
  may 2009.
\newblock \doi{10.1088/1742-6596/166/1/012016}.

\bibitem[Hannachi et~al.(2007)Hannachi, Jolliffe, and Stephenson]{Hannachi2007}
A.~Hannachi, I.~T. Jolliffe, and D.~B. Stephenson.
\newblock Empirical orthogonal functions and related techniques in atmospheric
  science: A review.
\newblock \emph{International Journal of Climatology}, 27\penalty0
  (9):\penalty0 1119--1152, 2007.
\newblock \doi{10.1002/joc.1499}.

\bibitem[Harris(1978)]{Harris1978}
F.J. Harris.
\newblock On the use of windows for harmonic analysis with the discrete fourier
  transform.
\newblock \emph{Proceedings of the {IEEE}}, 66\penalty0 (1):\penalty0 51--83,
  1978.
\newblock \doi{10.1109/proc.1978.10837}.

\bibitem[Hasselmann(1988)]{Hasselmann1988}
K.~Hasselmann.
\newblock {PIPs} and {POPs}: The reduction of complex dynamical systems using
  principal interaction and oscillation patterns.
\newblock \emph{Journal of Geophysical Research}, 93\penalty0 (D9):\penalty0
  11015, 1988.
\newblock \doi{10.1029/jd093id09p11015}.

\bibitem[Holmes et~al.(1997)Holmes, Lumley, Berkooz, Mattingly, and
  Wittenberg]{Holmes}
P.~J. Holmes, J.~L. Lumley, G.~Berkooz, J.~C. Mattingly, and R.~W. Wittenberg.
\newblock Low-dimensional models of coherent structures in turbulence.
\newblock \emph{Phys. Rep.}, 287\penalty0 (4):\penalty0 337--384, aug 1997.
\newblock \doi{10.1016/s0370-1573(97)00017-3}.

\bibitem[Holmes et~al.(1996)Holmes, Lumley, and Berkooz]{Holmes1996}
Philip Holmes, John~L. Lumley, and Gal Berkooz.
\newblock \emph{Turbulence, Coherent Structures, Dynamical Systems and
  Symmetry}.
\newblock Cambridge University Press, 1996.
\newblock \doi{10.1017/cbo9780511622700}.

\bibitem[Huang et~al.(2006)Huang, Zhou, and Zhou]{Huang2006}
J.~F. Huang, Y.~Zhou, and T.~Zhou.
\newblock Three-dimensional wake structure measurement using a modified {PIV}
  technique.
\newblock \emph{Experiments in Fluids}, 40\penalty0 (6):\penalty0 884--896, may
  2006.
\newblock \doi{10.1007/s00348-006-0126-9}.

\bibitem[Hussain and Hayakawa(1987)]{Hussain1987}
A.~K. M.~Fazle Hussain and M.~Hayakawa.
\newblock Eduction of large-scale organized structures in a turbulent plane
  wake.
\newblock \emph{Journal of Fluid Mechanics}, 180\penalty0 (-1):\penalty0 193,
  jul 1987.
\newblock \doi{10.1017/s0022112087001782}.

\bibitem[Kutz et~al.(2016)Kutz, Fu, and Brunton]{MultiDMD}
J.~N. Kutz, X.~Fu, and S.~L. Brunton.
\newblock Multiresolution dynamic mode decomposition.
\newblock \emph{SIAM J. Appl. Dyn. Syst.}, 15\penalty0 (2):\penalty0 713--735,
  jan 2016.
\newblock \doi{10.1137/15m1023543}.

\bibitem[Lumley(1970)]{Lumley1}
John~L. Lumley.
\newblock \emph{Stochastic Tools in Turbulence}.
\newblock DOVER PUBN INC, 1970.
\newblock ISBN 0486462706.

\bibitem[Lumley and Poje(1997)]{Lumley1997}
John~L. Lumley and Andrew Poje.
\newblock Low-dimensional models for flows with density fluctuations.
\newblock \emph{Physics of Fluids}, 9\penalty0 (7):\penalty0 2023--2031, jul
  1997.
\newblock \doi{10.1063/1.869321}.

\bibitem[Mallat(2009)]{Wavelet1}
S.G. Mallat.
\newblock \emph{A Wavelet Tour of Signal Processing}.
\newblock Elsevier LTD, Oxford, 2009.
\newblock ISBN 0123743702.

\bibitem[Maurel et~al.(2001)Maurel, Bor{\'{e}}e1, and Lumley2]{Maurel12001}
S.~Maurel, J.~Bor{\'{e}}e1, and J.L. Lumley2.
\newblock Extended proper orthogonal decomposition: Application to jet/vortex
  interaction.
\newblock \emph{Flow, Turbulence and Combustion}, 67\penalty0 (2):\penalty0
  125--136, 2001.
\newblock \doi{10.1023/a:1014050204350}.

\bibitem[Mendez et~al.(2018{\natexlab{a}})Mendez, Balabane, and
  Buchlin]{Mendez_ICNAM}
M.~A. Mendez, M.~Balabane, and J.-M Buchlin.
\newblock Multi-scale proper orthogonal decomposition ({mPOD}).
\newblock In \emph{AIP Conference Proceedings}, 2018{\natexlab{a}}.
\newblock \doi{10.1063/1.5043720}.

\bibitem[Mendez et~al.(2018{\natexlab{b}})Mendez, Scelzo, and
  Buchlin]{Mendez_Journal_2}
M.~A. Mendez, M.T. Scelzo, and J.-M. Buchlin.
\newblock Multiscale modal analysis of an oscillating impinging gas jet.
\newblock \emph{Experimental Thermal and Fluid Science}, 91:\penalty0 256--276,
  feb 2018{\natexlab{b}}.
\newblock \doi{10.1016/j.expthermflusci.2017.10.032}.

\bibitem[Mendez et~al.(2019)Mendez, Balabane, and Buchlin]{Mendez2019}
M.~A. Mendez, M.~Balabane, and J.-M. Buchlin.
\newblock Multi-scale proper orthogonal decomposition of complex fluid flows.
\newblock \emph{Journal of Fluid Mechanics}, 870:\penalty0 988--1036. Preprint
  available at \url{https://arxiv.org/abs/1804.09646}., may 2019.
\newblock \doi{10.1017/jfm.2019.212}.

\bibitem[Mendez et~al.(2017)Mendez, Raiola, Masullo, Discetti, Ianiro,
  Theunissen, and Buchlin]{Mendez2017}
M.A. Mendez, M.~Raiola, A.~Masullo, S.~Discetti, A.~Ianiro, R.~Theunissen, and
  J.-M. Buchlin.
\newblock {POD}-based background removal for particle image velocimetry.
\newblock \emph{Experimental Thermal and Fluid Science}, 80:\penalty0 181--192,
  jan 2017.
\newblock \doi{10.1016/j.expthermflusci.2016.08.021}.

\bibitem[Misiti et~al.(2015)Misiti, Misiti, Oppenheim, and Poggi]{MATLAB}
Michel Misiti, Yves Misiti, Georges Oppenheim, and Jean-Michel Poggi.
\newblock \emph{Wavelet Toolbox User's Guide}.
\newblock Mathworks, 2015.

\bibitem[Murata et~al.(2019)Murata, Fukami, and Fukagata]{Murata2019}
Takaaki Murata, Kai Fukami, and Koji Fukagata.
\newblock Nonlinear mode decomposition with convolutional neural networks for
  fluid~dynamics.
\newblock \emph{Journal of Fluid Mechanics}, 882, nov 2019.
\newblock \doi{10.1017/jfm.2019.822}.

\bibitem[Ninni and Mendez(2020)]{Ninni}
Davide Ninni and Miguel~Alfonso Mendez.
\newblock Modulo: A software for multiscale proper orthogonal decomposition of
  data.
\newblock \emph{in Preparation}, 2020.
\newblock Codes available at \url{https://github.com/mendezVKI/MODULO}.
  Accessed: 2019-11-22.

\bibitem[Noack et~al.(2016)Noack, Stankiewicz, Morzy{\'{n}}ski, and
  Schmid]{Noack_RDMD}
B.~R. Noack, W.~Stankiewicz, M.~Morzy{\'{n}}ski, and P.~J. Schmid.
\newblock Recursive dynamic mode decomposition of transient and post-transient
  wake flows.
\newblock \emph{J. Fluid Mech.}, 809:\penalty0 843--872, nov 2016.
\newblock \doi{10.1017/jfm.2016.678}.

\bibitem[Noack(2016)]{Focus}
Bernd~R. Noack.
\newblock From snapshots to modal expansions {\textendash} bridging low
  residuals and pure frequencies.
\newblock \emph{J. Fluid Mech.}, 802:\penalty0 1--4, aug 2016.
\newblock \doi{10.1017/jfm.2016.416}.

\bibitem[Noack et~al.(2003)Noack, Afanasiev, Morzy{\'{n}}ski, Tadmor, and
  Thiele]{NOACK2003}
Bernd~R. Noack, Konstantin Afanasiev, Marek Morzy{\'{n}}ski, Gilead Tadmor, and
  Frank Thiele.
\newblock A hierarchy of low-dimensional models for the transient and
  post-transient cylinder wake.
\newblock \emph{Journal of Fluid Mechanics}, 497:\penalty0 335--363, dec 2003.
\newblock \doi{10.1017/s0022112003006694}.

\bibitem[Pawar et~al.(2019)Pawar, Rahman, Vaddireddy, San, Rasheed, and
  Vedula]{Pawar2019}
S.~Pawar, S.~M. Rahman, H.~Vaddireddy, O.~San, A.~Rasheed, and P.~Vedula.
\newblock A deep learning enabler for nonintrusive reduced order modeling of
  fluid flows.
\newblock \emph{Physics of Fluids}, 31\penalty0 (8):\penalty0 085101, aug 2019.
\newblock \doi{10.1063/1.5113494}.

\bibitem[Penland and Magorian(1993)]{LIM1}
Cécile Penland and Theresa Magorian.
\newblock Prediction of ni\'{n}o 3 sea surface temperatures using linear
  inverse modeling.
\newblock \emph{Journal of Climate}, 6\penalty0 (6):\penalty0 1067--1076, 1993.
\newblock \doi{10.1175/1520-0442(1993)006<1067:PONSST>2.0.CO;2}.

\bibitem[Penland(1996)]{Penland1996}
C{\'{e}}cile Penland.
\newblock A stochastic model of {IndoPacific} sea surface temperature
  anomalies.
\newblock \emph{Physica D: Nonlinear Phenomena}, 98\penalty0 (2-4):\penalty0
  534--558, nov 1996.
\newblock \doi{10.1016/0167-2789(96)00124-8}.

\bibitem[Petersson et~al.(2012)Petersson, Wellander, Olofsson, Carlsson,
  Carlsson, Watz, Boetkjaer, Richter, Ald\'{e}n, Fuchs, and Bai]{Bo1}
Per Petersson, Rikard Wellander, Jimmy Olofsson, Henning Carlsson, Christian
  Carlsson, Bo~Beltoft Watz, Nicolas Boetkjaer, Mattias Richter, Marcus
  Ald\'{e}n, Laszlo Fuchs, and Xue-Song Bai.
\newblock Simultaneous high-speed piv and oh plif measurements and modal
  analysis for investigating flame-flow interaction in a low swirl flame.
\newblock In \emph{16th In Symp on Applications of Laser Techniques to Fluid
  Mechanics}, Lisbon, Portugal, July 2012.

\bibitem[Rabault et~al.(2019)Rabault, Kuchta, Jensen, R{\'{e}}glade, and
  Cerardi]{Rabault2019}
Jean Rabault, Miroslav Kuchta, Atle Jensen, Ulysse R{\'{e}}glade, and Nicolas
  Cerardi.
\newblock Artificial neural networks trained through deep reinforcement
  learning discover control strategies for active flow control.
\newblock \emph{Journal of Fluid Mechanics}, 865:\penalty0 281--302, feb 2019.
\newblock \doi{10.1017/jfm.2019.62}.

\bibitem[Raiola et~al.(2015)Raiola, Discetti, and Ianiro]{Raiola2015}
Marco Raiola, Stefano Discetti, and Andrea Ianiro.
\newblock On {PIV} random error minimization with optimal {POD}-based low-order
  reconstruction.
\newblock \emph{Experiments in Fluids}, 56\penalty0 (4), mar 2015.
\newblock \doi{10.1007/s00348-015-1940-8}.

\bibitem[Rao(2010)]{Rao}
S.~S. Rao.
\newblock \emph{Mechanical Vibrations}.
\newblock Pearson, 5th edition, 2010.

\bibitem[Reynolds and Hussain(1972)]{Reynolds1972}
W.~C. Reynolds and A.~K. M.~F. Hussain.
\newblock The mechanics of an organized wave in turbulent shear flow. part 3.
  theoretical models and comparisons with experiments.
\newblock \emph{Journal of Fluid Mechanics}, 54\penalty0 (2):\penalty0
  263--288, jul 1972.
\newblock \doi{10.1017/s0022112072000679}.

\bibitem[Rowley and Dawson(2017)]{Rowley}
C.~W. Rowley and S.~T.M. Dawson.
\newblock Model reduction for flow analysis and control.
\newblock \emph{Annu. Rev. Fluid Mech.}, 49\penalty0 (1):\penalty0 387--417,
  jan 2017.
\newblock \doi{10.1146/annurev-fluid-010816-060042}.

\bibitem[Rowley et~al.(2004)Rowley, Colonius, and Murray]{Rowley2004}
Clarence~W. Rowley, Tim Colonius, and Richard~M. Murray.
\newblock Model reduction for compressible flows using {POD} and galerkin
  projection.
\newblock \emph{Physica D: Nonlinear Phenomena}, 189\penalty0 (1-2):\penalty0
  115--129, feb 2004.
\newblock \doi{10.1016/j.physd.2003.03.001}.

\bibitem[Rowley et~al.(2009)Rowley, Mezi\'{c}, Bagheri, Schlatter, and
  Henningson]{Rowley2}
C.W. Rowley, I.~Mezi\'{c}, S.~Bagheri, P.~Schlatter, and D.S. Henningson.
\newblock Spectral analysis of nonlinear flows.
\newblock \emph{J. Fluid Mech.}, 641:\penalty0 115, nov 2009.
\newblock \doi{10.1017/s0022112009992059}.

\bibitem[Schmid(2010)]{Schmid}
P.J. Schmid.
\newblock Dynamic mode decomposition of numerical and experimental data.
\newblock \emph{J. Fluid Mech.}, 656:\penalty0 5--28, jul 2010.
\newblock \doi{10.1017/s0022112010001217}.

\bibitem[Sieber et~al.(2016)Sieber, Paschereit, and Oberleithner]{SPOD}
Moritz Sieber, C.~Oliver Paschereit, and Kilian Oberleithner.
\newblock Spectral proper orthogonal decomposition.
\newblock \emph{J Fluid Mech}, 792:\penalty0 798--828, mar 2016.
\newblock \doi{10.1017/jfm.2016.103}.

\bibitem[Siegel et~al.(2008)Siegel, Seidel, Fagley, Luchtenburg, Cohen, and
  Mclaughlin]{SIEGEL2008}
Stefan~G. Siegel, Jorgen Seidel, Casey Fagley, D.~M. Luchtenburg, Kelly Cohen,
  and Thomas Mclaughlin.
\newblock Low-dimensional modelling of a transient cylinder wake using double
  proper orthogonal decomposition.
\newblock \emph{Journal of Fluid Mechanics}, 610:\penalty0 1--42, aug 2008.
\newblock \doi{10.1017/s0022112008002115}.

\bibitem[Sirovich(1987)]{Siro1}
L.~Sirovich.
\newblock Turbulence and the dynamics of coherent structures: Part i. coherent
  structures.
\newblock \emph{Quart. Appl. Math}, 45\penalty0 (3):\penalty0 561--571, 1987.
\newblock \doi{https://doi.org/10.1090/qam/910462}.

\bibitem[Sirovich(1989)]{Siro2}
L.~Sirovich.
\newblock Chaotic dynamics of coherent structures.
\newblock \emph{Physica D}, 37\penalty0 (1):\penalty0 126--145, 1989.
\newblock \doi{https://doi.org/10.1016/0167-2789(89)90123-1}.

\bibitem[Smith(2007)]{DFT}
Julius~O. Smith.
\newblock \emph{Mathematics of the Discrete Fourier Transform (DFT): with Audio
  Applications}.
\newblock W3K Publishing; 2 edition, April 2007.
\newblock ISBN 978-0974560748.

\bibitem[Sung and Yoo(2001)]{Sung2001}
J~Sung and J~Y Yoo.
\newblock Three-dimensional phase averaging of time-resolved {PIV} measurement
  data.
\newblock \emph{Measurement Science and Technology}, 12\penalty0 (6):\penalty0
  655--662, may 2001.
\newblock \doi{10.1088/0957-0233/12/6/301}.

\bibitem[Tadmor et~al.(2010)Tadmor, Lehmann, Noack, and
  Morzy{\'{n}}ski]{Tadmor2010}
Gilead Tadmor, Oliver Lehmann, Bernd~R. Noack, and Marek Morzy{\'{n}}ski.
\newblock Mean field representation of the natural and actuated cylinder wake.
\newblock \emph{Physics of Fluids}, 22\penalty0 (3):\penalty0 034102, mar 2010.
\newblock \doi{10.1063/1.3298960}.

\bibitem[Taira et~al.(2017)Taira, Brunton, Dawson, Rowley, Colonius, McKeon,
  Schmidt, Gordeyev, Theofilis, and Ukeiley]{Taira}
K.~Taira, S.~L. Brunton, S.~T.~M. Dawson, C.~W. Rowley, T.~Colonius, B.~J.
  McKeon, O.~T. Schmidt, S.~Gordeyev, V.~Theofilis, and L.~S. Ukeiley.
\newblock Modal analysis of fluid flows: An overview.
\newblock \emph{{AIAA} J.}, 55\penalty0 (12):\penalty0 4013--4041, dec 2017.
\newblock \doi{10.2514/1.j056060}.

\bibitem[Theunissen(2010)]{RAF}
Raf Theunissen.
\newblock \emph{Adaptive resolution in PIV image analysis - Application to
  complex flows and interfaces}.
\newblock PhD thesis, von Karman Institute for Fluid Dynamics \& Technische
  Universiteit Delft, The Netherlands \& Vrije Universiteit Brussel, Belgium,,
  2010.

\bibitem[Towne et~al.(2018)Towne, Schmidt, and Colonius]{Towne2018}
Aaron Towne, Oliver~T. Schmidt, and Tim Colonius.
\newblock Spectral proper orthogonal decomposition and its relationship to
  dynamic mode decomposition and resolvent analysis.
\newblock \emph{Journal of Fluid Mechanics}, 847:\penalty0 821--867, may 2018.
\newblock \doi{10.1017/jfm.2018.283}.

\bibitem[Tu et~al.(2014)Tu, Rowley, Luchtenburg, Brunton, and Kutz]{Kutz2014}
Jonathan~H. Tu, Clarence~W. Rowley, Dirk~M. Luchtenburg, Steven~L. Brunton, and
  J.~Nathan Kutz.
\newblock On dynamic mode decomposition: Theory and applications.
\newblock \emph{Journal of Computational Dynamics}, 1\penalty0 (2):\penalty0
  391--421, dec 2014.
\newblock \doi{10.3934/jcd.2014.1.391}.

\bibitem[Uruba(2012)]{Vlacav1}
V{\'{a}}clav Uruba.
\newblock Decomposition methods in turbulence research.
\newblock \emph{{EPJ} Web of Conferences}, 25:\penalty0 01095, 2012.
\newblock \doi{10.1051/epjconf/20122501095}.

\bibitem[Uruba(2015)]{Vlacav2}
V{\'{a}}clav Uruba.
\newblock Near wake dynamics around a vibrating airfoil by means of {PIV} and
  oscillation pattern decomposition at reynolds number of 65 000.
\newblock \emph{Journal of Fluids and Structures}, 55:\penalty0 372--383, may
  2015.
\newblock \doi{10.1016/j.jfluidstructs.2015.03.011}.

\bibitem[Uruba and Proch{\'{a}}zka(2019)]{Vlacav3}
V{\'{a}}clav Uruba and Pavel Proch{\'{a}}zka.
\newblock On interpretation of spatiotemporal data decomposition.
\newblock In \emph{15th International Conference on Fluid Control, Measurements
  and Visualization}, 2019.

\bibitem[von Storch and Xu(1990)]{Storch1990}
Hans von Storch and Jinsong Xu.
\newblock Principal oscillation pattern analysis of the 30- to 60-day
  oscillation in the tropical troposphere.
\newblock \emph{Climate Dynamics}, 4\penalty0 (3):\penalty0 175--190, sep 1990.
\newblock \doi{10.1007/bf00209520}.

\bibitem[Welch(1967)]{Welch1967}
P.~Welch.
\newblock The use of fast fourier transform for the estimation of power
  spectra: A method based on time averaging over short, modified periodograms.
\newblock \emph{{IEEE} Transactions on Audio and Electroacoustics}, 15\penalty0
  (2):\penalty0 70--73, jun 1967.
\newblock \doi{10.1109/tau.1967.1161901}.

\bibitem[Westerweel and Scarano(2005)]{Westerweel2005}
Jerry Westerweel and Fulvio Scarano.
\newblock Universal outlier detection for {PIV} data.
\newblock \emph{Experiments in Fluids}, 39\penalty0 (6):\penalty0 1096--1100,
  aug 2005.
\newblock \doi{10.1007/s00348-005-0016-6}.

\bibitem[Williamson(1996)]{Williamson1996}
C~H~K Williamson.
\newblock Vortex dynamics in the cylinder wake.
\newblock \emph{Annual Review of Fluid Mechanics}, 28\penalty0 (1):\penalty0
  477--539, jan 1996.
\newblock \doi{10.1146/annurev.fl.28.010196.002401}.

\bibitem[Wu et~al.(1996)Wu, Sheridan, Welsh, and Hourigan]{Wu1996}
J.~Wu, J.~Sheridan, M.~C. Welsh, and K.~Hourigan.
\newblock Three-dimensional vortex structures in a cylinder wake.
\newblock \emph{Journal of Fluid Mechanics}, 312:\penalty0 201--222, apr 1996.
\newblock \doi{10.1017/s0022112096001978}.

\bibitem[Zhang et~al.(2000)Zhang, Zhou, and Antonia]{Zhang2000}
H.~J. Zhang, Y.~Zhou, and R.~A. Antonia.
\newblock Longitudinal and spanwise vortical structures in a turbulent near
  wake.
\newblock \emph{Physics of Fluids}, 12\penalty0 (11):\penalty0 2954, 2000.
\newblock \doi{10.1063/1.1309532}.

\bibitem[Zhou et~al.(2003)Zhou, Zhou, Yiu, and Chua]{Zhou2003}
T.~Zhou, Y.~Zhou, M.~W. Yiu, and L.~P. Chua.
\newblock Three-dimensional vorticity in a turbulent cylinder wake.
\newblock \emph{Experiments in Fluids}, 35\penalty0 (5):\penalty0 459--471, nov
  2003.
\newblock \doi{10.1007/s00348-003-0700-3}.

\end{thebibliography}

\end{document}